\documentclass[aps,pra,singlecolumn,superscriptaddress,showpacs,floatfix]{revtex4-2}
\DeclareUnicodeCharacter{2009}{\,}
\usepackage{amsthm}
\usepackage{palatino}
\usepackage{mathpazo}
\usepackage{bbm}
\usepackage{braket}
\usepackage{amsfonts}
\usepackage{amssymb}
\usepackage{amsmath}
\usepackage{latexsym}
\usepackage{amsthm}
\usepackage[usenames]{color}
\usepackage[colorlinks=true, linkcolor=red, urlcolor=red, citecolor=red]{hyperref}
\usepackage{todonotes}
\usepackage{bbm}
\usepackage{algorithm}
\usepackage{algpseudocode}
\usepackage{tikz}
\numberwithin{equation}{section}
\usepackage{cooltooltips}
\theoremstyle{definition}

\theoremstyle{definition}

\usepackage{color,graphicx}
\newcommand{\C}{\mathbb{C}}

\newcommand{\I}{\mathbb{I}} 

\newcommand{\X}{\mathcal{X}}
\newcommand{\Y}{\mathcal{Y}}
\newcommand{\D}{\mathcal{D}}

\newcommand{\Tr}{\textrm{Tr}} 
 
\newcommand{\Herm}{\mathrm{Herm}}

\renewcommand{\to}{\longrightarrow}
\algblock{ParFor}{EndParFor}
\algnewcommand\algorithmicparfor{\textbf{parallel\ for}}
\algnewcommand\algorithmicpardo{\textbf{do}}
\algnewcommand\algorithmicendparfor{\textbf{end\ parallel\ for}}
\algrenewtext{ParFor}[1]{\algorithmicparfor\ #1\ \algorithmicpardo}
\algrenewtext{EndParFor}{\algorithmicendparfor}

\begin{document}
\title{Modified security analysis of device-independent quantum key distribution with random key basis}
\author{Sawan Bhattacharyya}
\affiliation{Department of Computer Science and Engineering, University of Calcutta, Bidhan Nagar, Kolkata - 700106, India}
\email{sawanbhattacharyyakv@gmail.com}
\affiliation{Center for Quantum Engineering, Research, and Education, TCG CREST, Bidhan Nagar, Kolkata - 700091, India.}
\author{Turbasu Chatterjee}
\affiliation{Department of Computer Science, Virginia Tech, Blacksburg, VA 24060, United States}
\author{Pankaj Agrawal}
\affiliation{Center for Quantum Engineering, Research, and Education, TCG CREST, Bidhan Nagar, Kolkata - 700091, India.}
\author{Prasenjit Deb}
\affiliation{Center for Quantum Engineering, Research, and Education, TCG CREST, Bidhan Nagar, Kolkata - 700091, India.}

\date{\today}

\begin{abstract}
Security analysis is a critical part in any cryptographic protocol, may it be classical or quantum. Without security analysis, one cannot ensure the secrecy of the distributed keys. To perform a conclusive security analysis, it is very often necessary to frame the problem as an optimization problem. However, solving such optimization problems is quite challenging. In this article, we focus on the security analysis of device-independent quantum key distribution (DIQKD) with random key basis protocol. We show that the optimization cost of the existing security analysis can be reduced without compromising the key rate. In particular, we reframe the entire security analysis of this protocol as a strongly convex optimization problem and demonstrate that unlike the original security proof, optimization of Bob's measurement angles for finding a lower bound on Eve's uncertainty about Alice's key generation basis can be done with lesser cost. We derive an explicit form of the pessimistic error that arises while optimizing the measurement angles of both the parties. We also clarify a few parts of the original security proof, making the analysis more rigorous and complete .  
 
\end{abstract}
\maketitle


\section{Introduction}\label{sec1}
Cryptographic protocols \cite{mammeri2024cryptography} aid in establishing secure keys between two communicating parties, who later use the keys to access the information that they share between them. However, before implementing such protocols, it is necessary to perform their security analysis for avoiding any kind of potential unauthorized access of information during communication. In particular, through a conclusive security analysis we can find out the necessary and sufficient conditions for generating a positive secret key rate \cite{renner2006securityquantumkeydistribution} as well as its optimal lower bound. However, finding such bounds often involves optimization of cost functions that are derived by mapping the security analysis to an optimization problem. 
\paragraph*{}
Quantum cryptography offers non-trivial advantages over its classical counterpart due to the inherent randomness present in the state of the carrier of quantum information -- qubit \cite{Nielsen_Chuang_2010}. Moreover, the nonlocality \cite{locality} feature of quantum mechanics enables us to design quantum key distribution protocols that do not rely on any assumptions
about the internal working of the quantum devices used in the protocols. Such key distribution protocols are termed as device-independent quantum key distribution (DIQKD) \cite{acindiqkd}. Allowing relaxation of the security assumptions made in usual quantum key distribution(QKD) \cite{renner2006securityquantumkeydistribution}, DIQKD has made it possible to distribute secret keys using untrusted devices between two users embedded in an untrusted network. Due to this counterintuitive advantage over classical key distribution, this quantum cryptographic protocol is considered as the future of key distribution.
\paragraph*{}
In this quantum key distribution protocol, the nonlocal state shared between the communicating parties limits the information that can be obtained by any third party. However, certification of nonlocality between the legitimate parties requires verification of violation of Clauser-Horne-Shimony-Holt (CHSH) inequality. For such a purpose, a sufficiently large number of entangled states has to be shared between the parties. This requirement is also necessary for generation of secure keys. However, for the possibility of secure key distribution, mere violation of the CHSH inequality with a large number of shared highly entangled state is not sufficient -- the amount of violation must be above a certain threshold that depends on the protocol as well as on the security proofs\cite{Pironio_2009}. Until the discovery of three variants of DIQKD \cite{Schwonnek_2021},\cite{Xu_2022},\cite{Ho_2020} unfortunately, the best known threshold so far was beyond the range achievable by the present day state-of-the-art experiments despite of the methods, e.g., heralding-type solutions \cite{gisinheral},\cite{Heralded-qubit-amplifiers},\cite{Kolodynski2020deviceindependent}, local precertification \cite{precertification1},\cite{precertification2}, local Bell tests \cite{local_bell}, and so on, proposed to improve the robustness of DIQKD. One such variant of this quantum key distribution protocol is DIQKD with random key basis \cite{Schwonnek_2021}. As the name suggests, instead of a single measurement setting, more than one measurement settings are used for generating secret keys. The key idea is to add additional difficulty for the eavesdropper by increasing the uncertainty about the measurement settings used to generate secret keys.
In this article, we focus on the security analysis of this quantum key distribution protocol. We find that the optimization technique used in the existing security analysis to find secret key rate can be simplified, thereby reducing its computational cost. By framing the problem as a strongly convex optimization problem, we show that Bob's measurement angles can be optimized with lesser computational cost. We also find a closed form of the pessimistic error arising while optimizing the measurement angles of the parties Alice and Bob. Moreover, we address few other points in the original security analysis that demand clarification.
\paragraph*{}
The rest of the article is arranged as follows. In Section (\ref{sec2}), we briefly describe the quantum key distribution protocol that we analyze. Section (\ref{sec3}) contains the results and discussions. Finally, in Section (\ref{sec4}), we conclude our work.
\section{The protocol}\label{sec2}
In the original DIQKD protocol \cite{acindiqkd}, Eve knows the predetermined key generation basis. In particular, let Alice's and Bob's measurement settings are denoted by $X\in\{0, 1\}$ and $Y\in\{0, 1, 2\}$, respectively. The corresponding outcomes are given by $A_X\in\{0, 1\}$ and $B_Y\in\{0, 1\}$, respectively. They derive the secret key from the measurement rounds in which the settings $A=0$ and $B=0$, respectively, are chosen. The CHSH score is determined using the remaining measurement combinations. Unlike this setting, in DIQKD with random key basis protocol, the secret is generated using both the measurement settings of Alice. Therefore, to obtain correlated outcomes, Bob needs an extra measurement setting, which is the proposed modification of the original protocol. Hence, in this variant of DIQKD protocol, Alice and Bob choose the settings $X=Y=0$ and $X=Y=1$ for generating secret key. The key intuition behind this protocol follows from the fact that the requirement of incompatible measurements for violation of the CHSH inequality necessitates the Alice's key generation measurements to be incompatible when the CHSH score $S>2$. This condition imposes additional difficulty on Eve because now she has to guess the secret key from two randomly chosen incompatible measurements. For CHSH-based quantum key distribution protocols like DIQKD, different measurements will generate different amount of side information for Eve \cite{Schwonnek_2021}. Therefore, Eve will fail to maximize her side information with at least one of the measurements, which in turn, provides an advantage over protocols based only on a single key-generating measurement. Due to her ignorance about which measurement setting will be chosen by Alice for key generation \textit{a priori}, she will not be able to tailor her attack to the measurement in each round individually.
\paragraph*{}
The protocol can be divided into five parts -- measurements, sifting, parameter estimation, one-way error correction and verification, and finally privacy amplification. The measurement step is carried out with asymptotically large number of rounds $N$. In each measurement round, Alice and Bob choose their measurement settings $X\in\{0,1\}$ and $Y\in\{0,1,2,3\}$, respectively, following the probability distributions as defined: $P(X=0)=p$, $P(X=1)=1-p$, $P(Y=0)=qp$, $P(Y=1)=q(1-p)$ and 
$P(Y=2)=P(Y=3)=(1-q)/2$, where $0 \leq p$, $q \leq 1$. Once Alice and Bob perform measurements with their respective devices and settings, they obtain outcomes $A_X\in\{0, 1\}$ and $B_Y\in \{0, 1\}$, respectively.  They begin the second step of the protocol by announcing their measurement inputs over an authenticated public channel. Then using the inputs $Y\in\{0,1\}$ and $X = Y$ and the corresponding outcomes, they  generate a pair of raw key of size $\sim q(p^2 + (1-p)^2)N$. Alice and Bob use the measurement settings $X\in\{0,1\}$ and $Y\in\{2,3\}$ to extract a pair of parameter estimation data of size $\sim (1-q)N$. All other measurement data are discarded. The third step of this key distribution protocol aims at computing and estimating the CHSH value. To do so, Alice and Bob reveal their measurement outcomes from the parameter estimation data set. The CHSH parameter can be expressed as,
\begin{equation}
  S = \mbox{max}\{0, C_{12}-C_{02}- C_{03}-C_{13}\}  
\end{equation}
where $C_{XY}=P(A_X=B_Y|X,Y) - P(A_X\neq B_Y|X,Y)$ quantifies the correlations between measurement outcomes $A_X$ and $B_Y$. This step comprises the determining factor for continuation of the entire process. If $S>S_{tol}$, then Alice and Bob proceed to the next step, otherwise, they abort the protocol. Here $S_{tol}$ is a predetermined threshold value of the violation of the CHSH inequality. A violation of the inequality beyond the threshold value confirms that the generated raw keys are secure. After getting this confirmation, Alice and Bob perform one-way error correction and verification, followed by privacy amplification.
\paragraph*{}
Just like other DIQKD protocols, the security of this protocol can be quantified using the asymptotic secret key rate. However, due to the use of two measurement key bases for key generation, the key rate is defined in a way different from the existing DIQKD protocols,
\begin{equation}
    K_{\infty} = p_sr_{\infty}
\end{equation}
where $p_s := p^2 + (1-p)^2$ is the probability of Alice and Bob having matching key bases in the limit $q\rightarrow 1$. This key rate is the ratio of the extractable secret key length to the total number of measurement rounds $N$, which generally tends to infinity, i.e., $N\rightarrow\infty$. The second product term in the above equation is the secret fraction \cite{Schwonnek_2021} and it can be written in terms of entropic functions as,
\begin{eqnarray}
    r_{\infty} &:=& \lambda H(A_0|E) + (1-\lambda)H(A_1|E)\nonumber\\
    &&- \lambda h(Q_{A_0B_0}) - (1-\lambda) h(Q_{A_1B_1})
\end{eqnarray}
where $\lambda := p^2/p_s$ and $h(x):=-x\mbox{log}(x)-(1-x)\mbox{log}(1-x)$ is the binary entropy function. The quantity in the argument of the binary function quantifies the quantum bit-error rate (QBER) for $X,Y$ and is given by $Q_{A_XB_Y}:=P(A_X\neq B_Y|X,Y)$. The quantum side-information that Eve can gather just before the error-correction step is given by $E$. From the secret fraction we can evaluate how much information Eve can get about Alice's and Bob's measurement outcomes. The terms, conditional von Neumann entropy, comprising the first line of Eq.(3) measure the amount of uncertainty Eve has about Alice's measurement outcomes given her side-information $E$. The entropic terms in the second line of the equation quantify the information that Eve can steal during the error-correction step by decoding her side-information. If we denote $U=\lambda H(A_0|E) + (1-\lambda)H(A_1|E)$, then Alice and Bob can generate a positive secret key rate, provided they can put a reliable lower bound on $U$ using solely the CHSH violation that they observe. This task is quiet challenging, however, it has been shown that by employing a family of device-independent entropic uncertainty relations\cite{Berta2010}\cite{entropic_uncertianity2}, one can have\cite{Schwonnek_2021},
\begin{equation}
    \lambda H(A_0|E) + (1-\lambda)H(A_1|E)\geq C^{\star}(S)
\end{equation}
where $C^{\star}$ is a function of the CHSH score $S$. Using $U$ as a performance metric, it has been found that the uncertainty of Eve in this protocol is always higher than the original protocol for all $S\in (2,2\sqrt{2}]$ \cite{Schwonnek_2021}. In fact, given a fixed $S$, for $\lambda =\frac{1}{2}$, the bound on $U$ converges to its nearly optimal value, which is the fundamental upper limit of Eve's uncertainty \cite{Schwonnek_2021}.  In this work, our goal is to reduce the optimization cost of establishing a lower bound on Eve's uncertainty without compromising the secret key rate.
\section{Results and discussions}\label{sec3}

Computation of the function $C^{\star}(S)$ involves a refined version of Pinsker’s inequality \cite{markmwilde}, semi-definite optimization (SDP)\cite{sikora2012analyzing}, and an $\epsilon$-net approximation \cite{Schwonnek_2021}. In the original security analysis of the protocol, to find a lower bound on the function $C^{\star}(S)$, three optimization steps have been followed -- optimization of the joint state $\rho$ between Alice and Bob for fixed measurement angles on both sides, optimization of Bob's angle, and optimization of Alice's angles. All the steps in the original analysis assume the worst-case scenario so that the final value is a strict lower bound on the function. Here, we provide an improvised security analysis of the protocol.
\paragraph*{}
It has been found that Alice's and Bob's measurement angles $\phi_A^i  \text{ and } \phi_B^j$ appear as constraints in the optimization problem -- the SDP used to find out a lower bound on the function $C^{\star}(S)$. The work of \cite{Schwonnek_2021} showed that the Alice's angle $\phi_A^i$ can be optimized using $\epsilon$ - net approach. Here, we show that one can use a similar approach to optimize the Bob angle $\phi_B^j$, thus eliminating the polytope optimization used in the original protocol. The polytope optimization step introduced in the original security analysis initialised the polytope $\mathcal{P}$ as a rectangle. $\mathcal{P}$ contain the unit circle \cite{Schwonnek_2021} of all the point that depends on Bob's angle through $\textit{sin} \text{ and }\textit{cos}$ function. In this work, we extend the idea of the $\epsilon-$net approach to optimize Bob's angle, thereby making the overall optimization process less costly.

The standard cryptographic security analysis assume that the measurement angles of Alice and Bob are independent variables. To begin with, let us divide  the possible range of each angle, $[0,\frac{\pi}{2}]$, into a set of discrete points, $\phi_{A_k}^i$ for Alice and $\phi_{B_l}^j$ for Bob, and choose the midpoint of each segment as $k$ and $l$.
Recognizing that the optimal angle for either Alice or Bob could lie anywhere within the continuous range $[0,\frac{\pi}{2}]$, a pessimistic error approach has been adopted. This involves iteratively refining the angle for one party (either Alice or Bob) while keeping the other's angle fixed.
In each iteration, the SDP is being evaluated at the midpoints of the current segments for both angles. The segment that yields the smallest SDP value is then chosen for further subdivision. This process continues until a point corresponding to a local minimum is reached for the chosen angle. During this minimization for one angle, a pessimistic error estimate is subtracted from the SDP result to account for the continuous nature of the angle.
Once the optimal angle for one party is determined through this iterative refinement, this optimal value is then fixed. The same iterative optimization process is then applied to find the optimal angle for the other party, now considering the previously determined optimal angle as a constant. This sequential optimization allows them to find a potentially optimal solution for the SDP by iteratively minimizing with respect to one angle at a time.

Given the interval $I = [0, \pi/2]$ and a desired precision $\epsilon_0 > 0$, without loss of generality, an $\epsilon_0$-net for the product space $I \times I$ is a pair of finite sets of points $\{\phi_{A_{k}}^i\}_{k=1}^{S_A} \subset I$ and $\{\phi_{B_{l}}^j\}_{l=1}^{S_B} \subset I$ such that for any $(\phi_A^i, \phi_B^j) \in I \times I$, there exist $\phi_{A_{k}}^i$ and $\phi_{B_{l}}^j$ satisfying:
\begin{equation}
     |\phi_A^i - \phi_{A_{k}}^i| \le \epsilon_0 \quad
    |\phi_B^j - \phi_{B_{l}}^j| \le \epsilon_0
\end{equation}
where $S_A \text{ and } S_B$ are the number of segments in the interval $I$. Each segment is centralized around the angles $\phi_{A_{k}}^i \text{ and } \phi_{B_{l}}^j$ for $k^{th} \text{ and } l^{th}$ segments respectively. The pessimistic error term was introduced in the original protocol \cite{Schwonnek_2021} as a function of change $\Delta$ in $\epsilon_0\text{ and } \phi_A^i (\phi_B^j)$.    
In this work, a closed form of the same is derived.

Let us consider that $f(\phi)$ is a solution of the optimization problem for given angle.
Each discrete angle $\phi_{A_{k}}^i\text{ and } \phi_{B_{l}}^j$ is being separated by a distance of $2\epsilon_0$, thus representing two segments of the same width as,
\begin{equation}
\label{eqn5.3}
    \begin{aligned}
       I_A =  &\left[\phi_{A_{k}}^i - \epsilon_0, \phi_{A_{k}}^i + \epsilon_0 \right]\\
       I_B = &\left[\phi_{B_{l}}^j - \epsilon_0, \phi_{B_{l}}^j + \epsilon_0 \right]
    \end{aligned}
\end{equation}
The pessimistic error terms $\Delta(\epsilon_0,\phi_A^i)$ and $\Delta(\epsilon_0,\phi_B^j)$ provide an upper bound on the absolute difference between the values of the function at any point within these segments and at the representative discrete point. Hence, from the definition of $f(\phi)$, we have, 
\begin{equation}
\label{eqn5.4}
    \begin{aligned}        
       &|f(\phi_A^i) - f(\phi_{A_{k}}^i)| \leq \Delta\left (\epsilon_0,\phi_A^i\right)\\
       &|f(\phi_B^i) - f(\phi_{B_{l}}^j)| \leq \Delta\left (\epsilon_0,\phi_B^j\right),~~~~~~~\forall \phi_A^i \in I_A
       \text{ and }\forall \phi_B^j \in I_B \\
    \end{aligned}
\end{equation}

\subsection{Modification of the optimization problem accommodating Lipschitz continuity }
Proving the below mentioned theorem, we first modify the optimization problem.

\textit{Theorem} : If $ f(\phi) $ is a solution of an well-defined optimization problem, then it is Lipschitz continuous, i.e., there exists a constant $  L_f > 0 $ such that,
\begin{equation}\label{eqn4}
\| f(\phi_1) - f(\phi_2)\| \leq L_f \|\phi_1 - \phi_2\|.
\end{equation}
The proof is provided in the Appendix section \ref{F1}. To make the objective function continuously differentiable with respect to both the density operator and the measurement angles, the squared Frobenius norm is introduced. The squared Frobenius norm is smooth and continuously differentiable at every point\cite{Petersen2008}.
To make the objective function strongly convex, it is necessary to add a strongly convex regularisation term. 
Therefore, the optimization problem in the original security analysis of the protocol now becomes,
\begin{equation}
\boxed{
    \begin{aligned}
n^*(S_{ij}) = \inf \quad & \lambda \left\| \rho_{AB}^{ij} - \Lambda_0[\rho_{AB}^{ij}] \right\|_F^2 + (1-\lambda)\left\| \rho_{AB}^{ij} - \Lambda_1[\rho_{AB}^{ij}] \right\|_F^2 + \frac{\mu}{2} \|\rho_{AB}^{ij}\|_F^2 \\
\text{s.t.} \quad &\text{Tr}\left(\rho_{AB}^{ij} ~ CHSH(\phi_A^i,\phi_B^j)\right) = S_{ij} \\
& \phi_A^i, \phi_B^j \in [0,\pi/2],\\
    &\rho_{AB}^{ij} \succeq 0 \\ &\text{Tr}(\rho_{AB}^{ij}) = 1\\
\end{aligned}}
\end{equation}
The modified version of the optimization problem can be reformulated into an SDP using Schur complements \cite{ben2001lectures}. Let us first express each Frobenius norm terms into the corresponding inner product form as,
\begin{equation}
\label{eqn5.27}
    \begin{aligned}
        n^*(S_{ij}) = -\text{maximize} \quad & \lambda \mbox{Tr} ({\rho_0}^{\dagger}\rho_0)  + (1-\lambda)\mbox{Tr} ({\rho_1}^{\dagger}\rho_1) + \frac{\mu}{2} \mbox{Tr}({\rho_2}^{\dagger}\rho_2) \\
        \text{s.t.} \quad &\text{Tr}\left(\rho_{AB}^{ij} ~ CHSH (\phi_A^i,\phi_B^j)\right) = S_{ij}\\
& \phi_A^i, \phi_B^j \in [0,\pi/2],\\
    &\rho_{AB}^{ij} \succeq 0 \\ &\text{Tr}(\rho_{AB}^{ij}) = 1\\
    \end{aligned}
\end{equation}
where $\rho_0 = \rho_{AB}^{ij} - \Lambda_0[\rho_{AB}^{ij}] $ , $\rho_1 = \rho_{AB}^{ij} - \Lambda_1[\rho_{AB}^{ij}] $ and $\rho_2= \rho_{AB}^{ij} $\\
Each inner product $(\rho_k,\rho_k)$ for $k \in \{0,1,2\}$ in the above equation is quadratic in $\rho_k$. The standard SDP formulation requires that the objective function to be linear in its decision variables and the constraints to be in the form of linear matrix inequalities.
Let $t_k \geq (\rho_k,\rho_k) $, then using Schur's complement we get,
\begin{equation}\begin{aligned}
    \begin{pmatrix}
        t_k & \mbox{vec}(\rho_k)^{\dagger}\\ \mbox{vec}(\rho_k) & 1 
        \end{pmatrix}\succeq 0
\end{aligned}
\end{equation}
Since CHSH correlation operator is fixed for a given value of the angles $\phi_A^i$ and $\phi_B^j$, the standard SDP formulation of the optimization problem would be,
\begin{equation}
\label{eqn5.30}
  \boxed{  \begin{aligned}
        n^*(S_{ij}) = -&\text{maximize} \quad  \lambda t_0 + (1-\lambda)t_1 + \frac{\mu}{2} t_2 \\
        \text{s.t.} & \begin{pmatrix}
        t_0 & \mbox{vec}(\rho_0)^{\dagger}\\ \mbox{vec}(\rho_0) & 1 
        \end{pmatrix}\succeq 0, 
         \begin{pmatrix}
        t_1 & \mbox{vec}(\rho_1)^{\dagger}\\\mbox{vec}(\rho_1) & 1 
        \end{pmatrix}\succeq 0, 
\begin{pmatrix}
        t_2 & \mbox{vec}(\rho_2)^{\dagger}\\\mbox{vec}(\rho_2) & 1 
        \end{pmatrix}\succeq 0 \\ 
        \quad &\text{Tr}\left(\rho_{AB}^{ij} ~ CHSH(\phi_A^i,\phi_B^j)\right) = S_{ij},\\
& \phi_A^i, \phi_B^j \in [0,\pi/2],\\
    &\rho_{AB}^{ij} \succeq 0 \\ &\text{Tr}(\rho_{AB}^{ij}) = 1\\
    \end{aligned}}
\end{equation}

\subsection{Bounding the pessimistic error terms $\Delta\left(\epsilon_0,\phi_A^i\right) \text{ and }\Delta\left(\epsilon_0,\phi_B^j\right)$}

The solution of the modified optimization problem, $n^*(S_{ij})$, is thus established to be Lipschitz continuous. Now one can proceed towards formulating a closed form for the pessimistic error terms, $\Delta\left(\epsilon_0,\phi_A^i\right) \text{ and }\Delta\left(\epsilon_0,\phi_B^j\right)$.
The solution of the SDP gives an optimal value for the $k^{th}$ and $l^{th}$ segment centered around $\phi_{A_k}^i$ and $\phi_{B_l}^j$ for Alice and Bob, respectively. The solution of the SDP is based on fixed $\phi_A^i$ and $\phi_B^j$. Without loss of generality, we consider the function $f$ to be the solution of the optimization problem. Let denote two functions analogous to the solution of the modified optimization problem for Alice's and Bob's angles as, 
\begin{equation}
    f_A(\phi_A^i) = R_A \text{, }
    f_B(\phi_B^j) = R_B
\end{equation}
where $R_A$ and $R_B$ give the optimal value of the SDP for the given parameterized angles.\\
Now, since the functions defined above are Lipschitz continuous, one can have the following relation for each segment, where $\phi_{A_k}^i \text{ and } \phi_{B_l} ^j$ are the centres of the segments $k$ and $l$, respectively.
\begin{equation}
\label{eqn5.39}
    \begin{aligned}
        &| f_A(\phi_A^i) - f_A(\phi_{A_k}^i)| \leq L_A|\phi_A^i-\phi_{A_k}^i|\\
        &| f_B(\phi_B^j) - f_B(\phi_{B_l}^j)| \leq L_B|\phi_B^j-\phi_{B_l}^j|, 
        ~~~~~~~~~~~~~~~~~~\forall \phi_A^i \in I_A~
     \text{ and }~\forall \phi_B^j \in I_B
    \end{aligned}
\end{equation}
The above inequalities put upper bounds on the absolute deviation in the solution of the optimization problem. The supremum of the above functions are given as,
\begin{equation}
\label{eqn5.41}
    \begin{aligned}
        &M_A^i = sup_{\phi_A^i \in I_A } f_A(\phi_A^i)~~ \mbox{and}
        ~~M_B^j = sup_{\phi_B^j \in I_B } f_B(\phi_B^j)\\
    \end{aligned}
\end{equation}
Similarly, the infimum of the solutions are being given as,
\begin{equation}
\label{eqn5.42}
    \begin{aligned}
        &m_A^i = inf_{\phi_A^i \in I_A } f_A(\phi_A^i)~~\mbox{and}~~
        m_B^j = inf_{\phi_B^j \in I_B } f_B(\phi_B^j)\\
    \end{aligned}
\end{equation}
The Lipschitz continuity of the solution functions implies that they are continuous in the given interval. Therefore, the supremums are the corresponding maximums and the infimum are the corresponding minimums. The supremum and infimum guarantee that no number smaller than $M_A^i \text{ and } M_B^j$ can serve as an upper-bound, and equivalently, no number larger than $m_A^i \text{ and } m_B^j$ can act as lower bound for $ f_A(\phi_A^i) \text{ and } f_B(\phi_B^j)$, respectively.
The quantities that we are  interested in bounding are,
\begin{equation}
\label{eqn5.46n}
    \begin{aligned}
        sup_{\phi_A^i \in I_A} &\left| f_A(\phi_A^i) - f_A(\phi_{A_k}^i)\right|~~~~ \mbox{and}~~~~
        sup_{\phi_B^j \in I_B}\left| f_B(\phi_B^j) - f_B(\phi_{B_l}^j)\right| \\ 
    \end{aligned}
\end{equation}
where $\phi_{A_k}^i \text{ and } \phi_{B_l}^j $ represent the center of the segments $k$ and $l$, respectively. 
We find that the deviation gets maximised at the boundary of the respective segments (\textit{Theorem 3 } in Appendix \ref{Theorem3}). Hence, we have,
\begin{equation}
    \begin{aligned}
       sup_{\phi_A^i \in I_A}
 &| f_A(\phi_A^i) - f_A(\phi_{A_k}^i)|=  L_A\epsilon_0\\
       sup_{\phi_B^j \in I_B}
 &| f_B(\phi_B^j) - f_B(\phi_{B_l}^j)| = L_B\epsilon_0,~~~~~~~~~~~~
        \forall \phi_A^i \in I_A~
     \text{ and }~\forall \phi_B^j \in I_B
    \end{aligned}
\end{equation}
And from the definition of the pessimistic error terms
$\Delta\left (\epsilon_0,\phi_A^i\right)$ and $\Delta\left (\epsilon_0,\phi_B^j\right)$, we get,
\begin{equation}
\label{eqn5.52n}
    \begin{aligned}
        \Delta\left (\epsilon_0,\phi_A^i\right) = L_A\epsilon_0~~~~~~\mbox{and}~~~~~~
        \Delta\left (\epsilon_0,\phi_B^j\right) = L_B\epsilon_0
    \end{aligned}
\end{equation}
Since the functions $f_A(\phi_A^i)$ and $f_B(\phi_B^j)$ are differentiable in the intervals $I_A$ and $I_B$, respectively, we determine the values of $L_A$ and $L_B$ using the following Taylor series as follows,
\begin{equation}
    \begin{aligned}
        &f_A(\phi_A^i) = \sum_{n=0}^{\infty}\frac{f_A^n(\phi_{A}^{i_*})}{n!}(\phi_A^i - \phi_{A}^{i_*})^n\\
        &f_B(\phi_B^j) = \sum_{m=0}^{\infty}\frac{f_B^m(\phi_{B}^{j_*})}{m!}(\phi_B^j - \phi_{B}^{j_*})^m,~~~
    &\forall \phi_{A}^{i_*} \in I_A
    ~~\text{and }~~\forall \phi_{B}^{j_*} \in I_B 
    \end{aligned}
\end{equation}
Now in this case, linear approximations of $f_A(\phi_A^i)$ near $\phi_{A_k}^i$ and $f_B(\phi_B^j)$ near $\phi_{B_l}^j$ are needed for small $\epsilon_0$. One can neglect the higher-order derivatives terms because $f_A^1(\phi_{A_k}^i)(\phi_A^i - \phi_{A_k}^{i})$ and $f_B^1(\phi_{B_l}^j)(\phi_B^j - \phi_{B_l}^{j})$  dominate as $|\phi_A^i - \phi_{A_k}^{i}|$ and 
$|\phi_B^j - \phi_{B_l}^{j}|$ tends to 0 for $k^{th}$ and $l^{th}$ segments of Alice and Bob, respectively.
The necessity of incorporating the higher-order terms in the error analysis arises when $\epsilon_0$ is nonnegligible, the underlying functions exhibit strong non-linearity, or when a more precise and tighter bound on the error is desired. In such scenarios, only linear approximations are insufficient to accurately capture the function's behaviour within the interval of width $2\epsilon_0$. The contributions from higher derivatives that are neglected in the first-order analysis become non-negligible and must be accounted for to achieve the required accuracy in the error estimation.
Thus, using the first order linear approximation, one can get an approximation of $f_A(\phi_A^i)$ and $f_A(\phi_B^j)$ near $\phi_{A_k}^i$ and $\phi_{B_l}^j$
as,
\begin{equation}
    \begin{aligned}
        &f_A(\phi_A^i) = f_A(\phi_{A_k}^i) + f_A^1(\phi_{A_k}^i)(\phi_A^i - \phi_{A_k}^{i}) + \mathcal{O}((\phi_A^i - \phi_{A_k}^{i})^2)\\
        & |f_A(\phi_A^i) - f_A(\phi_{A_k}^i)| \leq |f_A^1(\phi_{A_k}^i)||(\phi_A^i - \phi_{A_k}^{i})| + \mathcal{O}((\phi_A^i - \phi_{A_k}^{i})^2) 
    \end{aligned}
\end{equation}
Similarly,
\begin{equation}
    |f_B(\phi_B^j) - f_B(\phi_{B_l}^j)| \leq |f_B^1(\phi_{B_l}^j)||(\phi_B^j - \phi_{B_l}^{j})| + \mathcal{O}((\phi_B^j - \phi_{B_l}^{j})^2)
\end{equation}
The higher-order terms, represented by $\mathcal{O}((\phi_A^i-\phi_{A_k}^i)^2)$ and $\mathcal{O}((\phi_B^j-\phi_{B_l}^j)^2)$, are indeed positive if the second derivatives are positive, indicating local convexity. While typically removing positive terms would weaken an inequality, the context of a pessimistic error bound requires careful consideration. The pessimistic error is designed to overestimate the potential deviation. By truncating the Taylor series and neglecting these positive higher-order terms, we essentially underestimate the actual deviations $|f_A(\phi_A^i)-f_A(\phi_{A_k}^i)|$ and $|f_B(\phi_B^j)-f_B(\phi_{B_l}^j)|$. Consequently, when these underestimated deviations are used to construct a pessimistic error (which is subtracted from the SDP result), the error itself becomes overestimated. This overestimation of the error leads to a more conservative (and potentially less tight) lower bound on the true optimal value of the SDP.
Thus, after first order linear approximation we get,
\begin{equation}
\label{eqn5.56n}
    \begin{aligned}
       sup_{\phi_A^i \in I_A}
 & |f_A(\phi_A^i) - f_A(\phi_{A_k}^i)| \leq |f_A^1(\phi_{A_k}^i)||(\phi_A^i - \phi_{A_k}^{i})|  \\
sup_{\phi_B^j \in I_B}
&|f_B(\phi_B^j) - f_B(\phi_{B_l}^j)| \leq |f_B^1(\phi_{B_l}^j)||(\phi_B^j - \phi_{B_l}^{j})| 
    \end{aligned}
\end{equation}
The above relations hold good $\forall ~ \phi_A^i,\phi_B^j\in I$ if,
\begin{equation}
    \begin{aligned}
        \Delta\left (\epsilon_0,\phi_A^i\right) =        max_{\phi_A^i \in I}|f_A^1(\phi_{A_k}^i)|
\epsilon_0~~~\mbox{and}~~~
        \Delta\left (\epsilon_0,\phi_B^j\right) = max_{\phi_A^i \in I}|f_B^1(\phi_{B_l}^j)|\epsilon_0
    \end{aligned}
\end{equation}
where $L_A = max_{\phi_A^i \in I}|f_A^1(\phi_A^i)| $ and $L_B = max_{\phi_A^i \in I}|f_B^1(\phi_B^j)|.$ Thus, we evaluate the pessimistic errors involved in the $\epsilon-$net approximation used for optimizing Alice's and Bob's angles.

\subsection{Alice's and Bob's measurement angles }

The work of \cite{Schwonnek_2021} shows that without loss of generality one can can reduce the problem to two-qubit space $\mathbb{C}_{4\times4}$ of Alice's and Bob's subsystem. From the result of \cite{twoprojectortheory} we can decompose a projector of higher dimension into projectors of dimension ${2\times 2}$ acting on either Alice's or Bob's state. Now the work\cite{Schwonnek_2021} proposed that the angles for Alice and Bob for choosing the projector and hence the observables are obtained from the \textit{spectrum} of $P^{0|0}_{\mathcal{X}} +P^{0|1}_{\mathcal{X}} $ and $P^{0|2}_{\mathcal{Y}} + P^{0|3}_{\mathcal{Y}}$, respectively \eqref{eqn45}. Using this result, we find the explicit expression for Alice's and Bob's measurement angles as,
\begin{equation}
\begin{aligned} \phi_A^i = 2 \arccos\left(\frac{\lambda_\mathcal{X}^1 - \lambda_\mathcal{X}^2}{2}\right)~~~\mbox{and}~~~
 \phi_B^j = 2 \arccos\left(\frac{\lambda_\mathcal{Y}^1 - \lambda_\mathcal{Y}^2}{2}\right),   
\end{aligned}
\end{equation}
where $\phi_A^i$ and $\phi_B^j$ are the angles corresponding to $ij^{th}$ two qubit space marked by $i^{th}$ and $j^{th}$ indices of Alice and Bob, respectively. $\lambda_\mathcal{X}^1 ,\lambda_\mathcal{X}^2$ and $\lambda_\mathcal{Y}^1,\lambda_\mathcal{Y}^2$ are the eigen value of $P^{0|0}_{\mathcal{X}} +P^{0|1}_{\mathcal{X}} $ and $P^{0|2}_{\mathcal{Y}} + P^{0|3}_{\mathcal{Y}}$, respectively. We find that such a mapping is unique (See \textit{Lemma 1 } in appendix \ref{Lemma1}).\\
\subsection{Dependence of CHSH operator on Alice's and Bob's angles}
The optimisation problem presented in equation \eqref{eqn5.26} exhibits a dependence on Alice's angle, denoted as $\phi_A^i$, through the channel $\Lambda_1$ as defined by equations \eqref{eqn20}, \eqref{eqn44}, \eqref{eqn45}, and \eqref{eqn46}. Additionally, it depends on the CHSH operator, as specified in equations \eqref{eqn97} and \eqref{eqn5.26}. Bob's angle, $\phi_B^j$, solely influences the CHSH operator. Notably, variations in the CHSH operator are of particular interest due to their direct impact on the feasible solution set of the optimisation problem. Furthermore, analysing the effect of changes in the feasible region holds greater significance than examining alterations in the channel $\Lambda_1$. This is because Alice's and Bob's parameters jointly determine the feasible region, allowing for a more unified analysis concerning variations in the parameters.

Given discrete points $\phi_{A_k}^i$ and $\phi_{B_l}^j$ located at the midpoint of their respective intervals, the inherent symmetry of the underlying function or data distribution implies that the maximum deviation from these discrete points will occur symmetrically around them. Consequently, a small perturbation $\epsilon$ from the midpoint will result in equal magnitudes of deviation, such that the deviation at $\phi_{A_k+\epsilon}^i$ is equivalent to the deviation at $\phi_{A_k-\epsilon}^i$, and similarly for $\phi_{B_l}^j$.
The spectral norm of the difference between the CHSH operators evaluated at $\phi_{A_k}^i$ and $\phi_{A_k \pm \epsilon}^i$ (and similarly for $\phi_{B_l}^j$ and $\phi_{B_l \pm \epsilon}^j$) is equal to the largest singular value of this difference operator. Let us define the maximum of the spectral norms of the differences in the CHSH operator as,
\begin{equation}
\label{eqn5.61n}
    \begin{aligned}
        \delta_p = \max\Big(& ||CHSH(\phi_{A}^i,\phi_B^j) - CHSH(\phi_{A}^i+\epsilon,\phi_B^j)||_{\infty} , \\
        & ||CHSH(\phi_{A}^i,\phi_B^j) - CHSH(\phi_{A}^i,\phi_{B}^j+\epsilon)||_{\infty} , \\
        & ||CHSH(\phi_{A}^i,\phi_B^j) - CHSH(\phi_{A}^i+\epsilon,\phi_{B}^j+\epsilon)||_{\infty} \Big)
    \end{aligned}
\end{equation}
Evaluating each term in the above equation \ref{change_of_angles_through_chsh}, we finally get,
\begin{equation}
\label{eqn5.70n}
    \begin{aligned}
        \delta_p \leq \max\Big(& \Big\| {\epsilon}\left( \cos\left(\phi_A^i\right) \sigma_x -\sin\left(\phi_A^i\right)\sigma_z  \right) \Big\|_{\infty}\cdot \Big\|  2\left(Q\left({\phi_B^j}\right) - \begin{pmatrix}
    1&0\\0&0
\end{pmatrix} \right) \Big\|_{\infty}, \\
        & \Big\|  2\left(Q\left({\phi_A^i}\right) - \begin{pmatrix}
    0&0\\0&1
\end{pmatrix} \right) \Big\|_{\infty}\cdot \Big\|{\epsilon}\left(\cos\left(\phi_B^j\right)\sigma_x - \sin\left(\phi_B^j\right)\sigma_z \right)   \Big\|_{\infty}, \\
        & \Big\| {\epsilon}\left(\cos\left(\phi_A^i\right)\sigma_x - \sin\left(\phi_A^i\right)\sigma_z \right) \Big\|_{\infty}\cdot \Big\| {\epsilon}\left(\cos\left(\phi_B^j\right)\sigma_x - \sin\left(\phi_B^j\right)\sigma_z \right)  \Big\|_{\infty} \Big)\\
        \leq \max &\left(\epsilon \cdot 2\sin\left(\frac{\phi_B^j}{2}\right), \quad 2\cos\left(\frac{\phi_A^i}{2}\right)\cdot\epsilon,\quad \epsilon^2 \right)\\
        \leq \quad  2\epsilon_0
    \end{aligned}
\end{equation}
The equality holds for the spectral norm of Kronecker products, therefore, this upper bound is exact. 
Now in the interval $I,$ $\sin\left(\frac{\phi_B^j}{2}\right)$ and $\cos\left(\frac{\phi_A^i}{2}\right)$ functions are monotonically increasing and decreasing respectively. The first term $2\sin\left(\frac{\phi_B^j}{2}\right)$ is increasing from $0 \text{ to } 1.414$ and the second term $2\cos\left(\frac{\phi_A^i}{2}\right)$ is decreasing from $2\epsilon \text{ to } 1.414$.
Thus, the maximum deviation of the CHSH operator is achieved when $\phi_A^i \text{ is fixed at } 0$ but $\phi_B^j \text{ is increased by }\epsilon_0.$

\subsection{Dependence of objective function on Alice's angle}
The objective function in \eqref{eqn5.26} is a function of the density operator describing the joint state between Alice and Bob and Alice's angle $\phi_A^i$. To analyse the dependency of the solution of the optimisation problem on $\phi_{A}^i$, we reinterpret the Frobenius norm terms in \eqref{eqn5.26} using dual norms. Since the Frobenius norm is self-dual, this framework retains the norm structure while emphasising its role as a maximiser of inner products. Now, as the objective function depends only upon Alice's angle, it is sufficient to consider only the perturbation of Alice's angle $\phi_A^i$. Moreover, in the objective function, only one term is a function of the Alice's angle; therefore, we consider only that term and analyse the perturbation. The absolute change in the objective function can be written as,
\begin{equation}
    \Big|h(\lambda,\phi_A^i,\rho_{AB}^{ij}) -  h(\lambda,\phi_A^i+\epsilon,\rho_{AB}^{ij})\Big|
\end{equation}
The first term can be expanded in the following way, 
\begin{equation}
\begin{aligned}
    h(\lambda,\phi_A^i,\rho_{AB}^{ij}) &=  (1-\lambda) \Tr\left(\left[ \rho_{AB}^{ij} - \left(\frac{1}{2}\{\rho_{AB}^{ij}, Q(\phi_A^i)\otimes \I \} - \frac{1}{2}[\rho_{AB}^{ij}, Q(\phi_A^i)\otimes \I ] \right)Q(\phi_A^i)\otimes \I \right.\right. \\
 & \left.\left. \qquad\qquad - \left(\frac{1}{2}\{\rho_{AB}^{ij}, (\I-Q(\phi_A^i)\otimes \I) \} - \frac{1}{2}[\rho_{AB}^{ij}, (\I-Q(\phi_A^i)\otimes \I) ] \right)\{\I-Q(\phi_A^i)\otimes \I\} \right]^2\right)
\end{aligned}
\end{equation}
In a similar way, we can expand the second term and finally get,
\begin{equation}
\boxed{    \begin{aligned}
        &\left|h(\lambda,\phi_A^i,\rho_{AB}^{ij}) - h(\lambda,\phi_A^i+\epsilon_0,\rho_{AB}^{ij})\right| \\
\leq{}&(1-\lambda)[4\epsilon_0+\mathcal{O}(\epsilon_0^2)]
    \end{aligned}}
\end{equation}
The tighter bound of 4$\epsilon_0$ in the previous work is based upon assumption that $\rho_{AB}^{ij}\text{ and } Q(\phi_A^i)\otimes \I$ commute.
Having evaluated the change in the objective function with respect to Alice's angle, we can now lower bound the maximal error in $k^{th}$ segment centered around $\phi_{A_k}^i$ and get the modified optimization problem as,
\begin{equation}
\boxed{
    \begin{aligned}
n^*(S_{ij}) \geq \inf \quad & \lambda \left\| \rho_{AB}^{ij} - \Lambda_0[\rho_{AB}^{ij}] \right\|_F^2 + (1-\lambda)\left\| \rho_{AB}^{ij} - \Lambda_1[\rho_{AB}^{ij}] \right\|_F^2 \\&+ \frac{\mu}{2} \|\rho_{AB}^{ij}\|_F^2 \text{ }-2(1-\lambda)\epsilon_0 \\
\text{s.t.} \quad &\text{Tr}\left(\rho_{AB}^{ij} ~CHSH(\phi_A^i,\phi_B^j)\right) = S_{ij} \text{ }-2\epsilon_0 \\
& \phi_A^i, \phi_B^j \in [0,\pi/2],\\
    &\rho_{AB}^{ij} \succeq 0 \\ &\text{Tr}(\rho_{AB}^{ij}) = 1
\end{aligned}
}
\end{equation}
\subsection{Convex lower bound of $C^*(S)$}
The optimization problem defining \(n^*(S)\) is given as,
\begin{equation}
\begin{aligned}
&n^*(S_{ij}) = \inf \left\{ 
\lambda \left\| \rho_{AB}^{ij} - \Lambda_0\left[\rho_{AB}^{ij}\right] \right\|_F^2 
+ (1-\lambda)\left\| \rho_{AB}^{ij} - \Lambda_1\left[\rho_{AB}^{ij}\right] \right\|_F^2 
+ \frac{\mu}{2} \left\| \rho_{AB}^{ij}\right\|_F^2 
\right\}\\    
&s.t. ~ \text{Tr}\left(\rho_{AB}^{ij}~ CHSH(\phi_A^i, \phi_B^j)\right) = S_{ij}.
\end{aligned}
\end{equation}
Now, the given objective function involving the Frobenius norm \(\left\| \cdot \right\|_F^2\) is strongly convex in \(\rho\). The term \(\frac{\mu}{2} \left\| \rho_{AB}^{ij} \right\|_F^2\) introduces \(\mu\)-strong convexity, and the linear constraints of the optimization problem preserve the strong convexity on the feasible set.
For fixed \(S_{ij}\), the objective function is strongly convex in \(\rho\). The parameterized problem's solution function \(n^*(S_{ij})\) inherits convexity. Moreover, strong convexity implies quadratic dependence on the perturbations in \(S_{ij}\).
Thus, \(n^*(S)\) is {strongly convex} in \(S\) for $S\in(2,2\sqrt{2}]$
For the measure \(\eta\) with,
\begin{equation}
\int_{2}^{2\sqrt{2}} \!\eta(dS') = 1, \quad \eta \geq 0, \quad \int_{2}^{2\sqrt{2}} \!\eta(dS') S' = S    
\end{equation}
Now application of Jensen's inequality for strongly convex functions yields,
\begin{equation}
\int_{2}^{2\sqrt{2}} \!\eta(dS') n^*(S') \geq n^*\left(\int_{2}^{2\sqrt{2}} \!\eta(dS') S'\right) + \frac{\mu}{2} ~\text{Var}_\eta(S')    
\end{equation}
Since \(\text{Var}_\eta(S') \geq 0\),
\begin{equation}
\int_{2}^{2\sqrt{2}} \!\eta(dS') n^*(S') \geq n^*(S).    
\end{equation}
Substituting \(\overline{C}(S') = n^*(S')\) into the original inequality, we get,
\begin{equation}
C^{*}(S) \geq \int_{2}^{2\sqrt{2}} \!\eta(dS') n^*(S') \geq n^*(S)
\end{equation}
Therefore, \(\overline{C}(S) = n^*(S)\) is a valid convex lower bound.

\section{Conclusion}\label{sec4}

In most of the QKD protocols, Alice and Bob single out one of their measurement settings for key generation. This strategy helps them improving the key rate of those protocols. In the original DIQKD protocol \cite{acindiqkd}, Alice and Bob chose a single measurement setting $\{A_0,B_0\}$ with high probability for key generation, whereas, other measurement settings with low probability for testing the channels. However, the security analysis of this protocol shows that there exists optimal attack such that $H(A_0|E)\leq H(A_1|E)$ \cite{Pironio_2009}. In other words, as Eve has the information about Alice's measurement settings, she could focus on minimizing her uncertainty about the key generating measurement $A_0$ at the cost of having higher uncertainty about the other setting. To circumvent this problem (to increase Eve's uncertainty about the key generation measurement setting), DIQKD with random key basis is proposed. In this variant of DIQKD protocol, as Alice uses two measurement settings for key generation, the uncertainty of Eve about Alice's measurement settings is now a combination of conditional entropies, $\lambda H(A_0|E) + (1-\lambda)H(A_1|E)$, where $\lambda$ is the probability for choosing the measurement $A_0$ and $E$ is Eve's side information. The main challenge in the security analysis of this protocol is to find a reliable lower bound on this quantity using only the observed CHSH violation. The authors in\cite{Schwonnek_2021} have framed the problem onto an optimization problem and found a lower bound $C^{\star}(S)$, such that $\lambda H(A_0|E) + (1-\lambda)H(A_1|E)\geq C^{\star}(S)$. The results produced in this manuscript show that the optimization cost involved in the security analysis of the DIQKD protocol can be reduced. In the original work, polytope optimization has been used for finding Bob's optimal angle in context to the main optimization problem. Our results show that reframing the security analysis as a strongly convex optimization problem can help in optimizing Bob's measurement angles using the same approach as used for Alice, i.e., $\epsilon$-net approximation. Now, as polytope optimization is costlier than $\epsilon$-net approximation, substituting the earlier with the later in the main problem results to lesser cost for the security analysis. Evaluation of the pessimistic error involved in the $\epsilon-$net approximation is an important task because in every iteration the error is being subtracted from the CHSH score $S$. However, so far, any explicit expression for the pessimistic  error has not been noticed in any literature. Here, we have derived an explicit expression for the pessimistic error. It is found to be the maximised value of the product of two terms, namely, desired precision of the angles $\phi_A^i$ and $\phi_B^j$ and the first-order derivative of the solution functions of the optimization problem. Our result shows how Alice and Bob can determine their measurement angles $\phi_A^i$ and $\phi_B^j$ from the spectrum of the projectors they chose for checking CHSH violation and generating secret keys. The dependence of CHSH operator on the angles of the parties were not clearly explained in the original security analysis. Our results also show how the operator varies with the change in the angles of Alice and Bob, determining the instance that results to maximum deviation in the CHSH operator. Moreover, from the results derived in this manuscript, it can be concluded that a convex lower bound on the uncertainty of Eve about Alice's key generation measurements can be derived analytically. We believe that our work has made the security analysis of the DIQKD with random key basis more clear and complete.

\section{Acknowledgment}
SB would like to acknowledge the University of Calcutta for providing the opportunity to work at TCG CREST. He is also grateful to team of CQuERE, TCG CREST, for fruitful discussion.\\

\bibliographystyle{alpha}
\bibliography{references}

\newpage
\onecolumngrid
\appendix
\section*{Appendix}
\section{Framework}
We first describe the framework for phrasing the security analysis as an optimization problem. The aim of the analysis is to establish a lower bound on the quantity $C^{*}(S)$ \cite{Schwonnek_2021}, where $ S$ is the observed CHSH value for the shared entangled state between the parties Alice and Bob, such that 
\begin{equation}
\label{eqna.1}
    \lambda H(A_0|E)_{\rho_{A'BEA_0}} + (1-\lambda)H(A_1|E)_{\rho_{A'BEA_1}} \geq C^{*}(S), 
\end{equation}
where $A_X$ is Alice's outcome for her measurement setting $X \in \{0,1\}$ and $H(A_{X}|E)_{\rho_{A'BEA_X}}$ signifies Eve's uncertainty about Alice's outcomes given her side-channel quantum information $E.$ The protocol assumes the measurement setting for Bob to be $Y\in\{0,1,2,3\}$. Let $\mathcal{X}$ and $\mathcal{Y}$ be the Hilbert spaces defining Alice's and Bob's subsystems, respectively. Now consider $O^{\mathcal{X}}_{x} \in \Herm(\mathcal{X})$ for $x \in \{0,1\}$ and $O^{\mathcal{Y}}_{y} \in \Herm(\mathcal{Y})$ for $y \in \{0,1,2,3\}$, respectively, to be the observables corresponding to Alice's and Bob's measurement settings. The observables can be explicitly written as $ \mathcal{A} = \{ O^{\mathcal{X}}_{0}, O^{\mathcal{X}}_{1}\}$ and 
$ \mathcal{B} = \{O^{\mathcal{Y}}_{0}, O^{\mathcal{Y}}_{1}, O^{\mathcal{Y}}_{2},O^{\mathcal{Y}}_{3}\}$. These can be written in spectral form as,
\begin{equation}
\label{eqn2}
   O^{\mathcal{X}}_{x} = \sum_{i}{\lambda^{x}_i}{P_{i}^{x}}, \quad
O^{\mathcal{Y}}_{y} = \sum_{j}{\lambda^{y}_j}{P_{j}^{y}}
\end{equation}
where $\lambda_i^x$ and  $\lambda_j^y$ are eigenvalues corresponding to the projectors $P_{i}^{x}$ and $P_{j}^{y}$, respectively. As the measurement outcomes are dichotomic, we can also express the observables as,
\begin{equation}
\label{eqn3}
   O^{\mathcal{X}}_{x} = P^{0|x}_{\mathcal{X}} - P^{1|x}_{\mathcal{X}}, \quad
O^{\mathcal{Y}}_{y} = P^{0|y}_{\mathcal{Y}} - P^{1|y}_{\mathcal{Y}}
\end{equation}
The measurements settings chosen by Alice and Bob to check the CHSH violation are $\{A_0, A_1\}$ and $\{B_2, B_3\}$, respectively. Therefore, the correlation measurement operators corresponding to the chosen measurement settings can be defined as, 
\begin{equation}
\begin{aligned}
\label{eqn3}
C^{O^{\mathcal{X}}_{0}} &= P^{0|0}_{\mathcal{X}} - P^{1|0}_{\mathcal{X}}~~~\mbox{and}~~~
C^{O^{\mathcal{X}}_{1}} = P^{0|1}_{\mathcal{X}} - P^{1|1}_{\mathcal{X}}, \\
C^{O^{\mathcal{Y}}_{2}} &= P^{0|2}_{\mathcal{Y}} - P^{1|2}_{\mathcal{Y}}~~~\mbox{and}~~~
C^{O^{\mathcal{Y}}_{3}} = P^{0|3}_{\mathcal{Y}} - P^{1|3}_{\mathcal{Y}}
\end{aligned}
\end{equation}
The correlation function defined in the main text, $C_{XY} = P(A_X=B_Y|X,Y) - P(A_X\neq B_Y|X,Y)$, is related to the correlation measurement operators via
$C_{XY} = \Tr(\rho~ P^{a|x}_{\mathcal{X}} \otimes P^{b|y}_{\mathcal{Y}})$, where $\rho$ is the state on which Alice and Bob perform the measurement. Hence, the CHSH operator can be written in terms of correlation measurement operators as 
 \cite{Schwonnek_2021},
\begin{eqnarray}
\label{eqna.10}
    {CHSH} &:=&  C^{O^{\mathcal{X}}_{1}} \otimes C^{O^{\mathcal{Y}}_{2}} - C^{O^{\mathcal{X}}_{0}} \otimes C^{O^{\mathcal{Y}}_{2}}\nonumber\\
    &-&C^{O^{\mathcal{X}}_{0}} \otimes C^{O^{\mathcal{Y}}_{3}} - C^{O^{\mathcal{X}}_{1}} \otimes C^{O^{\mathcal{Y}}_{3}}
\end{eqnarray}
Each term in the operator can be expressed in terms of the projectors defined above as \cite {markmwilde}, 
\begin{equation}
\begin{aligned}
\label{eqn11}
C^{O^{\mathcal{X}}_{x}} \otimes C^{O^{\mathcal{Y}}_{y}} &= (P^{0|x}_{\mathcal{X}} - P^{1|x}_{\mathcal{X}}) \otimes (P^{0|y}_{\mathcal{Y}} - P^{1|y}_{\mathcal{Y}}) \\
&= (P^{0|x}_{\mathcal{X}} \otimes P^{0|y}_{\mathcal{Y}}) - (P^{0|x}_{\mathcal{X}} \otimes P^{1|y}_{\mathcal{Y}}) - (P^{1|x}_{\mathcal{X}} \otimes P^{0|y}_{\mathcal{Y}}) + (P^{1|x}_{\mathcal{X}} \otimes P^{1|y}_{\mathcal{Y}}) \\
&= (P^{0|x}_{\mathcal{X}} \otimes P^{0|y}_{\mathcal{Y}} + P^{1|x}_{\mathcal{X}} \otimes P^{1|y}_{\mathcal{Y}}) - (P^{0|x}_{\mathcal{X}} \otimes P^{1|y}_{\mathcal{Y}} + P^{1|x}_{\mathcal{X}} \otimes P^{0|y}_{\mathcal{Y}}) 
\end{aligned}
\end{equation}
On the assumption that the state shared between Alice and Bob is a mixed state $\rho_{AB}$ in general, and Eve holds the purification of this state, the pure state is now described by $\Psi_{ABE}$. A more particular description of the state may include a fourth component -- Alice's outcome $A_X$, which is initially associated with a pure quantum state $\psi$, and hence the joint state between Alice, Bob, and Eve can be expressed as $\Psi_{ABEA_X}$. The dimension of Eve's system is unknown, and Eve has full control over all the devices, including Alice's and Bob's measurement devices. However, in this protocol, Eve can not determine beforehand the measurement setting used by Alice to generate the secret key -- Alice now uses two random bases for key generation.  

The purpose of optimising the objective function defined in (\ref{eqn18}) is to find a lower bound on the uncertainty of Eve about Alice's measurement outcomes. The conditional von Neumann entropy function defined in the Eq.(\ref{eqna.1}) can be decomposed as $H(A_X|E)_{\rho_{A'BEA_X}} = H(A_X)_{\rho_{A'BEA_X}} - \chi(A_X; E)$, where \( \chi(A_X; E) \) is the Holevo information \cite{Holevo+2019} between Alice and Eve.
Similarly, the von Neumann entropy can be expressed as $H(A_X)_{\rho_{A'BEA_X}} = I(A_X; B)_{\rho_{A'BEA_X}} + H(A_X|B)_{\rho_{A'BEA_X}}$, where 
$ I(A_X; B)_{\rho_{A'BEA_X}}$ and $H(A_X|B)_{\rho_{A'BEA_X}}$ are respectively the mutual information and conditional entropy between Alice and Bob. Using the above two relations, Eq. \eqref{eqna.1} can be rewritten as,
\begin{equation}
\begin{aligned}
\label{eqn13}
C^{*}(S) \leq &\ \lambda \left[I(A_0; B)_{\rho_{A'BEA_0}} + H(A_0|B)_{\rho_{A'BEA_0}} - \chi(A_0; E)\right] \\
&+ (1-\lambda)\left[I(A_1; B)_{\rho_{A'BEA_1}} + H(A_1|B)_{\rho_{A'BEA_1}} - \chi(A_1; E)\right]
\end{aligned}
\end{equation}
Under the assumption of optimal error correction (\( H(A_X|B_Y)_{\rho_{A'BEA_X}} \rightarrow 0 \)), the above inequality simplifies to:
   \begin{equation}
   \label{eqn14}
       C^{*}(S) \leq \lambda \left[I(A_0; B_{})_{\rho_{A'BEA_0}} - \chi(A_0; E)\right] + (1-\lambda)\left[I(A_1; B_{})_{\rho_{A'BEA_1}} - \chi(A_1; E)\right].
   \end{equation}
The quantum-classical state obtained after Alice's measurement is given by,
\begin{equation}
\label{eqn16}
    \rho_{A'BEA_X} = \sum_{i \in \{0,1\}}\rho_{\psi_i} \otimes \Tr_{A}((P^{i|x}_{\mathcal{X}} \otimes \mathbbm{1}_{BE})\Psi_{ABE}  (P^{i|x}_{\mathcal{X}} \otimes \mathbbm{1}_{BE})^*)
\end{equation}
where $A',B,E,A_X$ are Alice's subsystem, Bob's subsystem, Eve's subsystem, and Alice's measurement outcome associated with pure quantum state $\psi$, respectively. Using this state, if the conditional Von Neumann entropy $H(A_X|E)_{\rho_{A'BEA_X}}$ is calculated, then we find that it essentially captures the change in entropy of the joint state of Alice and Bob after and before Alice's measurement,
\begin{equation}
\begin{aligned}
\label{eqn17}
    H(A_X|E)_{\rho_{A'BEA_X}} &= H(A_XE)_{\rho_{A'BEA_X}} - H(E)_{\rho_{ABEA_X'}} \\
    &= H(A'B)_{\rho_{A'BEA_X}} - H(AB)_{\rho_{ABEA_X'}}\\
    &= \Delta H_X
\end{aligned}
\end{equation}
The optimization problem includes a constraint that the reduced density operator of the joint subsystem comprising Alice's and Bob's systems, denoted as $\rho_{AB}$, must be equal to the partial trace of the overall state $\rho_{ABEA_X'}$ over the environment ($E$) and Alice's ancilla ($A'_X$), i.e.,
$\rho_{AB} = \Tr_{EA'_X}(\rho_{ABEA_X'})$.
Now one can formally define the optimization problem as
\begin{equation}
\boxed{
\begin{aligned}
\label{eqn18}
    C^*(S) = \inf \quad &\lambda H(A_0|E)_{\rho_{A'BEA_0}} + (1-\lambda)H(A_1|E)_{\rho_{A'BEA_1}} \\
    \text{s.t.} \quad & \Tr(\rho_{AB} \cdot \text{CHSH}) = S\\
    &\rho_{AB} \succeq 0 \\ &\text{Tr}(\rho_{AB}) = 1\\
\text{or equivalently, }\\
C^*(S) = \inf \quad &\lambda \left[I(A_0; B_{})_{\rho_{A'BEA_0}} - \chi(A_0; E)\right] + (1-\lambda)\left[I(A_1; B_{})_{\rho_{A'BEA_1}} - \chi(A_1; E)\right] \\
    \text{s.t.} \quad & \Tr(\rho_{AB} \cdot \text{CHSH}) = S\\
    &\rho_{AB} \succeq 0 \\ &\text{Tr}(\rho_{AB}) = 1
\end{aligned}}
\end{equation}

\section{Reformulation of change in entropy $\Delta H_X$ in terms of Alice and Bob subsystems}
After formally defining the optimization problem, our first task is to reformulate the problem in terms of the Alice's and Bob's subsystems $A$ and $B$. The problem defined in \eqref{eqn18} is defined in terms of Alice's outcome $A_X$ and Eve's subsystem $E$. This setting has a serious flaw: inaccessibility to the Eve subsystem $E$. Although one can access Alice's outcome $A_X$ in the context of security analysis, access to the Eve subsystem is impractical and meaningless. The Eve's subsystem can be described by a vector space, whose dimension is unknown. If one has access to Eve's subsystem, one can configure the protocol accordingly to bypass Eve's intervention, which is not only practically infeasible but also makes no sense in the DIQKD framework.

The global state $\rho$ is a pure state; one may consider it as a bipartite state where $A_X$ and $E$ are constitute one part, and $A$ and $B$ the other. Now, from the result of \textit{Theorem 2.c} of \cite{araki}, one can easily find that the local entropy production $\Delta H_X$ on the two sides of the bipartition is essentially the same, 
\begin{equation}
    \Delta H_X = H(A'B)_{\rho_{A'BEA_X}} - H(AB)_{\rho_{ABEA_X'}}
\end{equation}
Hence, using Eq. \eqref{eqn17}, one can perform the security analysis in terms of accessible Alice and Bob subsystems.\cite{Schwonnek_2021}
The transformation can be seen as the transformation from the state $\rho_{ABEA_X'} \in \mathcal{R}$, where $\mathcal{R}$ defines the joint state describing Alice's system $A$, Bob's system $B$, Eve's system $E$, and the $A_X'$ being Alice's measurement outcome subsystem before the effect of her chosen observables associated with pure quantum state $\psi$ to $\rho_{A'BEA_X} \in \mathcal{H}$, where $\mathcal{H}$ denotes the same as a joint state describing Alice Bob and Eve system except for the fact that here $A'$ denotes the Alice subsystem after the transformation, and $A_X$ is the classical register storing the Alice's measurement outcomes. The work of \cite{Schwonnek_2021} shows that this transformation can indeed be defined through pinching channels defined for each observable as,
\begin{equation}
    \begin{aligned}
    \label{eqn20}
        \Lambda_0[\rho_{ABEA_0'}] = (P^{0|0}_{\mathcal{X}}\otimes \mathbbm{1})\rho_{ABEA_0'} (P^{0|0}_{\mathcal{X}}\otimes \mathbbm{1}) + (P^{1|0}_{\mathcal{X}}\otimes \mathbbm{1})\rho_{ABEA_0'} (P^{1|0}_{\mathcal{X}}\otimes \mathbbm{1})
        = \rho_{A'BEA_0}\\
        \Lambda_1[\rho_{ABEA_1'}] = (P^{0|1}_{\mathcal{X}}\otimes \mathbbm{1})\rho_{ABEA_1'} (P^{0|1}_{\mathcal{X}}\otimes \mathbbm{1}) + (P^{1|1}_{\mathcal{X}}\otimes \mathbbm{1})\rho_{ABEA_1'} (P^{1|1}_{\mathcal{X}}\otimes \mathbbm{1})=\rho_{A'BEA_1}
    \end{aligned}
\end{equation}
Now let us redefine the reduced joint state of Alice and Bob as,
\begin{equation}
    \begin{aligned}
        &\mbox{Tr}_{EA_X}(\Lambda[\rho_{ABEA_X'}])= \rho_{A'B}\\&\mbox{Tr}_{EA_X}(\rho_{ABEA_X'})=\rho_{AB}
    \end{aligned}
\end{equation}
\subsection{Pinching Channel}
The pinching channel \(\Lambda\) is defined as:
\begin{equation}
\Lambda[\rho] = \sum_i P_i \rho P_i    
\end{equation}
where \(\{P_i\}\) is a set of orthogonal projectors and \(\rho \in \) is a density operator.\\

\textit{Properties of Pinching Channels}
\begin{enumerate}
\item{Idempotence:} The pinching channel is idempotent, i.e.,
\begin{equation}
\label{eqn22}
\Lambda[\Lambda[\rho]] = \Lambda[\rho].    
\end{equation}
\textit{Proof:}
\begin{equation}
\Lambda[\Lambda[\rho]] = \sum_i P_i \left(\sum_j P_j \rho P_j\right) P_i.    
\end{equation}
Using the orthogonality of the projectors (\(P_i P_j = \delta_{ij} P_i\)), this simplifies to:
\begin{equation}
\Lambda[\Lambda[\rho]] = \sum_i P_i \rho P_i = \Lambda[\rho].
\end{equation}
Thus, \(\Lambda\) is idempotent.
\item Any pinching channel acts as an identity operator if \(\rho\) commutes with all \(P_i\), i.e.,
\begin{equation}
\Lambda[\rho] = \rho, ~~~~~~~~~\forall \rho ~~~~~if [\rho, P_i]=0
\end{equation}
\textit{Proof:}
If \(\rho\) commutes with \(P_i\), then \(P_i \rho = \rho P_i\). Substituting into the definition of \(\Lambda\):
\begin{equation}
\Lambda[\rho] = \sum_i P_i \rho P_i = \sum_i P_i P_i \rho = \sum_i P_i \rho = \rho,    
\end{equation}
where one can used \(P_i^2 = P_i\) (since \(P_i\) is a projector) and \(\sum_i P_i = I\).
\item Pinching channel does not increase trace distance between any two states, i.e., for any two states \(\rho\) and \(\sigma\),
\begin{equation}
\|\Lambda[\rho] - \Lambda[\sigma]\|_1 \leq \|\rho - \sigma\|_1.  \end{equation}
\textit{Proof:}
The trace distance is defined as:
\begin{equation}
\|\rho - \sigma\|_1 = Tr|\rho - \sigma|,
\end{equation}
where \(|A| = \sqrt{A^* A}\). The pinching channel \(\Lambda\) is a completely positive trace-preserving (CPTP) map, and all CPTP maps are contractive concerning the trace norm. Thus:
\begin{equation}
\|\Lambda[\rho] - \Lambda[\sigma]\|_1 \leq \|\rho - \sigma\|_1. \end{equation}
\item For any state \(\rho\), the diagonal terms of \(\rho\) in the basis of \(\{P_i\}\) remain unchanged after applying \(\Lambda\).\\
\textit{Proof:}
The diagonal terms of \(\rho\) in the basis of \(\{P_i\}\) are given by \(\Tr(P_i \rho)\). Applying \(\Lambda\):
\begin{equation}
\Tr(P_i \Lambda[\rho]) = \Tr\left(P_i \sum_j P_j \rho P_j\right) = \Tr(P_i \rho P_i) = \Tr(P_i \rho)
\end{equation}
one can use the trace's cyclic property and \(P_i P_j = \delta_{ij} P_i\). Thus, the diagonal terms are preserved.
\item Pinching channel increases the von Neumann entropy, ie.,
\begin{equation}
S(\Lambda[\rho]) \geq S(\rho),    
\end{equation}
where \(S(\rho) = -\Tr(\rho \log_2 \rho)\) is the von Neumann entropy.\\
\textit{Proof:}
The pinching channel \(\Lambda\) is unital (\(\Lambda[I] = I\)) and doubly stochastic. By the monotonicity of the von Neumann entropy under doubly stochastic maps, one can have:
\begin{equation}
S(\Lambda[\rho]) \geq S(\rho).    
\end{equation}
\item The eigenvalues of \(\Lambda[\rho]\) are majorized by the eigenvalues of \(\rho\), reflecting a redistribution of probabilities towards uniformity.\\
\textit{Proof:}
Let \(\lambda(\rho)\) and \(\lambda(\Lambda[\rho])\) denote the vectors of eigenvalues of \(\rho\) and \(\Lambda[\rho]\), respectively. The pinching channel \(\Lambda\) is an unital quantum channel, and by the Schur-Horn theorem, the eigenvalues of \(\Lambda[\rho]\) majorize the eigenvalues of \(\rho\):
\begin{equation}
\lambda(\Lambda[\rho]) \prec \lambda(\rho).    
\end{equation}
This means that the eigenvalues of \(\Lambda[\rho]\) are more uniformly distributed than those of \(\rho\).
\item For any measurable function \(f\), the pinching channel satisfies:
\begin{equation}
f(\Lambda[\rho]) = \Lambda[f(\rho)].    
\end{equation}
\textit{Proof:}
The pinching channel \(\Lambda\) acts as a projection onto the diagonal basis defined by \(\{P_i\}\). For any function \(f\), the action of \(f\) on \(\rho\) commutes with the pinching operation because \(f\) acts on the eigenvalues of \(\rho\), and \(\Lambda\) preserves the diagonal terms (eigenvalues) while removing off-diagonal terms. Thus:
\begin{equation}
f(\Lambda[\rho]) = \Lambda[f(\rho)].    
\end{equation}
\end{enumerate}
In quantum information theory, the dual (adjoint) of a map \( {\Phi} \), denoted \( {\Phi^*} \), is defined via the Hilbert-Schmidt inner product:
\begin{equation}
\label{eqn36}
\operatorname{Tr}(A \cdot \Phi[\rho]) = \operatorname{Tr}(\Phi^*[A] \cdot \rho), \quad \forall A, \rho \in \mathcal{}L(\mathcal{V}),
\end{equation}
where \( A \) and \( \rho \) are Hermitian operators on a complex Euclidean space \( \mathcal{V} \).
More generally, for any linear map \( \Phi \), the adjoint satisfies:
\begin{equation}
\label{eqn37}
\operatorname{Tr}(A \cdot \Phi[B]) = \operatorname{Tr}(\Phi^*[A] \cdot B), \quad \forall A, B \in L(\mathcal{V}),
\end{equation}
where \( A \) and \( B \) are Hermitian operators on a complex Euclidean space \( \mathcal{V} \).
Now since pinching channel ${\Lambda}$ are self adjoint from \eqref{eqn37} one  can have,
\begin{equation}
\label{eqn38}
    \Tr(A\cdot {\Lambda}[\rho]) = \Tr(\rho\cdot {\Lambda^*}[A]) = \Tr(\rho\cdot {\Lambda}[A]) 
    \quad \forall A,\rho \in L(\mathcal{V}) 
\end{equation}

\subsection{Entropy production}
Now, one can find a mathematical expression for the production of entropy $\Delta H_X$ as
\begin{eqnarray}
\label{eqn39}  
        \Delta H_X &=& H(A'B)_{\rho_{A'BEA_X}} - H(AB)_{\rho_{ABEA_X'}}\nonumber\\
        &=& H(\mbox{Tr}_{EA_X}(\Lambda[\rho_{ABEA_X'}])) -H(\mbox{Tr}_{EA_X'}(\rho_{ABEA_X'}))\nonumber\\
        &=& H(\Lambda[\rho_{AB}]) - H(\rho_{AB})
\end{eqnarray}
Using Eqs. (\eqref{eqn22},\eqref{eqn38}) and putting $A = \mbox{log}_2(\Lambda[\rho_{AB}])$ and $B=\rho_{AB}$, we get,
\begin{eqnarray}
\label{eqnB.23}
    \Delta H_X &=& -\Tr(\Lambda[\rho_{AB}] \log_2(\Lambda[\rho_{AB}])) + \Tr(\rho_{AB} \log_2 (\rho_{AB}))\nonumber\\
        &=& -\Tr(\rho_{AB} \Lambda^*(\log_2(\Lambda[\rho_{AB}]))) +\Tr(\rho_{AB}\log_2 (\rho_{AB}))\nonumber\\
        &=& -\Tr(\rho_{AB}\log_2(\Lambda[\rho_{AB})) +\Tr(\rho_{AB}\log_2 (\rho_{AB}))\nonumber\\
        &=& \Tr(\rho_{AB}\log_2 (\rho_{AB})) -Tr(\rho_{AB}\log_2(\Lambda[\rho_{AB}))\nonumber\\
        &=& \Tr(\rho_{AB} (\log_2 (\rho_{AB}) -\log_2(\Lambda[\rho_{AB}])))\nonumber\\
        &=& D(\rho_{AB}||(\Lambda[\rho_{AB}]))
\end{eqnarray}
Here the relative entropy $D$ is defined as $D(\rho||\sigma) = \Tr(\rho(\log(\rho)-\log(\sigma)))$ \cite{markmwilde}. 
Our main objective function in \eqref{eqn18} is a convex combination of conditional von Neumann entropies. Using \eqref{eqn17} and \eqref{eqnB.23} the objective function can thus be reformulated as,
\begin{equation}
    \begin{aligned}
    \label{eqn43}
        C^*(S) = \inf \quad &\lambda D(\rho_{AB}||(\Lambda_0[\rho_{AB})) + (1-\lambda)D(\rho_{AB}||(\Lambda_1[\rho_{AB}])) \\
    \text{s.t.} \quad &\text{Tr}\left(\rho_{AB} \cdot \text{CHSH}\right) = S\\
    &\rho_{AB} \succeq 0 \\ &\text{Tr}(\rho_{AB}) = 1
    \end{aligned}
\end{equation}

\section{Reduction of the optimization problem to two qubit space  $\mathbb{C}_{4\times4}$ of Alice and Bob}
Let $\rho_{A'BEA_x}$, $\rho_{ABEA_x'}$, $\in \D(\mathcal{P})$ for Hilbert space $\mathcal{P}$ over which the density operator space is defined. The dimension of $D(\mathcal{P})$ is unknown as the dimension of Eve's subsystem $E$ is unknown according to the standard assumptions of security analysis.
Analyzing the subsystem of Alice and Bob in a larger, unknown dimension is not only impractical but also complicates the analysis -- larger systems are inherently more susceptible to attacks due to their increased complexity. The work of \cite{Schwonnek_2021} shows that without loss of generality, one can can reduce the problem to two-qubit space $\mathbb{C}_{4\times4}$ of Alice and Bob. From the result of \cite{twoprojectortheory}, one can decompose a projector of higher dimension into projectors of dimension ${2\times 2}$ acting on either Alice's or Bob's local states. Therefore, the four pairs of projectors as in \eqref{eqn3} can be decomposed accordingly. Let $L_S^z$ be the set of pairwise-commuting projectors where $S\in \{A, B\}$ denotes the subsystem of either Alice or Bob and $z\in \{0,1,2,3\}$ corresponds to particular observables as stated earlier. Specifically $L_S^z = \{L_A^0,L_A^1,L_B^2,L_B^3\}$. Now let us consider the following projector of dimension ${2\times 2}$ and being parameterized by angle $\theta$ 
\begin{equation}
\label{eqn44}
    Q(\theta) = 
    \begin{aligned}
        \begin{pmatrix}
            \cos^2\left(\frac{\theta}{2}\right) & \cos\left(\frac{\theta}{2}\right)\sin\left(\frac{\theta}{2}\right) \\
            \cos\left(\frac{\theta}{2}\right)\sin\left(\frac{\theta}{2}\right)  & \sin^2\left(\frac{\theta}{2}\right)
        \end{pmatrix}
    \end{aligned}
\end{equation}
Having defined the local projectors as above, one can decompose $P^{0|0}_{\mathcal{X}}, P^{0|1}_{\mathcal{X}}, P^{0|2}_{\mathcal{Y}}$ and $P^{0|3}_{\mathcal{Y}}$ as,
\begin{equation}
\label{eqn45}
    \begin{aligned}
        P^{0|0}_{\mathcal{X}} = \bigoplus_i Q(0) \oplus L_A^0,~~ 
        P^{0|1}_{\mathcal{X}} = \bigoplus_i Q(\phi_A^i) \oplus L_A^1\\
        P^{0|2}_{\mathcal{Y}} = \bigoplus_j Q(0) \oplus L_B^2, ~~
        P^{0|3}_{\mathcal{Y}} = \bigoplus_j Q(\phi_B^j) \oplus L_B^3
    \end{aligned}
\end{equation}
The remaining projectors can be obtained using the completeness property and thus,
\begin{equation}
    \begin{aligned}
    \label{eqn46}
        P^{1|0}_{\mathcal{X}} = \mathbbm{1} - P^{0|0}_{\mathcal{X}},~~
        P^{1|1}_{\mathcal{X}} = \mathbbm{1} - P^{0|1}_{\mathcal{X}}\\
        P^{1|2}_{\mathcal{Y}} = \mathbbm{1} - P^{0|2}_{\mathcal{Y}},~~
        P^{1|3}_{\mathcal{Y}} = \mathbbm{1} - P^{0|3}_{\mathcal{Y}}
    \end{aligned}
\end{equation}
The angles $\{\phi_A^i\}$ and $\{\phi_B^j\}$ can be obtained from the spectrums of $P^{0|0}_{\mathcal{X}} +P^{0|1}_{\mathcal{X}} $ and $P^{0|2}_{\mathcal{Y}} + P^{0|3}_{\mathcal{Y}}$, respectively. As the projectors are block-diagonal operators, they are expressed as direct sum of commuting and noncommuting parts. The commuting parts contribute discrete and angle-independent eigenvalues, while the noncommuting parts contribute angle-dependent eigenvalues. Thus, the angles \(\phi_A^i, \phi_B^j\) can be uniquely determined from the non-integer eigenvalues in the spectrum.
Using Eq. (\ref{eqn44},\ref{eqn45}), we compute the angles $\phi_A^i$ and $\phi_A^i$ from the eigenvalues as,

\begin{equation}
\begin{aligned}
\label{eqn49}
 \cos(\phi_A/2) = \frac{\lambda_\mathcal{X}^1 - \lambda_\mathcal{X}^2}{2} \textrm{or, } \phi_A = 2 \arccos\left(\frac{\lambda_\mathcal{X}^1 - \lambda_\mathcal{X}^2}{2}\right).   \\ 
\cos(\phi_B/2) = \frac{\lambda_\mathcal{Y}^1 - \lambda_\mathcal{Y}^2}{2} \textrm{or, } \phi_A = 2 \arccos\left(\frac{\lambda_\mathcal{Y}^1 - \lambda_\mathcal{Y}^2}{2}\right).    
\end{aligned}
\end{equation}
where $\lambda_\mathcal{X}^1 = 1 + \cos(\phi_A^i/2)$, $\lambda_\mathcal{X}^2 = 1 - \cos(\phi_A^i/2)$ $\left(P^{0|2}_{\mathcal{Y}} + P^{0|3}_{\mathcal{Y}}\right)$, $\lambda_\mathcal{Y}^1 = 1 + \cos(\phi_B^j/2)$, and $\lambda_\mathcal{Y}^2 = 1 - \cos(\phi_B^j/2) $.
The mapping $\phi_N^{m} \mapsto 1 \pm \cos(\phi_{N}^m/2)$ where $N\in \{\mathcal{X},\mathcal{Y}\}$ and $m\in \{i,j\}$is bijective for $\phi_N^m \in [0, 2\pi)$ when restricted to non-integer eigenvalues. Moreover the inverse function $\textit{arccos}$ is unique in $[0, \pi]$.\\\\
\label{Lemma1}
\textbf{Lemma 1:} The mapping $\phi_N^m \mapsto 1 \pm \cos(\phi_{N}^m/2)$, where $N\in \{\mathcal{X},\mathcal{Y}\}$ and $m\in\{i,j\}$ is bijective for $\phi_N^m \in [0, \pi]$\\\\
\textit{Proof: }Let $h(\phi_N^m) = 1 \pm \cos(\phi_{N}^m/2)$ where $N\in\{\mathcal{X},\mathcal{Y}\}$ and $m\in\{i,j\}$\\

\textit{1. Proof of injectivity: } 
Let us assume that $\phi_N^1 \neq \phi_N^2 \in [0,\pi]$ where $\phi_N^1,\phi_N^2$ are arbitrarily two angles obtained from spectrum of the sum of either  $\left(P^{0|0}_{\mathcal{X}} +P^{0|1}_{\mathcal{X}}\right) $ or $\left(P^{0|2}_{\mathcal{Y}} + P^{0|3}_{\mathcal{Y}}\right)$
\begin{equation}
\cos(\phi_N^1/2) \neq \cos(\phi_N^2/2) \textrm{or, } \lambda_\mathcal{X}^1 \neq \lambda_\mathcal{X}^2 \quad \text{or} \quad (\lambda_\mathcal{Y}^1 \neq \lambda_\mathcal{Y}^2) \textrm{or, } h(\phi_N^1) \neq h(\phi_N^2)
\end{equation}
Thus $f$ is an injective function.\\

\textit{2. Proof of Surjectivity: }
The eigenvalues $\lambda_\mathcal{X}^1 , \lambda_\mathcal{X}^2$ or ($\lambda_\mathcal{Y}^1 , \lambda_\mathcal{Y}^2$) $ \in [0, 2]$ with $\lambda_\mathcal{X}^1 + \lambda_\mathcal{X}^2 = 2$ or ($\lambda_\mathcal{Y}^1 + \lambda_\mathcal{Y}^2$) and $\lambda_\mathcal{X}^1 > \lambda_\mathcal{X}^2$ or ($\lambda_\mathcal{Y}^1 > \lambda_\mathcal{Y}^2$) there exists a unique $\phi_N^m \in [0, \pi]$ where $m\in\{i,j\}$ such that the eigenvalues corresponding to $\phi_N^m$ are $\lambda_\mathcal{X}^1 , \lambda_\mathcal{X}^2$ or ($\lambda_\mathcal{Y}^1 , \lambda_\mathcal{Y}^2$)  for $N=A$ or $B$ respectively.\\
The condition of $\lambda_\mathcal{X}^1 + \lambda_\mathcal{X}^2 = 2$ or ($\lambda_\mathcal{Y}^1 + \lambda_\mathcal{Y}^2$) arises from a fundamental result of matrix algebra, \textit{trace of a matrix is the sum of its eigenvalues} and using this result we get, 
\begin{equation}
    \begin{aligned}
        \Tr(L) &= 1 + \cos^2(\phi_A^i/2) + \sin^2(\phi_A^i/2)\\
              &= 2\\
              &= \lambda_\mathcal{X}^1 + \lambda_\mathcal{X}^2
    \end{aligned}
\end{equation}
and equivalently 
\begin{equation}
    \lambda_\mathcal{Y}^1 + \lambda_\mathcal{Y}^2 = 2
\end{equation}
Given the above three criteria, one can have,
\begin{equation}
\begin{aligned}
\label{eqn53}
    c_\mathcal{X} = \frac{    \lambda_\mathcal{X}^1 - \lambda_\mathcal{X}^2 
}{2}\\
c_\mathcal{Y} = \frac{    \lambda_\mathcal{Y}^1 - \lambda_\mathcal{Y}^2 
}{2}    
\end{aligned}
\end{equation}
\textit{Compute $c_\mathcal{X}$:}
    Since $\lambda_\mathcal{X}^1 > \lambda_\mathcal{X}^2 $ and $\lambda_\mathcal{X}^1 , \lambda_\mathcal{X}^2 \in [0, 2]$, one can have $\lambda_\mathcal{X}^1 - \lambda_\mathcal{X}^2  > 0$ and $\lambda_\mathcal{X}^1 = 2-\lambda_\mathcal{X}^2 $. Thus,
    \begin{equation}
    c_\mathcal{X} = \frac{ 2- \lambda_\mathcal{X}^2 - \lambda_\mathcal{X}^2}{2} = \frac{2 - 2\lambda_\mathcal{X}^2}{2} = 1 - \lambda_\mathcal{X}^2    
    \end{equation}
    Since $\lambda_\mathcal{X}^1 > \lambda_\mathcal{X}^2$ and $\lambda_\mathcal{X}^1 + \lambda_\mathcal{X}^2 = 2$, we have $2\lambda_\mathcal{X}^2 < 2$, so $\lambda_\mathcal{X}^2 < 1$. Therefore, $c_\mathcal{X} = 1 - \lambda_\mathcal{X}^2 > 0$.
    Also, since $\lambda_\mathcal{X}^1 < 2$, one can have $\lambda_\mathcal{X}^1 - \lambda_\mathcal{X}^2 < 2 - \lambda_\mathcal{X}^2$. Since $\lambda_\mathcal{X}^2 \ge 0$, $\lambda_\mathcal{X}^1 - \lambda_\mathcal{X}^2 < 2$. Thus $c_\mathcal{X} < 1$.
    Therefore, $c \in (0, 1)$.\\\\
 \textit{Solve for the given angle $\phi_A^i$:}
Now using \eqref{eqn49} and \eqref{eqn53} one can have,
 \begin{equation}
    \phi_A = 2 \arccos(c_\mathcal{X})
 \end{equation}
    Since $c_\mathcal{X} \in (0, 1)$, $\arccos(c_\mathcal{X}) \in (0, \frac{\pi}{2})$. Therefore, $    \phi_A^i = 2 \arccos(c_\mathcal{X})
 \in (0, \pi)$.\\\\
\textit{3. Uniqueness: }
    The function $\arccos(c_\mathcal{X})$ is \textit{strictly decreasing} on the interval $[0, 1]$. Therefore, for each $c_\mathcal{X} \in (0, 1)$, there is a unique $\arccos(c_\mathcal{X}) \in (0, \frac{\pi}{2})$. Consequently, there is a unique $\phi = 2 \arccos(c_\mathcal{X}) \in (0, \pi)$.\\\\
\textit{4. Eigenvalues: }
    The eigenvalues are given by,
\begin{equation}
\begin{aligned}
\label{eqn56}
    \lambda_\mathcal{X}^1,\lambda_\mathcal{X}^2 =1 \pm \cos(\phi_{A}^i/2)  \\
    \lambda_\mathcal{Y}^1,\lambda_\mathcal{Y}^2 = 1 \pm \cos(\phi_{B}^j/2)
\end{aligned}    
\end{equation}
    Substituting $    \phi_A^i = 2 \arccos(c_\mathcal{X})
$, one can get:
    \begin{equation}
    \lambda_\mathcal{X}^1,\lambda_\mathcal{X}^2 = 1 \pm \cos(\arccos(c_\mathcal{X})) = 1 \pm c_\mathcal{X}
    \end{equation}
    Thus, $\lambda_\mathcal{X}^1 = 1 + c_\mathcal{X}$ and $\lambda_\mathcal{X}^2 = 1 - c_\mathcal{X}$.
    Substituting $ c_\mathcal{X} = \frac{\lambda_\mathcal{X}^1 - \lambda_\mathcal{X}^2 
}{2}$, one gets,
\begin{equation}
    \begin{aligned}
    \lambda_\mathcal{X}^1 = 1 + \frac{\lambda_\mathcal{X}^1 - \lambda_\mathcal{X}^2}{2}  \textrm{or, } 2\lambda_\mathcal{X}^1 = 2 + \lambda_\mathcal{X}^1 - \lambda_\mathcal{X}^2 \textrm{or, } \lambda_\mathcal{X}^1 = 2 - \lambda_\mathcal{X}^2\\
    \lambda_\mathcal{X}^2 = 1 - \frac{\lambda_\mathcal{X}^1 - \lambda_\mathcal{X}^2}{2} \textrm{or, } 2\lambda_\mathcal{X}^2 = 2 - \lambda_\mathcal{X}^1 + \lambda_\mathcal{X}^2 \textrm{or, } \lambda_\mathcal{X}^2 = 2 - \lambda_\mathcal{X}^1      
    \end{aligned}
\end{equation}
Therefore, for any non-integer pair $\lambda_\mathcal{X}^1, \lambda_\mathcal{X}^2 \in [0, 2]$ with $\lambda_\mathcal{X}^1 + \lambda_\mathcal{X}^2 = 2$ and $\lambda_\mathcal{X}^1 > \lambda_\mathcal{X}^2$, there exists a unique $\phi_A \in (0, \pi)$ such that the eigenvalues corresponding to $\phi_A^i$ are $\lambda_\mathcal{X}^1$ and $\lambda_\mathcal{X}^2$.
Similarly by substituting $\phi_B^j = 2arccos(c_\mathcal{Y})$ in \eqref{eqn56} we prove the surjectivity of the mapping. Since the map is both injective and surjective, this completes the proof that it is bijective.\\
\qed\\

\subsection{Computing lower bound of $C^*(S)$ }
Having established the functional relationship between the angles $\phi_A^i$ and $\phi_B^j$ and characterized the spectral properties of the operators $P^{0|0}_{\mathcal{X}} + P^{0|1}_{\mathcal{X}}$ (or equivalently, $P^{0|2}_{\mathcal{Y}} + P^{0|3}_{\mathcal{Y}}$), we now proceed to find a lower bound of the function $C^*(S)$. Let $\Lambda_{\mathcal{X}}$ and $\Lambda_{\mathcal{Y}}$ be the Pinching channels that act on Alice's and Bob's subsystems and decompose Alice's and Bob's subsystems into block structures as would be obtained from the effect of the projectors in Eq. \eqref{eqn45}. 

Now let $\Lambda_{\mathcal{XY}} = \Lambda_{\mathcal{X}} \otimes \Lambda_{\mathcal{Y}}$ be the channel acting on the combined state $\rho_{ABEA_X'}$, resulting to a joint state of Alice and Bob corresponding to the angles $\phi_A^i$ and $\phi_B^j$ and a commuting part obtained from either $L_A^0 \otimes L_A^1$ or $L_B^2 \otimes L_B^3$. Let ${P}$ be the set of all states that can be obtained from the operator $Q(\phi_A^i) \otimes Q(\phi_B^j)$ acting on the state $\rho_{ABEA_X'}$, 
\begin{equation}
\label{eqn81}
    {P} = \{\rho_{ABEA_X'}^{ij} \mid Q(\phi_A^i) \otimes Q(\phi_B^j)(\rho_{ABEA_X'}) = \rho_{ABEA_X'}^{ij} \}
\end{equation}
Thus,
\begin{equation}
    \Lambda_{\mathcal{XY}}[\rho_{ABEA_X'}] = \bigoplus_{ij} (\eta_{ij} \rho_{ABEA_X'}^{ij}) \oplus (\eta_{\text{comm}} \rho_{ABEA_X'}^{\text{comm}})
\end{equation}
where $\eta_{ij}$ are normalization factors obtained by normalizing each state $\rho_{ABEA_X'}^{ij} \in {P}$. Each of the states depends on the operators $Q(\phi_A^i) \otimes Q(\phi_B^j)$ and consequently on the angles obtained from respective spectra. The $\rho_{ABEA_X'}^{\text{comm}}$ are the projected blocks that commute with either Alice's subsystem or Bob's subsystem. Now, using the monotonicity property of relative entropy, we get from Eq. (B.24),
\begin{equation}
    \begin{aligned}
    \label{eqn83}
        C^*(S) = \inf \quad & \lambda D\left( \rho_{AB} \,||\, \Lambda_0[\rho_{AB}]\right) 
        + (1-\lambda)D\left( \rho_{AB} \,||\, \Lambda_1[\rho_{AB}]\right) \\
        & \geq \lambda D\left( \Lambda_{\mathcal{XY}}[\rho_{AB}] \,||\, \Lambda_0 \circ \Lambda_{\mathcal{XY}} [\rho_{AB}]\right) \\
        & + (1-\lambda)D\left( \Lambda_{\mathcal{XY}}[\rho_{AB}] \,||\, \Lambda_1 \circ \Lambda_{\mathcal{XY}} [\rho_{AB}]\right) \\
        &\geq \sum_{ij}\eta_{ij}\left[\lambda D\left( \rho_{AB}^{ij} \,||\,  \Lambda_0[\rho_{AB}^{ij}]\right) \right. \\
        & \qquad \left. + (1-\lambda)D\left(  \rho_{AB}^{ij} \,||\,  \Lambda_1[\rho_{AB}^{ij}] \right) \right] \\
        & \qquad + \eta_{\text{comm}}\left[\lambda D\left(  \rho_{AB}^{\text{comm}} \,||\,  \Lambda_0[\rho_{AB}^{\text{comm}}] \right) \right. \\
        & \qquad \left. + (1-\lambda)D\left( \rho_{AB}^{\text{comm}} \,||\,  \Lambda_1[\rho_{AB}^{\text{comm}}]\right) \right]\\
       & \text{Tr}(\rho_{AB}~{CHSH}) = S \\
    &\rho_{AB} \succeq 0 \\ &\text{Tr}(\rho_{AB}) = 1
    \end{aligned}
\end{equation}
\\
where,
  $\rho_{AB}^{ij} = \mbox{Tr}_{EA_X'}(\rho_{ABEA_X'}^{ij})$ \quad and \quad  
  $\Lambda[\rho_{AB}^{ij}] = \mbox{Tr}_{EA_X}(\Lambda[\rho_{ABEA_X'}^{ij}])$\\\\
  $\rho_{AB}^{\text{comm}} = \mbox{Tr}_{EA_X'}(\rho_{ABEA_X'}^{\text{comm}})$ \quad and \quad
  $\Lambda[\rho_{AB}^{\text{comm}}] = \mbox{Tr}_{EA_X}(\Lambda[\rho_{ABEA_X'}^{\text{comm}}])$
\\\\
When Alice's observables commute with that of and Bob,  the joint probabilities of the measurement outcomes can be factorized into local outcome probabilities. Using this fact, one can can similarly analyze the state $\rho_{ABEA_X'}^{\text{comm}}$, where $X \in \{0, 1\}$ represents the two measurement outcomes. Due to the classical nature of these probability distributions, we can can disregard the contribution from the part $\rho_{ABEA_X'}^{\text{comm}}$ in the violation of CHSH inequality, as it will lead to $\text{Tr}(\rho_{AB}^{\text{comm}}~{CHSH}) \leq 2$.
Now, from the constraint of the optimization problem in \eqref{eqn43}, we can establish the same for each state $\rho_{ABEA_X'}^{ij} \in \mathcal{P}$ as,
\begin{equation}
    \begin{aligned}
        \text{Tr}(\rho_{AB}~{CHSH}) &= S \\
        \Leftrightarrow \quad \text{Tr}(\rho_{AB}^{ij}~{CHSH}) &= S_{ij} \quad \text{and} \quad \sum_{ij} \eta_{ij} S_{ij} = S
    \end{aligned}
\end{equation}
Having established the constraint for each state $\rho_{ABEA_X'}^{ij} \in \mathcal{P}$, we can define the optimization problem for finding the lower bound of the function $C^*(S)$ as,
\begin{equation}
    \begin{aligned}
        \label{eqn85}
        C^*(S) \geq \inf \quad &\sum_{ij}\eta_{ij}\left[\lambda D\left( \rho_{AB}^{ij} \,||\,  \Lambda_0[\rho_{AB}^{ij}]\right) 
         + (1-\lambda)D\left(  \rho_{AB}^{ij} \,||\,   \Lambda_1[\rho_{AB}^{ij} ]\right) \right] \\
        \text{s.t.} \quad & \text{Tr}((\rho_{AB}^{ij}~{CHSH}) = S_{ij} \\
    &\rho_{AB}^{ij} \succeq 0 \\ &\text{Tr}(\rho_{AB}^{ij}) = 1\\
        \quad & \sum_{ij} \eta_{ij} \leq 1\\
        \quad & \sum_{ij} \eta_{ij}~ S_{ij} = S
    \end{aligned}
\end{equation}
The state $\rho_{ABEA_X'}$ is first passed through the channel $\Lambda_{\mathcal{XY}}$, which essentially decomposes the state into blocks along the principal diagonal as described in \eqref{eqn83}, before passing it through the pinching channel. Therefore, our objective function in \eqref{eqn85} is essentially lower bounded by the minimum values in each such block.
Thus,
\begin{equation}
    \begin{aligned}
        \label{eqn86}
        &\left[ \lambda D\left(  \rho_{AB}^{ij} \,||\,  \Lambda_0[\rho_{AB}^{ij}] \right)
         + (1-\lambda)D\left(  \rho_{AB}^{ij} \,||\,  \Lambda_1[\rho_{AB}^{ij}] \right) \right] 
        \geq C^*_{\mathbb{C}^{4 \times 4}}(S_{ij})
    \end{aligned}
\end{equation}
where,
\begin{equation}
    \begin{aligned}
        \label{eqn87}
        C^*_{\mathbb{C}^{4 \times 4}}(S_{ij}) = \inf \quad &\left[ \lambda D\left(  \rho_{AB}^{ij} \,||\,   \Lambda_0[\rho_{AB}^{ij}] \right)
        + (1-\lambda)D\left(  \rho_{AB}^{ij} \,||\,  \Lambda_1[\rho_{AB}^{ij}] \right) \right] \\
        \text{s.t.} \quad & \text{Tr}( \rho_{AB}^{ij}~{CHSH}) = S_{ij} \\
    &\rho_{AB}^{ij} \succeq 0 \\ &\text{Tr}(\rho_{AB}^{ij}) = 1\\
    \end{aligned}
\end{equation}
Thus using \eqref{eqn85},\eqref{eqn86} and \eqref{eqn87}, one can lower bound the function $C^*(S)$ as,
\begin{equation}
    \begin{aligned}
    \label{eqn88}
        C^*(S) &\geq \sum_{ij}\eta_{ij} \quad \text{inf} \quad\left( C^*_{\mathbb{C}^{4 \times 4}}(S_{ij})\right)\\
         \text{s.t.} \quad  & \text{Tr}( \rho_{AB}^{ij}~{CHSH}) = S_{ij} \\
    &\rho_{AB}^{ij} \succeq 0 \\ &\text{Tr}(\rho_{AB}^{ij}) = 1\\
        \quad & \sum_{ij} \eta_{ij} \leq 1\\
        \quad & \sum_{ij} \eta_{ij} S_{ij} = S
    \end{aligned}
\end{equation}
Now \eqref{eqn88} is independent of the angles $\phi_A^i$ and $\phi_B^j$ as each states obtained from \eqref{eqn81} are already optimized in \eqref{eqn87}. Therefore, we can reduce the optimization of $C^*(S)$  to optimizing only over to $\eta_{ij}$ as,
\begin{equation}
\boxed{
    \begin{aligned}
    \label{eqn89}
        C^*(S) &\geq \int_{S'=2}^{2\sqrt{2}}         C^*_{\mathbb{C}^{4 \times 4}}(S') \cdot \eta(dS')\\
        &\geq \int_{S'=2}^{2\sqrt{2}}\eta(dS')\cdot C^*_{\mathbb{C}^{4 \times 4}}(S')\\
        s.t \quad &\eta([2,2\sqrt{2}]) \leq 1 \\
        &\eta \geq 0\\
        &\int_{S'=2}^{2\sqrt{2}} \eta(dS')S' = S
    \end{aligned}}
\end{equation}
Here, essentially one single block is considered and  integrated over positive sub-normalized weights $\eta$ in the interval $S' = (2,2\sqrt{2}]$ with the conditions, 
\begin{equation}
C^*_{\mathbb{C}^{4 \times 4}}(S_{ij})=0~~ \forall S_{ij}\leq0    
\end{equation}
\subsection{Reformulating the CHSH operator in explicit matrix form}
Having established the optimization problem in terms of two-qubit space $\mathbb{C}_{4\times4}$ of Alice and Bob, we now find an explicit matrix representation of the CHSH operator in \eqref{eqn10} using \eqref{eqn11},
\begin{equation}
\begin{aligned}
    CHSH &=  C^{O^{\mathcal{X}}_{1}} \otimes C^{O^{\mathcal{Y}}_{2}} - C^{O^{\mathcal{X}}_{0}} \otimes C^{O^{\mathcal{Y}}_{2}} - C^{O^{\mathcal{X}}_{0}} \otimes C^{O^{\mathcal{Y}}_{3}} - C^{O^{\mathcal{X}}_{1}} \otimes C^{O^{\mathcal{Y}}_{3}}\\
    &=\left[(P^{0|1}_{\mathcal{X}} \otimes P^{0|2}_{\mathcal{Y}} + P^{1|1}_{\mathcal{X}} \otimes P^{1|2}_{\mathcal{Y}}) - (P^{0|1}_{\mathcal{X}} \otimes P^{1|2}_{\mathcal{Y}} + P^{1|1}_{\mathcal{X}} \otimes P^{0|2}_{\mathcal{Y}})\right]\\
    &-\left[(P^{0|0}_{\mathcal{X}} \otimes P^{0|2}_{\mathcal{Y}} + P^{1|0}_{\mathcal{X}} \otimes P^{1|2}_{\mathcal{Y}}) - (P^{0|0}_{\mathcal{X}} \otimes P^{1|2}_{\mathcal{Y}} + P^{1|0}_{\mathcal{X}} \otimes P^{0|2}_{\mathcal{Y}})\right]\\
    &-\left[(P^{0|0}_{\mathcal{X}} \otimes P^{0|3}_{\mathcal{Y}} + P^{1|0}_{\mathcal{X}} \otimes P^{1|3}_{\mathcal{Y}}) - (P^{0|0}_{\mathcal{X}} \otimes P^{1|3}_{\mathcal{Y}} + P^{1|0}_{\mathcal{X}} \otimes P^{0|3}_{\mathcal{Y}})\right]\\
    &-\left[(P^{0|1}_{\mathcal{X}} \otimes P^{0|3}_{\mathcal{Y}} + P^{1|1}_{\mathcal{X}} \otimes P^{1|3}_{\mathcal{Y}}) - (P^{0|1}_{\mathcal{X}} \otimes P^{1|3}_{\mathcal{Y}} + P^{1|1}_{\mathcal{X}} \otimes P^{0|3}_{\mathcal{Y}})\right]\\    
\end{aligned}
\end{equation}
Now using the explicit matrix form of the projectors in \eqref{eqn45} and \eqref{eqn46}, the CHSH operator can further be decomposed as,
\begin{equation}
\label{eqn92}
\begin{aligned}
CHSH &= \left[\left(Q(\phi_A^i) \otimes Q(0)\right) + \left((\mathbbm{1} - Q(\phi_A^i)) \otimes (\mathbbm{1} - Q(0))\right) -
  \left(Q(\phi_A^i) \otimes (\mathbbm{1} - Q(0))\right) - \left((\mathbbm{1} - Q(\phi_A^i)) \otimes Q(0)\right)\right]\\
&- \left[\left(Q(0) \otimes Q(0)\right) + \left(\mathbbm{1} - Q(0)) \otimes (\mathbbm{1} - Q(0))\right) - \left(Q(0) \otimes (\mathbbm{1} - Q(0)\right)) - \left(\mathbbm{1} - Q(0)) \otimes Q(0)\right)\right]\\
&- \left[\left(Q(0) \otimes Q(\phi_B^j)\right) + \left( (\mathbbm{1} - Q(0)) \otimes (\mathbbm{1} - Q(\phi_B^j))\right) -  \left(Q(0) \otimes (\mathbbm{1} - Q(\phi_B^j))\right) - \left((\mathbbm{1} - Q(0)) \otimes Q(\phi_B^j)\right)\right]\\ &- \left[\left(Q(\phi_A^i) \otimes Q(\phi_B^j)\right) + \left((\mathbbm{1} - Q(\phi_A^i)) \otimes (\mathbbm{1} - Q(\phi_B^j))\right) -  \left(Q(\phi_A^i) \otimes (\mathbbm{1} - Q(\phi_B^j))\right) - \left((\mathbbm{1} - Q(\phi_A^i)) \otimes Q(\phi_B^j)\right)\right]
\end{aligned}
\end{equation}
Solving the tensor products in each term above, one can explicitly write the CHSH operator as,

\begin{equation}
\label{eqn97}
  CHSH =   \begin{bmatrix}
    A & B \\
    C & D
  \end{bmatrix}
\end{equation}
where 
\begin{equation}
\begin{aligned}
\label{eqn98}
A = \begin{pmatrix}
\cos(\phi_A^i) - 1 - \cos(\phi_B^j) - \cos(\phi_A^i) \cos(\phi_B^j) & -\sin(\phi_B^j) - \cos(\phi_A^i) \sin(\phi_B^j) \\
-\sin(\phi_B^j) - \cos(\phi_A^i) \sin(\phi_B^j) & -\cos(\phi_A^i) + 1 + \cos(\phi_B^j) + \cos(\phi_A^i) \cos(\phi_B^j)
\end{pmatrix}
\end{aligned}
\end{equation}
\begin{equation}
\begin{aligned}
\label{eqn99}
 B = \begin{pmatrix}
\sin(\phi_A^i) - \sin(\phi_A^i) \cos(\phi_B^j) & -\sin(\phi_A^i) \sin(\phi_B^j)\\ -\sin(\phi_A^i) \sin(\phi_B^j) & \sin(\phi_A^i) \cos(\phi_B^j) - \sin(\phi_A^i)
\end{pmatrix}
\end{aligned}
\end{equation}
\begin{equation}
\begin{aligned}
\label{eqn100}
 C = \begin{pmatrix}
\sin(\phi_A^i) - \sin(\phi_A^i) \cos(\phi_B^j) & -\sin(\phi_A^i) \sin(\phi_B^j)\\-\sin(\phi_A^i) \sin(\phi_B^j) & \sin(\phi_A^i) \cos(\phi_B^j) - \sin(\phi_A^i)
\end{pmatrix}
\end{aligned}
\end{equation}
\begin{equation}
\begin{aligned}
\label{eqn101}
 D = \begin{pmatrix}
-\cos(\phi_A^i) + 1 + \cos(\phi_B^j) + \cos(\phi_A^i) \cos(\phi_B^j) & \sin(\phi_B^j) + \cos(\phi_A^i) \sin(\phi_B^j)\\\sin(\phi_B^j) + \cos(\phi_A^i) \sin(\phi_B^j) & \cos(\phi_A^i) - 1 - \cos(\phi_B^j) - \cos(\phi_A^i) \cos(\phi_B^j)
\end{pmatrix}
\end{aligned}
\end{equation}
Here individual operators  $A,B,C$ and $D$ are Hermitian and  $B= C^*$ implies that CHSH operator in \eqref{eqn97} is also Hermitian.
Now from \textit{Lemma 1,} the mapping $\phi_N^m \mapsto 1 \pm \cos(\phi_{N}^m/2)$ is bijective for $\phi_N^m \in [0, \pi]$. Since \textit{sin} and \textit{cosine} functions are monotonous and continuous for $\phi_N^m \in [0, \pi/2]$, the mapping is also bijective in the subinterval of $[0,\pi/2]$.
In the first quadrant, both \textit{sin} and \textit{cosine} functions are positive; therefore, restricting the arguments of these functions to the first quadrant ensures that the CHSH operator in \eqref{eqn97} does not change. Moreover, since both \textit{sin} and \textit{cosine} functions have unique values in the first quadrant,  the optimization problem in Eq. \eqref{eqn87} for each block can be reformulated with the explicit matrix representation of CHSH operator in \eqref{eqn97} as, 
\begin{equation}
    \boxed{\begin{aligned}
    \label{eqn102}
        C^*_{\mathbb{C}^{4 \times 4}}(S_{ij}) = \inf \quad &\left[\lambda D\left(  \rho_{AB}^{ij} \,||\,  \Lambda_0[\rho_{AB}^{ij}] \right) 
     + (1-\lambda)D\left( \rho_{AB}^{ij} \,||\,   \Lambda_1[\rho_{AB}^{ij}] \right) \right] \\
        \text{s.t.} \quad 
        &\text{Tr}\left(\rho_{AB}^{ij} \cdot     \begin{bmatrix}
    
A & B \\
C & D
\end{bmatrix}\right) = S_{ij}\\
    &\rho_{AB}^{ij} \succeq 0 \\ &\text{Tr}(\rho_{AB}^{ij}) = 1\\
& \phi_A^i, \phi_B^j \in [0,\pi/2]
    \end{aligned}}
\end{equation}

\section{Formulation of the objective function in terms of trace norm using a modified form of Pinsker's inequality}

Having established the objective function in terms of a two-qubit vector space $\C_{4\times4}$ describing a single block and expressed the CHSH operator in its explicit matrix form \eqref{eqn97}, we now focus on finding a lower bound on the relative entropies described in Eq. \eqref{eqn102} through a modified form of Pinsker's inequality\cite{Schwonnek_2021}.
The modified version of Pinsker's inequality, as shown in the original work, is stated in the following theorem,

\textbf{Theorem 1: }
  Let \( Q \) be a projector, not necessarily of rank 1, defines the action of a pinching channel $\Lambda$  acting on a state $\rho$ as $\Lambda[\rho] = Q\rho Q + (\mathbbm{1} - Q)\rho(\mathbbm{1}-Q)$. Then the quantum relative entropy between $\rho$ and $\Lambda[\rho]$ is lower bounded as,
\begin{equation}
    D(\rho || \Lambda[\rho]) \geq \mbox{log}_2(2) - h\left(\frac{1-||\rho - \Lambda[\rho]||_1}{2}\right)
\end{equation}
where $h(p) = -\sum_{i=0}^{1}p_i \mbox{log}_2(p_i)  $ is the binary Shannon entropy.\\\\
\textit{Proof : } The pinching channel $\Lambda$ is defined as,
    \begin{equation}
    \label{eqn3.1}
    \begin{aligned}
        \Lambda[\rho] &= Q\rho Q + (\mathbbm{1} - Q)\rho(\mathbbm{1}-Q)\\
        &= Q\rho Q+ (\rho - Q\rho)(\mathbbm{1}-Q)\\
        &= Q\rho Q + (\rho   - \rho Q  -Q \rho + Q \rho Q)\\
        &= 2Q \rho Q + \rho - \{Q,\rho\} 
    \end{aligned}
    \end{equation}
Now applying \textit{Theorem 1} in Eq. \eqref{eqn102}, we can have,
\begin{equation}
\label{eqn3.3}
    \begin{aligned}
        C^*_{\mathbb{C}^{4 \times 4}}(S_{ij}) &= \inf \quad \left[ \lambda D\left(   \rho_{AB}^{ij} \,||\, \Lambda_0[\rho_{AB}^{ij}] \right)
        + (1-\lambda)D\left(  \rho_{AB}^{ij} \,||\,   \Lambda_1[\rho_{AB}^{ij}] \right) \right] \\
        &\geq \left[ \lambda \left( \log_2(2) - h\left(\frac{1- ||   \rho_{AB}^{ij} -  \Lambda_0[\rho_{AB}^{ij}] ||_1}{2} \right) \right) \right. \\
        &\qquad \left. + (1-\lambda)\left( \log_2(2) - h\left(\frac{1- ||  \rho_{AB}^{ij} -  \Lambda_1[\rho_{AB}^{ij}] ||_1}{2} \right) \right) \right] \\
        &\geq  \left[ \lambda \log_2(2) - \lambda h\left(\frac{1- ||  \rho_{AB}^{ij} -  \Lambda_0[\rho_{AB}^{ij}] ||_1}{2} \right) \right. \\
        &\qquad \left. + (1-\lambda)\log_2(2) - (1-\lambda) h\left(\frac{1- ||  \rho_{AB}^{ij} -   \Lambda_1[\rho_{AB}^{ij}] ||_1}{2} \right) \right] \\
       & \geq  \left[ \log_2(2) - \lambda h\left(\frac{1- || \rho_{AB}^{ij} -  \Lambda_0[\rho_{AB}^{ij}] ||_1}{2} \right) \right. \\
        &\qquad \left. - (1-\lambda) h\left(\frac{1- || \rho_{AB}^{ij} -  \Lambda_1[\rho_{AB}^{ij}] ||_1
        }{2} \right) \right] \\
        \text{s.t.} \quad & \text{Tr}\left(\rho_{AB}^{ij} \cdot \begin{bmatrix} A & B \\ C & D \end{bmatrix}\right) = S_{ij}
    \end{aligned}
\end{equation}
Let us now define two arguments $a$ and $b$ as,
\begin{equation}
\label{eqn3.4}
    \begin{aligned}     
a &= \frac{1 - \left\| \rho_{AB}^{ij} - \Lambda_0[\rho_{AB}^{ij}] \right\|_1}{2},\\
b &= \frac{1 - \left\| \rho_{AB}^{ij} - \Lambda_1[\rho_{AB}^{ij}] \right\|_1}{2}
    \end{aligned}
\end{equation}
As $\left\| \cdot \right\|_1 \leq 1$ is valid for normalized quantum states, one can have $a, b \in [0, 0.5]$. This ensures that the entropy functions $h(a)$ and $h(b)$ are well-defined.
The binary entropy function, defined as $h(p) = -p\log_2 p - (1-p)\log_2(1-p)$ for $p \in [0, 1]$, is a concave function from \textit{Theorem 2.1} in \cite{rasa2015concavityentropies}. Thus from Jensen's inequality \cite{Boyd_Vandenberghe_2004}, for $\lambda \in [0, 1]$ and $a, b \in [0, 1]$, we have:
\begin{equation}
\label{eqn3.5}
\begin{aligned}
\lambda h(a) + (1-\lambda)h(b) \leq h(\lambda a + (1-\lambda)b)\\     -\left[ \lambda h(a) + (1-\lambda)h(b) \right] \geq -h(\lambda a + (1-\lambda)b)  
\end{aligned}
\end{equation}
Finally, using Eqns. \eqref{eqn3.3}, \eqref{eqn3.4}, and \eqref{eqn3.5}, we get the lower the bound on $ C^*_{\mathbb{C}^{4 \times 4}}(S_{ij})$ as,
\begin{equation}
\begin{aligned}
\label{eqnD.6}
     C^*_{\mathbb{C}^{4 \times 4}}(S_{ij}) &\geq \log_2(2) - h(\lambda a + (1-\lambda)b)\\
&\geq \log_2(2) - h\left( \frac{1}{2} - \frac{1}{2} n\right)
\end{aligned}
\end{equation}
where 
\begin{equation}
\label{eqn3.7}
n= 
    \begin{aligned}
        \left( \lambda \left\| \rho_{AB}^{ij} - \Lambda_0[\rho_{AB}^{ij}] \right\|_1\ + (1-\lambda)\left\| \rho_{AB}^{ij} - \Lambda_1[\rho_{AB}^{ij}] \right\|_1 \right)
    \end{aligned}
\end{equation}
and it represents a convex combination of trace norms.
Now, one may assume that there exists a function $n^*(S_{ij})\geq n$, which gives an upper bound on the given convex combination of trace norms \eqref{eqn3.7}. Hence, we can get the upper bound by solving the below optimization problem, 
\begin{equation}
\label{eqn3.8}\boxed{
\begin{aligned}
    n^*(S_{ij}) = \inf \quad &\left[\left( \lambda \left\| \rho_{AB}^{ij} - \Lambda_0[\rho_{AB}^{ij}] \right\|_1\ + (1-\lambda)\left\| \rho_{AB}^{ij} - \Lambda_1[\rho_{AB}^{ij}] \right\|_1 \right) \right] \\
        \text{s.t.} \quad
        &\text{Tr}\left(\rho_{AB}^{ij} \cdot     \begin{bmatrix}
        A & B \\
C & D
\end{bmatrix}\right) = S_{ij}\\
    &\rho_{AB}^{ij} \succeq 0 \\ &\text{Tr}(\rho_{AB}^{ij}) = 1\\
& \phi_A^i, \phi_B^j \in [0,\pi/2]\end{aligned}}
\end{equation}
\section{Semi definite programming formulation of the objective function for fixed $\phi_A^i$ and $\phi_B^j$  }
A semidefinite program (SDP) is an optimisation problem of a linear function defined over a positive semidefinite variable, subjected to affine constraints as \cite{sikora2012analyzing},
\begin{equation}
\label{eqn4.1}
    \begin{aligned}
        \alpha = \text{maximize} \quad & \langle A, X \rangle \\
        \text{s.t.} \quad & \Phi(X) = B \\
        & X \in \text{Pos}(\mathcal{X})
    \end{aligned}
\end{equation}
where,
\begin{equation}
\label{eqn4.2}
    \begin{aligned}
        & X \in \C^\Sigma, \quad Y \in \C^\Omega \\
        & A \in \text{Herm}(X), \quad B \in \text{Herm}(Y) \\
        & \Phi : \text{Herm}(X) \rightarrow \text{Herm}(Y) \\
        & (A, B, \Phi) \text{ problem's data} \\
        & \langle A, X \rangle \text{ is the objective function} \\
        & \Phi(X) = B, \text{ and } X \in \text{Pos}(\mathcal{X}) \\
        & \alpha \text{ is the optimal value}
    \end{aligned}
\end{equation}
The trace norm of a quadratic matrix M can be represented in the form of a trace of two additional matrices P and Q \cite{Boyd_Vandenberghe_2004} as,
\begin{equation}
\label{eqn4.3}
\begin{aligned}
     ||M||_1 &= inf \quad \frac{1}{2} \Tr(P + Q)\\
    &s.t.  \begin{pmatrix}
        P & M \\
        M^* & Q
    \end{pmatrix} \succeq 0 
\end{aligned}
\end{equation}
Now the terms $\rho_{AB}^{ij} - \Lambda_0[\rho_{AB}^{ij}]$ and $\rho_{AB}^{ij} - \Lambda_1[\rho_{AB}^{ij}]$ in Eq. \eqref{eqn3.8} can be decomposed using Eq. \eqref{eqn3.1} and Eq. \eqref{eqn20} as,
\begin{equation}
\label{eqn4.4}
    \begin{aligned}
        \rho_{AB}^{ij} - &\Lambda_0[\rho_{AB}^{ij}] \\
        &= \rho_{AB}^{ij} - \left(2Q(0)\otimes\I~\rho_{AB}^{ij}~Q(0)\otimes\I + \rho_{AB}^{ij} - \{ Q(0)\otimes \I,\rho_{AB}^{ij}\}\right) \\
        &= \rho_{AB}^{ij} - \left(2Q(0)\otimes \I~\rho_{AB}^{ij}~Q(0)\otimes \I + \rho_{AB}^{ij} -  \left(Q(0)\otimes \I~ \rho_{AB}^{ij} - \rho_{AB}^{ij} ~Q(0)\otimes \I\right) \right) \\
        &=  - 2Q(0)\otimes \I~\rho_{AB}^{ij}~Q(0)\otimes \I +  \left(Q(0)\otimes \I~\rho_{AB}^{ij} + \rho_{AB}^{ij} ~Q(0)\otimes \I\right)\\
        &=    \left(Q(0)\otimes \I~\rho_{AB}^{ij} + \rho_{AB}^{ij} ~Q(0)\otimes \I\right)- 2Q(0)\otimes \I~\rho_{AB}^{ij}~Q(0)\otimes \I \\
    \end{aligned}
\end{equation}
In a similar fashion, we can also expand the other term in Eq. (D.8) as,
\begin{equation}
\label{eqn4.5}
    \begin{aligned}
        \rho_{AB}^{ij} - &\Lambda_1[\rho_{AB}^{ij}] 
        = \left(Q(\phi_A^i)\otimes \I~\rho_{AB}^{ij} + \rho_{AB}^{ij} ~ Q(\phi_A^i)\otimes \I\right)- 2Q(\phi_A^i)\otimes \I~\rho_{AB}^{ij}~Q(\phi_A^i)\otimes \I
    \end{aligned}
\end{equation}
The SDP formulation of the objective function can now be formally done using the results of \eqref{eqn3.8},\eqref{eqn4.1},\eqref{eqn4.2},\eqref{eqn4.3},\eqref{eqn4.4}, and \eqref{eqn4.5} as,
\begin{equation}
\label{eqn4.6}\boxed{
    \begin{aligned}
        n^*(S_{ij}) = - \text{maximize} & \quad \frac{\lambda}{2} \langle (P_0 + Q_0),X_0 \rangle + \frac{(1 - \lambda)}{2}\langle (P_1 + Q_1),X_1 \rangle\\
        \text{s.t.} \quad & \text{Tr}\left(\rho_{AB}^{ij} ~\begin{bmatrix}A & B \\ C & D\end{bmatrix}\right) = S_{ij}, \\
& \begin{pmatrix}P_0 & M_0 \\ M_0^* & Q_0\end{pmatrix} \succeq 0, \quad \begin{pmatrix}P_1 & M_1 \\ M_1^* & Q_1\end{pmatrix} \succeq 0, \\
& Q(\phi_A^i) = \begin{bmatrix}q_{11} & q_{12} \\ q_{12}^* & q_{22}\end{bmatrix}, \quad q_{11} + q_{22} = 1, \quad q_{12}^2 \leq q_{11}q_{22}, \\
& \rho_{AB}^{ij} \succeq 0, \quad \text{Tr}(\rho_{AB}^{ij}) = 1    \end{aligned}}
\end{equation}
where
\begin{equation}
\begin{aligned}
      & M_0 = M_0^* =  \left(Q(0)\otimes \I~\rho_{AB}^{ij} + \rho_{AB}^{ij} ~Q(0)\otimes \I\right)- 2Q(0)\otimes \I~\rho_{AB}^{ij}~Q(0)\otimes \I \\
    &M_1 = M_1^* =\left(Q(\phi_A^i)\otimes \I~\rho_{AB}^{ij} + \rho_{AB}^{ij} ~ Q(\phi_A^i)\otimes \I\right)- 2Q(\phi_A^i)\otimes \I~\rho_{AB}^{ij}~Q(\phi_A^i)\otimes \I
\end{aligned}
\end{equation}
The projector $Q(\phi_A^i)$ follows the positive semi-definite condition \eqref{eqn44}.

\section{Optimization of the angles $\phi_A^i$ and $\phi_B^j$ using $\epsilon$ - net approximation} 

Alice's and Bob's angles $ \phi_A^i  \text{ and } \phi_B^j $ appear as constraints in the optimisation problem in Eq. \eqref{eqn3.8}. The objective function has been formulated in standard SDP form in Eq. \eqref{eqn4.6} taking the assumption of fixed angles. In this section, we optimize the angles $\phi_A^i  \text{ and } \phi_B^j $ using $\epsilon$ - net approach. The work of \cite{Schwonnek_2021} showed that the Alice's angle $\phi_A^i$ can be optimised using this approach. Here, we show that we can use the same approach to optimise Bob's angle $\phi_B^j$, thus eliminating the polytope optimisation in the security analysis as done \cite{Schwonnek_2021}.

 Given the interval $I = [0, \pi/2]$ and a desired precision $\epsilon_0 > 0$, without loss of generality, an $\epsilon_0$-net for the product space $I \times I$ is a pair of finite sets of points $\{\phi_{A_{k}}^i\}_{k=1}^{S_A} \subset I$ and $\{\phi_{B_{l}}^j\}_{l=1}^{S_B} \subset I$ such that for any $(\phi_A^i, \phi_B^j) \in I \times I$, there exist $\phi_{A_{k}}^i$ and $\phi_{B_{l}}^j$ satisfying:
\begin{equation}
     |\phi_A^i - \phi_{A_{k}}^i| \le \epsilon_0 \quad
    |\phi_B^j - \phi_{B_{l}}^j| \le \epsilon_0
\end{equation}
where $S_A \text{ and } S_B$ are the number of segments in the interval $I$ for Alice and Bob, respectively. Each segment is centralized around the angles $\phi_{A_{k}}^i \text{ and } \phi_{B_{l}}^j$ for $k^{th} \text{ and } l^{th}$ segments, respectively.
The values of both the angles $\phi_A^i\text { and }\phi_B^j$ are needed to solve any instance of the SDP and hence, it is solved for each discrete points $\phi_{A_{k}}^i$ and $\phi_{B_{l}}^j$. An error term known as pessimistic error \cite {Schwonnek_2021} is being subtracted from each SDP's result. The error term accounts for the variation of the optimal value. The pessimistic error term  $\Delta$ is introduced in the original work\cite{Schwonnek_2021} as a function of $\epsilon_0$ and the angle $\phi_A^i\text{ or} \phi_B^j$. Iteratively, the segment that gives the smallest value of the objective function in \eqref{eqn3.8} is chosen until a point is reached that corresponds to the global minima. Here, we try to derive a closed form of the pessimistic error.

Let us consider that $f(\phi)$ is a solution of the optimization problem in \eqref{eqn3.8} for a given value of the angle $\phi$. Now, the discrete angles $\phi_{A_{k}}^i\text{ and } \phi_{B_{l}}^j$ corresponding to Alice's and Bob's measurement settings are being separated by a distance of $2\epsilon_0$. Thus each angle represents a segment of the same width as,
\begin{equation}
\label{eqn5.3}
    \begin{aligned}
       I_A = \left[\phi_{A_{k}}^i - \epsilon_0, \phi_{A_{k}}^i + \epsilon_0 \right]~~~\mbox{and}~~~
       I_B = \left[\phi_{B_{l}}^j - \epsilon_0, \phi_{B_{l}}^j + \epsilon_0 \right]
    \end{aligned}
\end{equation}
Hence, the pessimistic error terms $\Delta(\epsilon_0,\phi_A^i)$ and $\Delta(\epsilon_0,\phi_B^j)$ provide an upper bound on the absolute difference between the values of the function at any point within these segments and the optimal one, i.e., 
\begin{equation}
\label{eqn5.4}
    \begin{aligned}        
       &|f(\phi_A^i) - f(\phi_{A_{k}}^i)| \leq \Delta\left (\epsilon_0,\phi_A^i\right)\\
       &|f(\phi_B^i) - f(\phi_{B_{l}}^j)| \leq \Delta\left (\epsilon_0,\phi_B^j\right),~~~~~~
       \forall \phi_A^i \in I_A~
       \text{ and }~\forall \phi_B^j \in I_B 
    \end{aligned}
\end{equation}
The iterative process involves selecting the segment that yields the minimum value of the objective function  in Eq. \eqref{eqn3.8}, continuing until the global minima is reached. 

\subsection{Lipschitz Continuity of \texorpdfstring{$f(\phi)$}{f(phi)}}
\label{F1}
\textit{Definition : }A function $f$ from $S \subset \mathbb{R}^n$ into $\mathbb{R}^m$ is Lipschitz continuous at $x \in S$ if there is a
constant $L > 0$ such that
\begin{equation}
\|f(y) - f(x)\| \leq L\|y - x\| \quad 
\end{equation}
for all $y \in S$ sufficiently near $x$.\cite{thomson2008elementary}\\

\textbf{Theorem 2: }{If $f(\phi)$ is the solution of a well-behaved optimisation problem, then it is Lipschitz continuous }\\

\textit{Problem setup
    }
    To prove our claim, we first formally define the optimisation problem. Let $f(\phi)$  be defined as the solution to the following optimization problem,
    \begin{equation}
    \begin{aligned}
        f(\phi) = &\text{min } g(x, \phi)\\
        & \text{s.t. }  x \in \mathcal{X}
    \end{aligned}
    \end{equation}
    where
         \( \mathcal{X} \subseteq \mathbb{R}^n \) is a compact and convex set,
         \( g(x, \phi) \) is a continuously differentiable function in both \( x \) and \( \phi \), and
         \( g(x, \phi) \) is strongly convex in \( x \) with modulus \( \mu > 0 \), i.e., for all \( x_1, x_2 \in \mathcal{X} \) and \( \phi \in \Phi \),
        \begin{equation}\label{eqn2}
        g(x_1, \phi) \geq g(x_2, \phi) + \nabla_x g(x_2, \phi)^T (x_1 - x_2) + \frac{\mu}{2} \|x_1 - x_2\|^2    
        \end{equation} and
         \( \nabla_x g(x, \phi) \) is Lipschitz continuous in \( \phi \) with constant \( L_\phi \), i.e.,
        \begin{equation}\label{eqn3}
        \|\nabla_x g(x, \phi_1) - \nabla_x g(x, \phi_2)\| \leq L_\phi \|\phi_1 - \phi_2\|. 
\end{equation}
    The optimisation problem is well-behaved in the sense that the solution \( f(\phi) \) exists and is unique for all \( \phi \in \Phi \) and the optimality condition \( \nabla_x g(f(\phi), \phi) = 0 \) holds.\\\\
    
\textit{Proof of Lipschitz Continuity of $f(\phi)$}
The aim of the proof is to show that \( f(\phi) \) is Lipschitz continuous, that is there exists a constant \( L_f > 0 \) such that:
\begin{equation}\label{eqn4}
\|f(\phi_1) - f(\phi_2)\| \leq L_f \|\phi_1 - \phi_2\|.
\end{equation}
Let us consider that for any two \( \phi_1, \phi_2 \in \Phi \) there are two solutions $f(\phi_1)$ and $f(\phi_2)$. Therefore, we have,  
\begin{equation}\label{eqn5}
\nabla_x g(f(\phi_1), \phi_1) = 0 \quad \text{and} \quad \nabla_x g(f(\phi_2), \phi_2) = 0.  
\end{equation}
Now since \( g(f(\phi), \phi) \) is strongly convex in \( f(\phi) \) with modulus \( \mu > 0 \), one can can have
\begin{equation}\label{eqn6}
g(f(\phi_1), \phi_1) \geq g(f(\phi_2), \phi_2) + \nabla_x g(f(\phi_2), \phi_2)^T (f(\phi_1) - f(\phi_2)) + \frac{\mu}{2} \|f(\phi_1) - f(\phi_2)\|^2
\end{equation}
and thus from Eq. \ref{eqn5} we get,
\begin{equation} \label{eqn7}
g(f(\phi_1), \phi_2) \geq g(f(\phi_2), \phi_2) + \frac{\mu}{2} \|f(\phi_1) - f(\phi_2)\|^2.
\end{equation}
Now, as the the lower bound set up by the quadratic norm is symmetric in terms of the angles, we have, \begin{equation}\label{eqn8}
g(f(\phi_2), \phi_1) \geq g(f(\phi_1), \phi_1) + \frac{\mu}{2} \|f(\phi_2) - f(\phi_1)\|^2.    
\end{equation}
Adding the two inequalities \ref{eqn7} and \ref{eqn8} and rearranging the terms we get,
\begin{equation}\label{eqn10}
\mu \|f(\phi_1) - f(\phi_2)\|^2 \leq g(f(\phi_1), \phi_2) - g(f(\phi_1), \phi_1) + g(f(\phi_2), \phi_1) - g(f(\phi_2), \phi_2).
\end{equation}
Using the above equation and considering the fact that \( \nabla_x g(x, \phi) \) in \( \phi \) is Lipschitz continuous, one can have the following two inequalities
\begin{equation}
\begin{aligned}
|g(f(\phi_1), \phi_2) - g(f(\phi_1), \phi_1)| &\leq L_\phi \|f(\phi_1)\| \|\phi_2 - \phi_1\|, \\
|g(f(\phi_2), \phi_1) - g(f(\phi_2), \phi_2)| &\leq L_\phi \|f(\phi_2)\| \|\phi_1 - \phi_2\|.
\label{eqn11}
\end{aligned}
\end{equation}
Since the feasible solution set  \( \mathcal{X} \) is compact, \ref{eqn10} and \ref{eqn11} \( \|f(\phi)\| \) jointly yield, 
\begin{equation}
\mu \|f(\phi_1) - f(\phi_2)\|^2 \leq 2 L_\phi M \|\phi_1 - \phi_2\|.    
\end{equation}
where $M$ is a constant.
Hence, simple algebra shows,
\begin{equation}
\|f(\phi_1) - f(\phi_2)\| \leq \sqrt{\frac{2 L_\phi M}{\mu}} \|\phi_1 - \phi_2\|    
\end{equation}
Thus, \( f(\phi) \) is Lipschitz continuous with constant,
\begin{equation}
L_f = \sqrt{\frac{2 L_\phi M}{\mu}}
\end{equation}
\qed\\

\subsection{Modification of the optimization problem}

In order to make the objective function  in \eqref{eqn3.8} continuously differentiable with respect to both the density operator \( \rho_{AB}^{ij} \) and the parameters \( \phi_A^i  \), we introduce the squared Frobenius norm. The squared Frobenius norm is smooth and continuously differentiable at every point \cite{Petersen2008}. For any operator \( X \), this norm is defined as,
\begin{equation}
\begin{aligned}
    &\left\| X \right\|_F = \sqrt{\Tr(X^* X)}\\
    &\left\| X \right\|_F^2 = \Tr(X^* X)
\end{aligned}    
\end{equation}
where \( X^* \) is the Hermitian conjugate of \( X \). 
Replacing the $l_1$ norm in the original objective function with the squared Frobenius norm and adding
a convex regularization term $ \frac{\mu}{2}||\rho_{AB}^{ij}||_F^2 $, we  define the modified optimisation problem as,
\begin{equation}
\label{eqn5.26}\boxed{
    \begin{aligned}
n^*(S_{ij}) = \inf \quad & \lambda \left\| \rho_{AB}^{ij} - \Lambda_0[\rho_{AB}^{ij}] \right\|_F^2 + (1-\lambda)\left\| \rho_{AB}^{ij} - \Lambda_1[\rho_{AB}^{ij}] \right\|_F^2 + \frac{\mu}{2} \|\rho_{AB}^{ij}\|_F^2 \\
\text{s.t.} \quad &\text{Tr}\left(\rho_{AB}^{ij}~CHSH(\phi_A^i,\phi_B^j)\right) = S_{ij} \\
& \phi_A^i, \phi_B^j \in [0,\pi/2],\\
    &\rho_{AB}^{ij} \succeq 0 \\ &\text{Tr}(\rho_{AB}^{ij}) = 1\\
\end{aligned}}
\end{equation}
This modified version of the optimisation problem can be reformulated into an SDP using Schur complements. Let us first decompose each Frobenius norm term into the corresponding inner product form as,
\begin{equation}
\label{eqn5.27}
    \begin{aligned}
        n^*(S_{ij}) = -\text{maximize} \quad & \lambda \mbox{Tr}({\rho_0}^{\dagger}\rho_0) + (1-\lambda)\mbox{Tr}({\rho_1}^{\dagger} \rho_1) + \frac{\mu}{2} \mbox{Tr}({\rho_2}^{\dagger} \rho_2) \\
        \text{s.t.} \quad &\text{Tr}\left(\rho_{AB}^{ij}~CHSH(\phi_A^i,\phi_B^j)\right) = S_{ij}\\
& \phi_A^i, \phi_B^j \in [0,\pi/2],\\
    &\rho_{AB}^{ij} \succeq 0 \\ &\text{Tr}(\rho_{AB}^{ij}) = 1\\
    \end{aligned}
\end{equation}
where $\rho_0 = \rho_{AB}^{ij} - \Lambda_0[\rho_{AB}^{ij}] $ , $\rho_1 = \rho_{AB}^{ij} - \Lambda_1[\rho_{AB}^{ij}] $ and $\rho_2 = \rho_{AB}^{ij} $\\
Here, each inner product term $ (\rho_k,\rho_k)$ for $k \in \{0,1,2\}$ in \eqref{eqn5.27} is quadratic in $\rho_k$. The standard SDP formulation requires that the objective function is linear in its decision variables and the constraints are in the form of linear matrix inequalities.
Let $t_k \geq (\rho_k,\rho_k)$, then using Schur's complement\cite{ben2001lectures} one can have,
\begin{equation}\begin{aligned}
    \begin{pmatrix}
        t_k & \mbox{vec}(\rho_k)^{\dagger}\\ \mbox{vec}(\rho_k) & 1 
        \end{pmatrix}\succeq 0
\end{aligned}
\end{equation}
Since $CHSH$ is fixed for given values of $\phi_A^i$ and $\phi_B^j$, the standard SDP formulation would be as,
\begin{equation}
\label{eqn5.30}
  \boxed{  \begin{aligned}
        n^*(S_{ij}) = -&\text{maximize} \quad  \lambda t_0 + (1-\lambda)t_1 + \frac{\mu}{2} t_2 \\
        \text{s.t.} & \begin{pmatrix}
        t_0& \mbox{vec}(\rho_0)^{\dagger}\\ \mbox{vec}(\rho_0) & 1 
        \end{pmatrix}\succeq 0 \\
        & \begin{pmatrix}
        t_1& \mbox{vec}(\rho_1)^{\dagger}\\ \mbox{vec}(\rho_1) & 1 
        \end{pmatrix}\succeq 0 \\
        & \begin{pmatrix}
        t_2& \mbox{vec}(\rho_2)^{\dagger}\\ \mbox{vec}(\rho_2) & 1 
        \end{pmatrix}\succeq 0 \\
        \quad &\text{Tr}\left(\rho_{AB}^{ij} ~ CHSH(\phi_A^i,\phi_B^j)\right) = S_{ij},\\
& \phi_A^i, \phi_B^j \in [0,\pi/2],\\
    &\rho_{AB}^{ij} \succeq 0 \\ &\text{Tr}(\rho_{AB}^{ij}) = 1\\
    \end{aligned}}
\end{equation}

In the below, we formally prove that the modified optimisation problem satisfies the conditions for being Lipschitz continuous.\\\\
\textbf{Lemma 4:} The modified objective function is continuously differentiable with respect to the density operator $ \rho_{AB}^{ij}$  and the parameter $ \phi_A^i $.\\
\textit{Proof: }
 The squared Frobenius norm $ \|X\|_F^2 = \mbox{Tr}(X^\dagger X) $ is smooth and infinitely differentiable. As the channels $\Lambda_0$ and  $\Lambda_1$ are linear maps, the action $ \Lambda_x[\rho] $ is linear in $\rho$. The regularization term $ \frac{\mu}{2} \|\rho\|_F^2 $ is quadratic and smooth. Hence, the objective function in the modified optimization problem is continuously differentiable with respect to both the density operator $ \rho_{AB}^{ij} $ and the parameters $ \phi_A^i $. \qed\\\\
\textbf{Lemma 5:} The modified objective function is strongly convex in $\rho_{AB}^{ij}$\\\\
\textit{Proof: } The objective function in Eq.\eqref{eqn5.26} includes the term $\frac{\mu}{2} \|\rho_{AB}^{ij}|_F^2$, which is \( \mu \)-strongly convex. For any \( \rho_{AB}^{ij}~\mbox{and}~\rho_{AB}^{kl} \):
\begin{equation}
    \begin{aligned}
        n\left(\rho_{AB}^{ij},(\phi_A^i )\right) \geq n\left(\rho_{AB}^{kl},(\phi_A^i )\right) + \nabla_\rho n\left(\rho_{AB}^{kl},(\phi_A^i )\right)^T (\rho_{AB}^{ij} - \rho_{AB}^{kl}) + \frac{\mu}{2} \|(\rho_{AB}^{ij} - \rho_{AB}^{kl})\|_F^2.
    \end{aligned}
\end{equation}
To rigorously establish the definition of strong convexity for the function, let us first calculate the gradient of the objective function $\nabla_\rho n\left(\rho_{AB}^{kl},(\phi_A^i )\right)$.
The Fréchet derivative of $||A||_F^2 \text{ is } 2A$\cite{watrous2018theory}. Thus,
\begin{equation}
\label{eqn5.32}
    \begin{aligned}
        &\nabla_\rho n\left(\rho_{AB}^{kl},(\phi_A^i )\right)\\ &= \nabla_\rho \left( \lambda \left\| \rho_{AB}^{kl} - \Lambda_0[\rho_{AB}^{kl}] \right\|_F^2 + (1-\lambda)\left\| \rho_{AB}^{kl} - \Lambda_1[\rho_{AB}^{kl}] \right\|_F^2 + \frac{\mu}{2} \|\rho_{AB}^{kl}\|_F^2
\right)\\
&=\nabla_\rho\left(\lambda \left\| \rho_{AB}^{kl} - \Lambda_0[\rho_{AB}^{kl}] \right\|_F^2\right) + \nabla_\rho\left((1-\lambda)\left\| \rho_{AB}^{kl} - \Lambda_1[\rho_{AB}^{kl}] \right\|_F^2\right) \\&\qquad+ \nabla_\rho\left(\frac{\mu}{2} \|\rho_{AB}^{kl}\|_F^2\right)\\
&=\lambda2\left( \rho_{AB}^{kl} - \Lambda_0[\rho_{AB}^{kl}]\right)\cdot\nabla_\rho\left(\rho_{AB}^{kl} - \Lambda_0[\rho_{AB}^{kl}]\right) \\& \qquad+ (1-\lambda)2\left(\rho_{AB}^{kl} - \Lambda_1[\rho_{AB}^{kl}]\right)\cdot\nabla_\rho\left(\rho_{AB}^{kl} - \Lambda_1[\rho_{AB}^{kl}]\right) + \mu\rho\\
&=\lambda2\left(\rho_{AB}^{kl} - \Lambda_0[\rho_{AB}^{kl}]\right)\cdot\left(\I - \Lambda_0^*\right) + (1-\lambda)2\left(\rho_{AB}^{kl} - \Lambda_1[\rho_{AB}^{kl}]\right)\cdot\left(\I -\Lambda_1^* \right) + \mu\rho\\
&=\lambda2\left(\rho_{AB}^{kl} - \Lambda_0[\rho_{AB}^{kl}]-\Lambda_0^*\left(\rho_{AB}^{kl} -\Lambda_0[\rho_{AB}^{kl}]\right)\right)\\&\qquad+ (1-\lambda)2\left(\rho_{AB}^{kl} - \Lambda_1[\rho_{AB}^{kl}]-\Lambda_1^*\left(\rho_{AB}^{kl} - \Lambda_1[\rho_{AB}^{kl}]\right)\right)+\mu\rho\\
&=\lambda2\left(\rho_{AB}^{kl} - \Lambda_0[\rho_{AB}^{kl}]\right)+ (1-\lambda)2\left(\rho_{AB}^{kl} - \Lambda_1[\rho_{AB}^{kl}]\right)+\mu\rho
    \end{aligned}
\end{equation}
Hence the objective function $n\left(\rho_{AB}^{ij},(\phi_A^i )\right)$ is strongly convex with modulus \( \mu > 0 \).\\

\qed\\
\textbf{Lemma 6:} The feasible set $\mathcal{D}$ of all density operator and the interval $I=[0,\frac{\pi}{2}]$ is compact and convex \\
\textit{Proof: }
A set \(\mathcal{S}\) is {convex} if for any \(x, y \in \mathcal{S}\) and \(\lambda \in [0, 1]\), the convex combination \(\lambda x + (1-\lambda)y \in \mathcal{S}\). In finite-dimensional spaces, a set is {compact} if it is {closed} (contains all its limit points) and {bounded} (fits within some finite-radius ball). The constraints are, 
$\text{Tr}\left(\rho_{AB}^{ij}~  CHSH(\phi_A^i,\phi_B^j)\right) = S_{ij}$, $\phi_A^i, \phi_B^j \in [0,\pi/2]$, $\rho_{AB}^{ij} \succeq 0$, $\text{Tr}(\rho_{AB}^{ij}) = 1$\\

a) Proof of convexity:
For \(\rho_1, \rho_2 \succeq 0\) and \(\lambda \in [0, 1]\),
    \begin{equation}
        \rho = \lambda \rho_1 + (1-\lambda)\rho_2 \succeq 0, \quad \text{Tr}(\rho) = \lambda \text{Tr}(\rho_1) + (1-\lambda)\text{Tr}(\rho_2) = 1.
    \end{equation}
Thus the set of density operators \(\rho \succeq 0\) with \(\text{Tr}(\rho) = 1\) is convex.\cite{watrous2018theory},\cite{markmwilde}\\
\begin{itemize}
\item Convexity of Angle Intervals: For \(\phi_1, \phi_2 \in [0, \pi/2]\),
\begin{equation}
 \lambda \phi_1 + (1-\lambda)\phi_2 \in [0, \pi/2]
\end{equation}
Thus the intervals \([0, \pi/2]\) for \(\phi_A^i, \phi_B^j\) are convex.\\

\item Linear Trace Constraint: For \(\rho_1, \rho_2\) satisfying \(\text{Tr}(\rho_1\cdot CHSH(\phi_A^i,\phi_B^j) ) = S_{ij}\) and \(\text{Tr}(\rho_2\cdot CHSH(\phi_A^i,\phi_B^j)) = S_{ij}\),
    \begin{equation}
         \text{Tr}\left((\lambda \rho_1 + (1-\lambda)\rho_2) \cdot CHSH(\phi_A^i,\phi_B^j) \right) = \lambda S_{ij} + (1-\lambda)S_{ij} = S_{ij}.
    \end{equation}
Thus the constraint \(\text{Tr}(\rho \cdot CHSH(\phi_A^i,\phi_B^j)) = S_{ij}\) is {linear in \(\rho\)} if \(M\) is fixed.\\

\item Combined Convexity: 
The Cartesian product of convex sets (density matrices, angle intervals) under linear constraints is convex.
\end{itemize}

b) Proof of compactness:
\begin{itemize}
\item Closedness: The set of all density operator $\mathcal{D}$ is closed because of positive semidefiniteness (\(\rho \succeq 0\)) is preserved under limits. The trace condition \(\text{Tr}(\rho) = 1\) is preserved under limits. The constraint \(\text{Tr}(\rho M) = S_{ij}\) is closed (as the preimage of a closed set under a continuous function)\cite{watrous2018theory}.\\

The interval \([0, \pi/2]\) is a closed interval in \(\mathbb{R}\).\\
\item Boundedness: The Frobenius norm satisfies \(\|\rho\|_F \leq \sqrt{\text{Tr}(\rho^2)} \leq \sqrt{\text{Tr}(\rho)} = 1\).Thus, the set of all density operators $\mathcal{D}$ is bounded.\cite{watrous2018theory}\\

The interval \([0, \pi/2]\) is bounded in \(\mathbb{R}\).
\end{itemize}
\qed\\
\textbf{Lemma 7: }{$\nabla_\rho n\left(\rho_{AB}^{ij},(\phi_A^i )\right)$ is Lipschitz continuous in \( \phi_A^i \)} for $x \in \{0,1\}$\\
\textit{Proof}:
     From \eqref{eqn5.32}, one can have the gradient of $\nabla_\rho n\left(\rho_{AB}^{ij},(\phi_A^i )\right)$ .
     For a given segment \( \phi_A^i \) parameterizes \( \Lambda_0, \Lambda_1 \) smoothly because for a particular segment its fixed, then $\nabla_\rho n\left(\rho_{AB}^{ij},(\phi_A^i )\right)$ depends smoothly on \( \phi_A^i \).
     Since \( \cos(\phi_A^i) \), \( \sin(\phi_A^i) \), and their derivatives are bounded, $\nabla_\rho n\left(\rho_{AB}^{ij},(\phi_A^i )\right)$ is Lipschitz continuous  in \( \phi_A^i \) as,
     \begin{equation}
                 \|\nabla_\rho n\left(\rho_{AB}^{ij},(\phi_A^i )\right) - \nabla_\rho n\left(\rho_{AB}^{kl},(\phi_A^k )\right)\| \leq L_{\phi_A} \|\phi_A^i - \phi_A^k\|.
     \end{equation}
 where $\rho_{AB}^{ij}$ and $\rho_{AB}^{kl}$ are being two arbitrary states and $\phi_A^i \text{ and } \phi_A^k$ are associated angles.\\
Hence {$\nabla_\rho n\left(\rho_{AB}^{ij},(\phi_A^i )\right)$ is lipschitz continuous in \( \phi_A^i \)} for $x \in \{0,1\}$.\\\\
\qed\\
\textbf{Lemma 8:} Existence of Unique optima, $\nabla_\rho n\left(n^*(S_{ij}),(\phi_A^i )\right) = 0$\\
\textit{Proof: }\\
Strong convexity ensures a unique minimiser \(n^*(S_{ij}) \) and Compactness of the feasible set guarantees existence.
With smoothness and strong convexity, the solution satisfies:
\begin{equation}
    \nabla_\rho n\left(n^*(S_{ij}),(\phi_A^i )\right) = 0
\end{equation}
Convex function:
A function $g:I\to\mathbb{R}$, where $I$ is an interval (or any convex set in general) in $\mathbb{R}$, is said to be {convex} if for any two points $x_a,x_b\in I$ and for any $t\in[0,1]$, the following inequality holds,
\begin{equation}
\label{eqn5.38n}
    g(tx_a+(1-t)x_b)\leq tg(x_a)+(1-t)g(x_b)
\end{equation}
The function $g$ is defined on an interval $I$. For any $x_a,x_b\in I$ and $t\in[0,1]$, the point $\phi=tx_a+(1-t)x_b$ also lies within the domain $I$.$\phi$ represents any point on the line segment connecting $x_a$ and $x_b$.\\{Now the term $tx_a+(1-t)x_b$} represents a convex combination of $x_a$ and $x_b$. As $t$ varies from $0$ to $1$, this expression covers all points on the line segment connecting $x_a$ and $x_b$.
For $t=0$, we get $x_b$, $t=1$, we get $x_a$ and for $0<t<1$, we get a point strictly between $x_a$ and $x_b$.\\
Now, finally {the term $tg(x_a)+(1-t)g(x_b)$}represents the $y$-coordinate of the point on the secant line connecting the points $(x_a,g(x_a))$ and $(x_b,g(x_b))$ at the $x$-coordinate $\phi=tx_a+(1-t)x_b$.\\\\
\qed\\
\textbf{Lemma 9: }Given a convex function $g$ defined on a domain containing distinct points $x_1$ and $x_2$, for any $t\in[0,1]$, let $\phi=tx_1+(1-t)x_2$ be a convex combination of $x_1$ and $x_2$. The $y$-coordinate of the point on the secant line passing through $(x_1,g(x_1))$ and $(x_2,g(x_2))$ at the $x$-coordinate $\phi$ is given by the convex combination of the function values, $tg(x_1)+(1-t)g(x_2)$.\\\\
\textit{Proof: }Let us have the convex function $g$ in \eqref{eqn5.38n}.For any two points $x_a$ and $x_b$ in the domain $I$ of $g$, 
the Secant line passing through $(x_a,g(x_a))$ and $(x_b,g(x_b))$ is,
\begin{equation}
\label{eqn5.39n}
    y-g(x_a)=\frac{g(x_b)-g(x_a)}{x_b-x_a}(x-x_a)
\end{equation}
Now, let $x= tx_a+(1-t)x_b$. Then,
\begin{equation}
\begin{aligned}
x-x_a &= tx_a+(1-t)x_b-x_a\\
&= (t-1)x_a+(1-t)x_b\\
&= (1-t)(x_b-x_a)
\end{aligned}
\end{equation}
Substituting this into the equation of the secant line, in \eqref{eqn5.39n}
\begin{equation}
    \begin{aligned}
y-g(x_a) &= \frac{g(x_b)-g(x_a)}{x_b-x_a}(1-t)(x_b-x_a)\\
y-g(x_a) &= (1-t)(g(x_b)-g(x_a))\\
y &= g(x_a)+(1-t)g(x_b)-(1-t)g(x_a)\\
y &= g(x_a)-(1-t)g(x_a)+(1-t)g(x_b)\\
y &= (1-(1-t))g(x_a)+(1-t)g(x_b)\\
y &= tg(x_a)+(1-t)g(x_b)
\end{aligned}
\end{equation}
{Thus from \textit{Lemma 9,} the inequality in \eqref{eqn5.38n} formally states that the value of the function $g$ at any point $x$ between $x_a$ and $x_b$ is less than or equal to the corresponding $y$-value on the secant line connecting $(x_a,g(x_a))$ and $(x_b,g(x_b))$.\qed\\

\subsection{Bounding the pessimistic error terms $\Delta\left(\epsilon_0,\phi_A^i\right) \text{ and }\Delta\left(\epsilon_0,\phi_B^j\right)$}
The solution of the modified optimisation problem in \eqref{eqn5.26}, $n^*(S_{ij})$, is thus established to be Lipschitz continuous. Now we proceed towards formulating a closed form for the pessimistic error terms $\Delta\left(\epsilon_0,\phi_A^i\right) \text{ and }\Delta\left(\epsilon_0,\phi_B^j\right)$.
The solution of the SDP in \eqref{eqn5.30} gives an optimal value for the $k^{th}$ and $l^{th}$ segment
centered around $\phi_{A_k}^i$ and $\phi_{B_l}^j$, respectively. As the solution of the SDP is based on fixed $\phi_A^i, \phi_B^j$, to find the optimal value of the objective function over the whole interval $I$, we need a slightly different approach from the one given in \cite{Schwonnek_2021} because here we optimize both the angles  using the $\epsilon$-net approach. The process begins by solving the SDP for each segment of Alice (Bob), parameterised by discretised angles $\phi_{A_k}^i ( \phi_{B_l}^j)$. Subsequently, the segment that produced the minimum value is iteratively refined. This refinement involves further subdivision of the segment's angle range, and the SDP is re-evaluated. This process continues until a global minimum is found for Alice or Bob. Then the optimal value with respect to the angles of other party is determined by applying the same iterative refinement process, but with the angle of the first party now fixed at their globally optimal value. Let us consider that $f_A(\phi_A^i)$ and $f_B(\phi_B^j)$ are the solutions of 
the SDP in \eqref{eqn5.30} or the optimization problem in \eqref{eqn5.26} for the fixed angles $\phi^i_A$ and $\phi^j_B$. Since it is proved that the solutions are Lipschitz continuous, we have the following relations for each segment where $\phi_{A_k}^i \text{ and } \phi_{B_l} ^j$ are the centres of the segments $k$ and $l$, respectively.
\begin{equation}
\label{eqn5.39}
    \begin{aligned}
        &| f_A(\phi_A^i) - f_A(\phi_{A_k}^i)| \leq L_A|\phi_A^i-\phi_{A_k}^i|\\
        &| f_B(\phi_B^j) - f_B(\phi_{B_l}^j)| \leq L_B|\phi_B^j-\phi_{B_l}^j|,~~~~~~~
        \forall \phi_A^i \in I_A~~
     \text{ and }~~\forall \phi_B^j \in I_B
    \end{aligned}
\end{equation}
The above inequalities depict that the absolute value of the difference between the solutions for two different angles is upper bounded by the absolute value of the difference between the respective angles times a constant. 
The supremum of the functions $f_A(\phi_A^i)$ and $f_B(\phi_B^j)$ are given as,
\begin{equation}
\label{eqn5.41}
    \begin{aligned}
        &M_A^i = sup_{\phi_A^i \in I_A } f_A(\phi_A^i)\\
        &M_B^j = sup_{\phi_B^j \in I_B } f_B(\phi_B^j)\\
    \end{aligned}
\end{equation}
Similarly, the infimum are given as,
\begin{equation}
\label{eqn5.42}
    \begin{aligned}
        &m_A^i = inf_{\phi_A^i \in I_A } f_A(\phi_A^i)\\
        &m_B^j = inf_{\phi_B^j \in I_B } f_B(\phi_B^j)\\
    \end{aligned}
\end{equation}
The Lipschitz continuity of the functions implies that they are continuous in the given interval. Therefore, the supremums in \eqref{eqn5.41} are the corresponding maxima and the infimums in \eqref{eqn5.42} are the corresponding minima. The supremum and infimum guarantee that no number smaller than $M_A^i (M_B^j)$ can serve as an upper bound, and equivalently, no number larger than $m_A^i (m_B^j)$ can be the lower bound for $ f_A(\phi_A^i)(f_B(\phi_B^j))$.
The quantity that one is interested in bounding is,
\begin{equation}
\label{eqn5.46n}
    \begin{aligned}
        sup_{\phi_A^i \in I_A} &\left| f_A(\phi_A^i) - f_A(\phi_{A_k}^i)\right| \\
        sup_{\phi_B^j \in I_B}&\left| f_B(\phi_B^j) - f_B(\phi_{B_l}^j)\right| \\ 
    \end{aligned}
\end{equation}
The deviation in \eqref{eqn5.39} is being maximised at the boundary of the respective segments. \\
\label{Theorem3}
\textbf{Theorem 3: } The values of $| f_A(\phi_A^i) - f_A(\phi_{A_k}^i)|$
and $\left| f_B(\phi_B^j) - f_B(\phi_{B_l}^j)\right|$ are maximum when $\phi_A^i = \phi_{A_k}^i \pm \epsilon_0$ and $\phi_B^i = \phi_{B_k}^i \pm \epsilon_0$\\\\
\textit{Proof: }
Before proving the theorem, we focus on some important properties of the function $f(\phi)$. The results of \textit{Theorem 2,} and \textit{Lemma 4} to \textit{Lemma 8} show that the functions $f_A(\phi_A^i) \text{ and } f_B(\phi_B^j)$ are Lipschitz continuous in the respective intervals $I_A$ and $I_B$. The functions are strictly monotonous in the intervals due to the monotonicity of the $\textit{sin}$ and $\textit {cosine}$ functions. Moreover, using the second derivative of the functions, one can find that these are concave in the respective intervals. Having mentioned the properties of the solution functions, we now prove the theorem.

Let us assume that the quantity  $| f_A(\phi_A^i)-f_A(\phi_{A_k}^i)|$ achieves its maximum at some $\phi_A^{*i} \in I_A$ such that $|\phi_A^{*i} - \phi_{A_k}^i| < \epsilon_0$. From the definition of the segments, we can write the interval as,  $I_A = [\phi_{A_k}^i-\epsilon_0, \phi_{A_k}^i+\epsilon_0]$. This means $\phi_A^{*i}$ lies strictly within the open interval $(\phi_{A_k}^i-\epsilon_0, \phi_{A_k}^i+\epsilon_0)$.
Since $f_A(\phi_A^i)$ is strictly monotonic on the closed interval $I_A$, its maximum and minimum values on this interval must occur at the endpoints, $\phi_{A_k}^i-\epsilon_0$ and $\phi_{A_k}^i+\epsilon_0$.
Now, if we consider the difference $|f_A(\phi_A^i)-f_A(\phi_{A_k}^i)|$, its absolute values at the endpoints are,
\begin{equation}
\begin{aligned}
|f_A(\phi_{A_k}^i-\epsilon_0)-f_A(\phi_{A_k}^i)|\\
|f_A(\phi_{A_k}^i+\epsilon_0)-f_A(\phi_{A_k}^i)|
\end{aligned}    
\end{equation}
Now, for any $\phi_A^{*i} \in (\phi_{A_k}^i-\epsilon_0, \phi_{A_k}^i+\epsilon_0)$ and $\phi_A^{*i} \neq \phi_{A_k}^i$, due to the strict monotonicity of $f_A$, we have two cases;
\begin{itemize}
\item If $f_A$ is strictly increasing, then for $\phi_{A_k}^i-\epsilon_0 < \phi_A^{*i} < \phi_{A_k}^i+\epsilon_0$, we have $f_A(\phi_{A_k}^i-\epsilon_0) < f_A(\phi_A^{*i}) < f_A(\phi_{A_k}^i+\epsilon_0)$. This implies that either $|f_A(\phi_{A_k}^i+\epsilon_0)-f_A(\phi_{A_k}^i)| > |f_A(\phi_A^{*i})-f_A(\phi_{A_k}^i)|$ or $|f_A(\phi_{A_k}^i-\epsilon_0)-f_A(\phi_{A_k}^i)| > |f_A(\phi_A^{*i})-f_A(\phi_{A_k}^i)|$ (or both).
\item If $f_A$ is strictly decreasing, then for $\phi_{A_k}^i-\epsilon_0 < \phi_A^{*i} < \phi_{A_k}^i+\epsilon_0$, we have $f_A(\phi_{A_k}^i-\epsilon_0) > f_A(\phi_A^{*i}) > f_A(\phi_{A_k}^i+\epsilon_0)$. Again, this implies that either $|f_A(\phi_{A_k}^i+\epsilon_0)-f_A(\phi_{A_k}^i)| > |f_A(\phi_A^{*i})-f_A(\phi_{A_k}^i)|$ or $|f_A(\phi_{A_k}^i-\epsilon_0)-f_A(\phi_{A_k}^i)| > |f_A(\phi_A^{*i})-f_A(\phi_{A_k}^i)|$ (or both).
\end{itemize}
In both cases (strictly increasing or strictly decreasing), the maximum deviation $|f_A(\phi_A^i)-f_A(\phi_{A_k}^i)|$ cannot occur strictly within the interval $(\phi_{A_k}^i-\epsilon_0, \phi_{A_k}^i+\epsilon_0)$. Therefore, the maximum must occur at one of the endpoints, $\phi_A^i = \phi_{A_k}^i-\epsilon_0$ or $\phi_A^i = \phi_{A_k}^i+\epsilon_0$, which can be written as $\phi_A^i = \phi_{A_k}^i \pm \epsilon_0$.
A similar argument holds for $|f_B(\phi_B^j)-f_B(\phi_{B_l}^j)|$, with the maximum occurring at $\phi_B^j = \phi_{B_l}^j \pm \epsilon_0$.\\\\
\qed\\
We now combine the results of \eqref{eqn5.46n}, \eqref{eqn5.39} and \textit{Theorem 3} as,
\begin{equation}
    \begin{aligned}
       sup_{\phi_A^i \in I_A}
 &| f_A(\phi_A^i) - f_A(\phi_{A_k}^i)|=  L_A\epsilon_0\\
       sup_{\phi_B^j \in I_B}
 &| f_B(\phi_B^j) - f_B(\phi_{B_l}^j)| = L_B\epsilon_0\\
        \forall &\phi_A^i \in I_A\\
     \text{ and }\forall & \phi_B^j \in I_B
    \end{aligned}
\end{equation}
And from the definition of the pessimistic error terms
$\Delta\left (\epsilon_0,\phi_A^i\right)$ and $\Delta\left (\epsilon_0,\phi_B^j\right)$ in \eqref{eqn5.4}, we have
\begin{equation}
\label{eqn5.52n}
    \begin{aligned}
        \Delta\left (\epsilon_0,\phi_A^i\right) = L_A\epsilon_0\\
        \Delta\left (\epsilon_0,\phi_B^j\right) = L_B\epsilon_0
    \end{aligned}
\end{equation}
Since $f_A(\phi_A^i)$ and $f_B(\phi_B^j)$ are differentiable in the interval $I$, the values of $L_A$ and $L_B$ can be evaluated using Taylor series expansion of the functions as,
\begin{equation}
    \begin{aligned}
        &f_A(\phi_A^i) = \sum_{n=0}^{\infty}\frac{f_A^n(\phi_{A}^{i_*})}{n!}(\phi_A^i - \phi_{A}^{i_*})^n\\
        &f_B(\phi_B^j) = \sum_{m=0}^{\infty}\frac{f_B^m(\phi_{B}^{j_*})}{m!}(\phi_B^j - \phi_{B}^{j_*})^m\\
    &\forall \phi_{A}^{i_*} \in I_A\\
    \text{and }&\forall \phi_{B}^{j_*} \in I_B 
    \end{aligned}
\end{equation}
Now in this case, one needs a linear approximation of $f_A(\phi_A^i)$ near $\phi_{A_k}^i$ and $f_B(\phi_B^j)$ near $\phi_{B_l}^j$ for small $\epsilon_0$. One can neglect the higher-order derivatives terms as $f_A^1(\phi_{A_k}^i)(\phi_A^i - \phi_{A_k}^{i})$ and $f_B^1(\phi_{B_l}^j)(\phi_B^j - \phi_{B_l}^{j})$  dominate as $|\phi_A^i - \phi_{A_k}^{i}|$ and 
$|\phi_B^j - \phi_{B_l}^{j}|$ tend to 0 for $k^{th}$ and $l^{th}$ segments, respectively. The necessity of incorporating higher-order terms in the error analysis arises when $\epsilon_0$ is significant, the underlying functions exhibit strong non-linearities, or when a more precise and tighter bound on the error is desired. In such scenarios, relying solely on linear approximations becomes insufficient to accurately capture the function's behavior within the interval of width $2\epsilon_0$. The contributions from higher derivatives, which are neglected in a first-order analysis, become substantial and must be accounted for to achieve the required accuracy in the error estimation.
Thus, using the first order linear approximation one can get an approximation of $f_A(\phi_A^i)$ and $f_A(\phi_B^j)$ near $\phi_{A_k}^i$ and $\phi_{B_l}^j$
as,
\begin{equation}
    \begin{aligned}
        &f_A(\phi_A^i) = f_A(\phi_{A_k}^i) + f_A^1(\phi_{A_k}^i)(\phi_A^i - \phi_{A_k}^{i}) + \mathcal{O}((\phi_A^i - \phi_{A_k}^{i})^2)\\
        & |f_A(\phi_A^i) - f_A(\phi_{A_k}^i)| \leq |f_A^1(\phi_{A_k}^i)||(\phi_A^i - \phi_{A_k}^{i})| + \mathcal{O}((\phi_A^i - \phi_{A_k}^{i})^2) 
    \end{aligned}
\end{equation}
Similarly,
\begin{equation}
    |f_B(\phi_B^j) - f_B(\phi_{B_l}^j)| \leq |f_B^1(\phi_{B_l}^j)||(\phi_B^j - \phi_{B_l}^{j})| + \mathcal{O}((\phi_B^j - \phi_{B_l}^{j})^2)
\end{equation}
The higher-order terms, represented by $\mathcal{O}((\phi_A^i-\phi_{A_k}^i)^2)$ and $\mathcal{O}((\phi_B^j-\phi_{B_l}^j)^2)$, are indeed positive if the second derivatives are positive (indicating local convexity). While typically removing positive terms would weaken an inequality, the context of a pessimistic error bound requires careful consideration. The pessimistic error is designed to overestimate the potential deviation. By truncating the Taylor series and neglecting these higher-order positive terms, we are essentially underestimating the actual deviation $|f_A(\phi_A^i)-f_A(\phi_{A_k}^i)|$ and $|f_B(\phi_B^j)-f_B(\phi_{B_l}^j)|$. Consequently, when these underestimated deviations are used to construct a pessimistic error (which is subtracted from the SDP result), the error itself becomes overestimated. This overestimation of the error leads to a more conservative (and potentially less tight) lower bound on the true optimal value of the SDP.
Thus after first order linear approximation, we get,
\begin{equation}
\label{eqn5.56n}
    \begin{aligned}
       sup_{\phi_A^i \in I_A}
 & |f_A(\phi_A^i) - f_A(\phi_{A_k}^i)| \leq |f_A^1(\phi_{A_k}^i)||(\phi_A^i - \phi_{A_k}^{i})|  \\
sup_{\phi_B^j \in I_B}
&|f_B(\phi_B^j) - f_B(\phi_{B_l}^j)| \leq |f_B^1(\phi_{B_l}^j)||(\phi_B^j - \phi_{B_l}^{j})| 
    \end{aligned}
\end{equation}
Therefore, using Eqs. \eqref{eqn5.56n} and \eqref{eqn5.52n}, we get the pessimistic error to be, 
\begin{equation}
    \begin{aligned}
        \Delta\left (\epsilon_0,\phi_A^i\right) =        max_{\phi_A^i \in I}|f_A^1(\phi_{A_k}^i)|
\epsilon_0\\
        \Delta\left (\epsilon_0,\phi_B^j\right) = max_{\phi_A^i \in I}|f_B^1(\phi_{B_l}^j)|\epsilon_0
    \end{aligned}
\end{equation}
where $L_A = max_{\phi_A^i \in I}|f_A^1(\phi_A^i)| $ and $L_B = max_{\phi_A^i \in I}|f_B^1(\phi_B^j)|. $

\subsection{Relating the change in $\phi_A^i$ and $\phi_B^j$ to the optimization problem through CHSH operator}
\label{change_of_angles_through_chsh}
 
The feasible set in \eqref{eqn5.26} can be relaxed by allowing all the density operators $\sigma_{AB}^{ij}$ that achieve CHSH violation greater than or equal to $S_{ij}$ \cite{Schwonnek_2021},
\begin{equation}
\begin{aligned}
  S_{ij}^{\phi_A^i\cup\phi_B^j}
  &=\bigl\{\sigma\;\big|\;\exists\,\phi_A\text{ with }\phi_B=\phi_B^j:\;
      \Tr[\sigma\,CHSH(\phi_A,\phi_B^j)]\ge S_{ij}\bigr\}\\
  &\quad\cup
   \bigl\{\sigma\;\big|\;\exists\,\phi_B\text{ with }\phi_A=\phi_A^i:\;
      \Tr[\sigma\,CHSH(\phi_A^i,\phi_B)]\ge S_{ij}\bigr\}.
\end{aligned}
\end{equation}
The above equation takes the union of two sets. The first set accounts for all those density operators $\sigma_{AB}^{ij}$ in the block specified by index $i,j$ that achieve CHSH violation greater than $S_{ij}$ for fixed $\phi_B^j$. The second set similarly accounts for the case when $\phi_A^i$ is fixed .
Now, consider the following set,
\begin{equation}
    \begin{aligned}
  M_{ij,\epsilon}^{\phi_A^i\cup\phi_B^j}
  &:=
  \bigcup_{|\delta_A|\le\epsilon_0}
    S_{ij}^{\phi_{A}^i+\delta_A\;\cup\;\phi_B^j}
  \;\cup\;
  \bigcup_{|\delta_B|\le\epsilon_0}
    S_{ij}^{\phi_A^i\;\cup\;\phi_{B}^j+\delta_B}
  \quad\cup\;
  \bigcup_{\substack{|\delta_A|\le\epsilon_0\\|\delta_B|\le\epsilon_0}}
    S_{ij}^{\phi_{A}^i+\delta_A\;\cup\;\phi_{B}^j+\delta_B}\,.
\end{aligned}
\end{equation}
In the aforementioned equation, the analysis focuses on a set of density operators derived by introducing a parameter $\epsilon$ to the initial angles $\phi_A^i$ and $\phi_B^j$, both individually and concurrently. The defining characteristic of these modified density operators is that the expected value of the CHSH operator evaluated on them remains greater than or equal to the initial CHSH value $S_{ij}$.

Given discrete points $\phi_{A_k}^i$ and $\phi_{B_l}^j$ located at the midpoint of their respective intervals, the inherent symmetry of the underlying function or data distribution implies that the maximum deviation from these discrete points will occur symmetrically around them. Consequently, a small perturbation $\epsilon$ from the midpoint will result in equal magnitudes of deviation, such that the deviation at $\phi_{A_k+\epsilon}^i$ is equivalent to the deviation at $\phi_{A_k-\epsilon}^i$, and similarly for $\phi_{B_l}^j$.
The spectral norm of the difference between the CHSH operators evaluated at $\phi_{A_k}^i$ and $\phi_{A_k \pm \epsilon}^i$ (and similarly for $\phi_{B_l}^j$ and $\phi_{B_l \pm \epsilon}^j$) is equal to the largest singular value of this difference operator. Let us define the maximum of the spectral norms of the differences in the CHSH operator as,
\begin{equation}
\label{eqn5.61n}
    \begin{aligned}
        \delta_p = \max\Big(& ||CHSH(\phi_{A}^i,\phi_B^j) - CHSH(\phi_{A}^i+\epsilon,\phi_B^j)||_{\infty} , \\
        & ||CHSH(\phi_{A}^i,\phi_B^j) - CHSH(\phi_{A}^i,\phi_{B}^j+\epsilon)||_{\infty} , \\
        & ||CHSH(\phi_{A}^i,\phi_B^j) - CHSH(\phi_{A}^i+\epsilon,\phi_{B}^j+\epsilon)||_{\infty} \Big)
    \end{aligned}
\end{equation}
This spectral norm corresponds to the largest singular value of these difference operators. For the Hermitian CHSH operator, this largest singular value coincides with the absolute value of the largest eigenvalue of the respective difference operators. The spectral norm can be found as, 
\begin{equation}
\label{eqn5.62n}
    \begin{aligned}
       & \Big\|CHSH(\phi_{A}^i,\phi_B^j) - CHSH(\phi_{A}^i+\epsilon,\phi_B^j)\Big\|_{\infty} \\
        =& \Big\|\begin{pmatrix}
\cos(\phi_A^i + \epsilon) - \cos\phi_A^i &  \sin(\phi_A^i + \epsilon) - \sin\phi_A^i \\
 \sin(\phi_A^i + \epsilon) - \sin\phi_A^i & -\cos(\phi_A^i + \epsilon) - \cos\phi_A^i
\end{pmatrix}
\otimes
\begin{pmatrix}
-(1 - \cos\phi_B^j) & \sin\phi_B^j \\
\sin\phi_B^j & 1 - \cos\phi_B^j
\end{pmatrix}\Big\|_{\infty}
    \end{aligned}
\end{equation}
Now, one can use some simple trigonometric relations to decompose the relation $\Delta \cos = \cos(\phi_A^i + \epsilon) - \cos\phi_A^i$ and $\Delta \sin = \sin(\phi_A^i + \epsilon) - \sin\phi_A^i$.
\begin{equation}
\label{eqn5.63n}
    \begin{aligned}
&\Delta \cos = \cos(\phi_A^i + \epsilon) - \cos(\phi_A^i) = -2 \sin\left(\frac{(\phi_A^i + \epsilon) + \phi_A^i}{2}\right) \sin\left(\frac{(\phi_A^i + \epsilon) - \phi_A^i}{2}\right) = -2 \sin\left(\phi_A^i + \frac{\epsilon}{2}\right) \sin\left(\frac{\epsilon}{2}\right)\\
&\Delta \sin = \sin(\phi_A^i + \epsilon) - \sin(\phi_A^i) = 2 \cos\left(\frac{(\phi_A^i + \epsilon) + \phi_A^i}{2}\right) \sin\left(\frac{(\phi_A^i + \epsilon) - \phi_A^i}{2}\right) = 2 \cos\left(\phi_A^i + \frac{\epsilon}{2}\right) \sin\left(\frac{\epsilon}{2}\right)\\
&1 - \cos\phi_B^j = 2\sin^2\left(\frac{\phi_B^j}{2}\right)\\
&\sin\phi_B^j  = 2\sin\left(\frac{\phi_B^j}{2}\right)\cos\left(\frac{\phi_B^j}{2}\right)
    \end{aligned}
\end{equation}
From \eqref{eqn5.63n} and \eqref{eqn5.62n},
\begin{equation}
\label{eqn5.64n}
    \begin{aligned}
       &\Big\|CHSH(\phi_{A}^i,\phi_B^j) - CHSH(\phi_{A}^i+\epsilon,\phi_B^j)\Big\|_{\infty} \\
        =& \Big\|\begin{pmatrix}
-2 \sin\left(\phi_A^i + \frac{\epsilon}{2}\right) \sin\left(\frac{\epsilon}{2}\right) & 2 \cos\left(\phi_A^i + \frac{\epsilon}{2}\right) \sin\left(\frac{\epsilon}{2}\right) \\
 2 \cos\left(\phi_A^i + \frac{\epsilon}{2}\right) \sin\left(\frac{\epsilon}{2}\right) & 2 \sin\left(\phi_A^i + \frac{\epsilon}{2}\right) \sin\left(\frac{\epsilon}{2}\right)
\end{pmatrix}
\otimes
\begin{pmatrix}
-2\sin^2\left(\frac{\phi_B^j}{2}\right) & 2\sin\left(\frac{\phi_B^j}{2}\right)\cos\left(\frac{\phi_B^j}{2}\right) \\
2\sin\left(\frac{\phi_B^j}{2}\right)\cos\left(\frac{\phi_B^j}{2}\right) & 2\sin^2\left(\frac{\phi_B^j}{2}\right)
\end{pmatrix}\Big\|_{\infty}\\
=&\Big\|2\sin\frac{\epsilon}{2}\left(\cos(\phi_A^i + \frac{\epsilon}{2})\sigma_x - \sin(\phi_A^i + \frac{\epsilon}{2})\sigma_z\right) \otimes  2\left(Q\left({\phi_B^j}\right) - \begin{pmatrix}
    1&0\\0&0
\end{pmatrix} \right)  \Big\|_{\infty}\\
=& \Big\|2\sin\left(\frac{\epsilon}{2}\right)\left[\left( \cos\left(\phi_A^i\right)\cos\left(\frac{\epsilon}{2}\right) - \sin\left(\phi_A^i\right)\sin\left(\frac{\epsilon}{2}\right)  \right)\sigma_x -\left( \sin\left(\phi_A^i\right)\cos\left(\frac{\epsilon}{2}\right) + \cos\left(\phi_A^i\right)\sin\left(\frac{\epsilon}{2}\right)  \right)\sigma_z \right] \\ & \qquad \otimes  2\left(Q\left({\phi_B^j}\right) - \begin{pmatrix}
    1&0\\0&0
\end{pmatrix} \right) \Big\|_{\infty}
    \end{aligned}
\end{equation}
Using Taylor expansions for $\sin\{\phi\}$ and $\cos\{\phi\}$,
$\sin x = x - \frac{x^3}{3!} + \frac{x^5}{5!} - \cdots$ and $
    \cos x = 1 - \frac{x^2}{2!} + \frac{x^4}{4!} - \cdots$ $
\text{If } x \ll 1,
\cos x \approx 1 ~\mbox{and} \sin x \approx x $.
Also using $\cos a + \sin b\cdot x \approx \cos a ~\mbox{and}
\sin a + \cos b\cdot x \approx \sin a$ for $ x \ll 1$, one can have
\begin{equation}
\label{eqn5.65n}
\begin{aligned}
    &\sin\left(\frac{\epsilon}{2}\right) \approx \frac{\epsilon}{2}\\
    &\left( \cos\left(\phi_A^i\right)\cos\left(\frac{\epsilon}{2}\right) - \sin\left(\phi_A^i\right)\sin\left(\frac{\epsilon}{2}\right)  \right) \approx  \cos\left(\phi_A^i\right)\\
    &\left( \sin\left(\phi_A^i\right)\cos\left(\frac{\epsilon}{2}\right) + \cos\left(\phi_A^i\right)\sin\left(\frac{\epsilon}{2}\right)  \right) \approx  \sin\left(\phi_A^i\right)
\end{aligned}
\end{equation}
Now, combining the results of \eqref{eqn5.64n} and linear approximation for $\epsilon>0$ in \eqref{eqn5.65n} we get,
\begin{equation}
\label{eqn5.66n}   \begin{aligned}
&\Big\|CHSH(\phi_{A}^i,\phi_B^j) - CHSH(\phi_{A}^i+\epsilon,\phi_B^j)\Big\|_{\infty} \\
        \approx &\Big\| {\epsilon}\left( \cos\left(\phi_A^i\right) \sigma_x -\sin\left(\phi_A^i\right)\sigma_z  \right) \otimes  2\left(Q\left({\phi_B^j}\right) - \begin{pmatrix}
    1&0\\0&0
\end{pmatrix} \right) \Big\|_{\infty}
    \end{aligned}
\end{equation}
Similarly, one can find the spectral norm when only $\phi_B^j$ is being perturbated by $\epsilon$ as,
\begin{equation}
\label{eqn5.67n}
\begin{aligned}
     &||CHSH(\phi_{A}^i,\phi_B^j) - CHSH(\phi_{A}^i,\phi_{B}^j+\epsilon)||_{\infty} \\
 =&\Big\|\begin{pmatrix}
        2\cos^2(\frac{\phi_A^i}{2}) & 2\sin\left(\frac{\phi_A^i}{2}\right)\cos\left(\frac{\phi_A^i}{2}\right) \\2\sin\left(\frac{\phi_A^i}{2}\right)\cos\left(\frac{\phi_A^i}{2}\right) &-2\cos^2(\frac{\phi_A^i}{2})
           \end{pmatrix}  \otimes \begin{pmatrix}
         2\sin\left(\phi_B^j + \frac{\epsilon}{2}\right)\sin\left(\frac{\epsilon}{2}\right) & -2\cos\left(\phi_B^j + \frac{\epsilon}{2}\right)\sin\left(\frac{\epsilon}{2}\right)\\-2\cos\left(\phi_B^j + \frac{\epsilon}{2}\right)\sin\left(\frac{\epsilon}{2}\right) & -2\sin\left(\phi_B^j + \frac{\epsilon}{2}\right)\sin\left(\frac{\epsilon}{2}\right)
    \end{pmatrix} \Big\|_{\infty} \\
=& \Big\|  2\left(Q\left({\phi_A^i}\right) - \begin{pmatrix}
    0&0\\0&1
\end{pmatrix} \right) \\& \qquad \otimes -2\sin\left(\frac{\epsilon}{2}\right)\left[\left( \cos\left(\phi_B^j\right)\cos\left(\frac{\epsilon}{2}\right) - \sin\left(\phi_B^j\right)\sin\left(\frac{\epsilon}{2}\right)  \right)\sigma_x -\left( \sin\left(\phi_B^j\right)\cos\left(\frac{\epsilon}{2}\right) + \cos\left(\phi_B^j\right)\sin\left(\frac{\epsilon}{2}\right)  \right)\sigma_z \right]  \Big\|_{\infty}\\
\approx& \Big\|  2\left(Q\left({\phi_A^i}\right) - \begin{pmatrix}
    0&0\\0&1
\end{pmatrix} \right) \otimes {\epsilon}\left(\cos\left(\phi_B^j\right)\sigma_x - \sin\left(\phi_B^j\right)\sigma_z \right)   \Big\|_{\infty}
 \end{aligned}
\end{equation}
The spectral norm when both $\phi_A^i$ and $\phi_B^j$ perturbated by $\epsilon$ is found to be,
\begin{equation}
\label{eqn5.68n}
    \begin{aligned}
        &||CHSH(\phi_{A}^i,\phi_B^j) - CHSH(\phi_{A}^i+\epsilon,\phi_{B}^j+\epsilon)||_{\infty}\\
        =&\Big\| 
        \begin{pmatrix}
            2\sin\left(\phi_{A}^i + \frac{\epsilon}{2}\right)\sin\left(\frac{\epsilon}{2}\right) & -2\cos\left(\phi_{A}^i + \frac{\epsilon}{2}\right)\sin\left(\frac{\epsilon}{2}\right) \\
        -2\cos\left(\phi_{A}^i + \frac{\epsilon}{2}\right)\sin\left(\frac{\epsilon}{2}\right) & -2\sin\left(\phi_{A}^i + \frac{\epsilon}{2}\right)\sin\left(\frac{\epsilon}{2}\right)
        \end{pmatrix} \otimes \begin{pmatrix}
        2\sin\left(\phi_{B}^j + \frac{\epsilon}{2}\right)\sin\left(\frac{\epsilon}{2}\right)
&-2\cos\left(\phi_{B}^j + \frac{\epsilon}{2}\right)\sin\left(\frac{\epsilon}{2}\right)\\-2\cos\left(\phi_{B}^j + \frac{\epsilon}{2}\right)\sin\left(\frac{\epsilon}{2}\right) &-2\sin\left(\phi_{B}^j + \frac{\epsilon}{2}\right)\sin\left(\frac{\epsilon}{2}\right)
        \end{pmatrix}       \Big\|_{\infty}\\
=&\Big\|-2\sin\left(\frac{\epsilon}{2}\right)\left[\left( \cos\left(\phi_A^i\right)\cos\left(\frac{\epsilon}{2}\right) - \sin\left(\phi_A^i\right)\sin\left(\frac{\epsilon}{2}\right)  \right)\sigma_x -\left( \sin\left(\phi_A^i\right)\cos\left(\frac{\epsilon}{2}\right) + \cos\left(\phi_A^i\right)\sin\left(\frac{\epsilon}{2}\right)  \right)\sigma_z \right] \\&\qquad\otimes -2\sin\left(\frac{\epsilon}{2}\right)\left[\left( \cos\left(\phi_B^j\right)\cos\left(\frac{\epsilon}{2}\right) - \sin\left(\phi_B^j\right)\sin\left(\frac{\epsilon}{2}\right)  \right)\sigma_x -\left( \sin\left(\phi_B^j\right)\cos\left(\frac{\epsilon}{2}\right) + \cos\left(\phi_B^j\right)\sin\left(\frac{\epsilon}{2}\right)  \right)\sigma_z \right]\Big\|_{\infty}\\
\approx& \Big\| {\epsilon}\left(\cos\left(\phi_A^i\right)\sigma_x - \sin\left(\phi_A^i\right)\sigma_z \right) \otimes {\epsilon}\left(\cos\left(\phi_B^j\right)\sigma_x - \sin\left(\phi_B^j\right)\sigma_z \right)  \Big\|_{\infty}
    \end{aligned}
\end{equation}
Hence, combining the results of \eqref{eqn5.66n}, \eqref{eqn5.67n} and \eqref{eqn5.68n} into \eqref{eqn5.61n}, we finally get,
\begin{equation}
    \begin{aligned}
        \delta_p \approx \max\Big(& \Big\| {\epsilon}\left( \cos\left(\phi_A^i\right) \sigma_x -\sin\left(\phi_A^i\right)\sigma_z  \right) \otimes  2\left(Q\left({\phi_B^j}\right) - \begin{pmatrix}
    1&0\\0&0
\end{pmatrix} \right) \Big\|_{\infty}, \\
        & \Big\|  2\left(Q\left({\phi_A^i}\right) - \begin{pmatrix}
    0&0\\0&1
\end{pmatrix} \right) \otimes {\epsilon}\left(\cos\left(\phi_B^j\right)\sigma_x - \sin\left(\phi_B^j\right)\sigma_z \right)   \Big\|_{\infty}, \\
        & \Big\| {\epsilon}\left(\cos\left(\phi_A^i\right)\sigma_x - \sin\left(\phi_A^i\right)\sigma_z \right) \otimes {\epsilon}\left(\cos\left(\phi_B^j\right)\sigma_x - \sin\left(\phi_B^j\right)\sigma_z \right)  \Big\|_{\infty} \Big)
    \end{aligned}
\end{equation}
One can provide an upper bound to the above relation using the property of submultiplicity of spectral norm\cite{watrous2018theory} as,
\begin{equation}
\label{eqn5.70n}
    \begin{aligned}
        \delta_p \leq \max\Big(& \Big\| {\epsilon}\left( \cos\left(\phi_A^i\right) \sigma_x -\sin\left(\phi_A^i\right)\sigma_z  \right) \Big\|_{\infty}\cdot \Big\|  2\left(Q\left({\phi_B^j}\right) - \begin{pmatrix}
    1&0\\0&0
\end{pmatrix} \right) \Big\|_{\infty}, \\
        & \Big\|  2\left(Q\left({\phi_A^i}\right) - \begin{pmatrix}
    0&0\\0&1
\end{pmatrix} \right) \Big\|_{\infty}\cdot \Big\|{\epsilon}\left(\cos\left(\phi_B^j\right)\sigma_x - \sin\left(\phi_B^j\right)\sigma_z \right)   \Big\|_{\infty}, \\
        & \Big\| {\epsilon}\left(\cos\left(\phi_A^i\right)\sigma_x - \sin\left(\phi_A^i\right)\sigma_z \right) \Big\|_{\infty}\cdot \Big\| {\epsilon}\left(\cos\left(\phi_B^j\right)\sigma_x - \sin\left(\phi_B^j\right)\sigma_z \right)  \Big\|_{\infty} \Big)\\
        \leq \max &\left(\epsilon \cdot 2\sin\left(\frac{\phi_B^j}{2}\right), \quad 2\cos\left(\frac{\phi_A^i}{2}\right)\cdot\epsilon,\quad \epsilon^2 \right)\\
        \leq \quad  2\epsilon_0
    \end{aligned}
\end{equation}
The equality holds for the spectral norm of Kronecker products, therefore, this upper bound is exact. 
Now in the interval $I,$ $\sin\left(\frac{\phi_B^j}{2}\right)$ and $\cos\left(\frac{\phi_A^i}{2}\right)$ functions are monotonically increasing and decreasing respectively. The first term $2\sin\left(\frac{\phi_B^j}{2}\right)$ is increasing from $0 \text{ to } 1.414$ and the second term $2\cos\left(\frac{\phi_A^i}{2}\right)$ is decreasing from $2\epsilon \text{ to } 1.414$.
Thus, the maximum deviation of the CHSH operator is achieved when $\phi_A^i \text{ is fixed at } 0$ but $\phi_B^j \text{ is increased by }\epsilon_0.$ 
Thus every state $\rho_{AB}^{ij}$ that would attain CHSH value $S_{ij}$ at $\phi_A^i = 0, \phi_B^j=\phi_{B_l}^j$  $\left(\phi_A^i = 0, \phi_B^j=\phi_{B}^j\pm\epsilon_0 \right)$ would attain CHSH value $S_{ij}-2\epsilon_0$ at $\phi_A^i = 0, \phi_B^j=\phi_{B}^j\pm\epsilon_0$ $\left(\phi_A^i = 0, \phi_B^j=\phi_{B_l}^j \right)$.
Thus, we finally have,
\begin{equation}
\begin{aligned}
  &\Tr\!\bigl(\rho~ CHSH(0,\phi_B^j)\bigr) = S_{ij},
  \quad
  \Tr\!\bigl(\rho~CHSH(0,\phi_B^j + \epsilon_0)\bigr) = S_{ij} - 2\epsilon_0,
  \quad
  \Tr\!\bigl(\rho~CHSH(0,\phi_B^j - \epsilon_0)\bigr) = S_{ij} - 2\epsilon_0,
  \\
  \text{or}\quad
  &\Tr\!\bigl(\rho~CHSH(0,\phi_B^j)\bigr) = S_{ij} - 2\epsilon_0,
  \quad
  \Tr\!\bigl(\rho~ CHSH(0,\phi_B^j + \epsilon_0)\bigr) = S_{ij},
  \quad
  \Tr\!\bigl(\rho~ CHSH(0,\phi_B^j - \epsilon_0)\bigr) = S_{ij},
  \\
  &\forall \rho\in M_{ij,\epsilon}^{\phi_B^j}.
\end{aligned}
\end{equation}
where \begin{equation}
    \begin{aligned}
&M_{ij,\epsilon}^{\phi_B^j}\\
 &= \underbrace{\bigl\{\rho\in M_{ij,\epsilon}^{\phi_A^i\cup\phi_B^j}\;\bigm|\;\Tr(\rho~CHSH(0,\phi_B^j))=S_{ij}\bigr\}}_{\text{Case A}}
\;\cup\;
\underbrace{\bigl\{\rho\in M_{ij,\epsilon}^{\phi_A^i\cup\phi_B^j}\;\bigm|\;\exists s\in\{+1,-1\}:\Tr(\rho~CHSH(0,\phi_B^j+s\epsilon_0))=S_{ij}\bigr\}}_{\text{Case B}}.
    \end{aligned}
\end{equation}
Thus, for each segment $I_B$ and $I_A$ being centralised around discrete $\phi_{B_l}^j$ and $\phi_{A_k}^i$ for $l^{th}$ and $k^{th}$ segment, respectively, performing the optimisation with relaxed constraint value $S-2\epsilon_0$ would take into account the dependency on $\phi_A^i$ and $\phi_B^j$ through the maximum deviation in the CHSH value.

\subsection{Relating the change in $\phi_A^i$ to the optimization problem }
The objective function in \eqref{eqn5.26} is a function of the density operator describing the joint state between Alice and Bob and Alice's angle $\phi_A^i$.
To analyse the dependency of the solution of the optimisation problem on $\phi_{A}^i$, we reinterpret the Frobenius norm terms in \eqref{eqn5.26} using dual norms. Dual norm is an indispensable tool in quantum information theory and semi-definite programming because they bridge the limits of quantum operations, e.g., distinguishability, entanglement, and convex optimisation via conic duality, enabling tractable solutions having physical interpretability. Since the Frobenius norm is self-dual, this framework retains the norm structure while emphasising its role as a maximiser of inner products. Specifically, translating the squared Frobenius differences into their dual characterisation reveals how $\phi_{A}^i$ governs alignment conditions between the states $\rho_{AB}^{ij}$ and the maps $\Lambda_0, \Lambda_1$. The dual norm's supremum property enables bounding critical trade-offs in the objective function, such as the balance between state fidelity $(\lambda, 1-\lambda)$ and distinguishability $\mu$, which directly depend on $\phi_{A}^i$. \\\\
\textbf{Theorem 4: }The Frobenius norm \(\|\cdot\|_{F}\) on \(m\times n\) matrices is self-dual.\\\\
\textit{Proof:} Let \(A\in\mathbb{C}^{m\times n}\) and $A\in L(\X,\Y)$ for some Hilbert spaces $\X,\Y$.
The Frobenius norm is given as\cite{watrous2018theory},
\begin{equation}
\|A\|_{F} =\sqrt{\Tr\bigl(A^{\dagger}A\bigr)} =\sqrt{\langle A,\,A\rangle} ,    
\end{equation}
Now, from the definition of the  dual of a norm,
\begin{equation}  
\label{eqn5.74n}
\|A\|_{F,*} =\{\sup\;\langle Y,\,A\rangle: {\|Y\|_{F}\le1}\}
\end{equation}
where $Y\in L(\Y,X)$
Now from Cauchy–Schwarz inequality for \(\langle\cdot,\cdot\rangle_{F}\)
\begin{equation}
\label{eqn5.75n}
\langle Y,\,A\rangle_{F}
\;\le\;\|Y\|_{F}\,\|A\|_{F}
\;\le\;\|A\|_{F}    
\end{equation}
Thus from \eqref{eqn5.74n} and \eqref{eqn5.75n},
\begin{equation}
    \|A\|_{F,*}\le\|A\|_{F}
\end{equation}
The equality will hold only when $Y$ is aligned with $A$.
Let \(Y=A/\|A\|_{F}\), then \(\|Y\|_{F}=1\) and
\begin{equation}
\langle Y,\,A\rangle_{F}
=\frac{1}{\|A\|_{F}}\,\langle A,\,A\rangle_{F}
=\frac{\|A\|_{F}^{2}}{\|A\|_{F}}
=\|A\|_{F}.    
\end{equation}
Therefore the supremum is attained when $Y$ aligned with $A$ and we have,
\begin{equation}
\|A\|_{F,*}=\|A\|_{F}  
\end{equation}
\qed\\
Now using the definition of channel $\Lambda_0$ and $\Lambda_1$ in \eqref{eqn20}, one can expand the modified optimisation problem in \eqref{eqn5.26} as,
\begin{equation}
\label{eqn5.79n}
    \begin{split}
n^*(S_{ij}) = \inf \quad & \lambda \Tr\Big(\left[\rho_{AB}^{ij} - \Lambda_0[\rho_{AB}^{ij}] \right]^*\left[\rho_{AB}^{ij} - \Lambda_0[\rho_{AB}^{ij}] \right]\Big) \\
&+ (1-\lambda) \Tr\left(\left[ \rho_{AB}^{ij} - \Lambda_1[\rho_{AB}^{ij}] \right]^*\left[ \rho_{AB}^{ij} - \Lambda_1[\rho_{AB}^{ij}] \right]\right) \\
&+ \frac{\mu}{2} \Tr\left(\left[\rho_{AB}^{ij}\right]^*\left[\rho_{AB}^{ij}\right]\right)  \\
=\inf \quad & \lambda \Tr\Big(\left[\rho_{AB}^{ij} - \Lambda_0[\rho_{AB}^{ij}] \right]^2\Big) + (1-\lambda) \Tr\left(\left[ \rho_{AB}^{ij} - \Lambda_1[\rho_{AB}^{ij}] \right]^2\right) \\
&+ \frac{\mu}{2} \Tr\left(\left[\rho_{AB}^{ij}\right]^2\right)  \\
= \inf \quad & \lambda \Tr\Big(\left[\rho_{AB}^{ij} - \left(Q(0)\otimes \I~\rho_{AB}^{ij}~Q(0)\otimes \I + \{\I-Q(0)\otimes \I\}~\rho_{AB}^{ij}~\{\I-Q(0)\otimes \I\} \right)\right]^2\Big) \\
&+ (1-\lambda) \Tr\left(\left[ \rho_{AB}^{ij} -  \left(Q(\phi_A^i)\otimes \I~\rho_{AB}^{ij}~Q(\phi_A^i)\otimes \I \right.\right.\right. \\
&\left.\left.\left. \qquad \qquad + \{\I-Q(\phi_A^i)\otimes \I\}~\rho_{AB}^{ij}~\{\I-Q(\phi_A^i)\otimes \I\} \right) \right]^2\right)\\
&+ \frac{\mu}{2} \Tr\left(\left[\rho_{AB}^{ij}\right]^2\right)  \\
\text{s.t.} \quad & \begin{aligned}
&\text{Tr}\left(\rho_{AB}^{ij} ~ CHSH(\phi_A^i,\phi_B^j)\right) = S_{ij} \\
& \phi_A^i, \phi_B^j \in [0,\pi/2],\\
&\rho_{AB}^{ij} \succeq 0 \\
&\text{Tr}(\rho_{AB}^{ij}) = 1
\end{aligned}
    \end{split}
\end{equation}
We can also express the above optimization problem in terms of commutator and anti-commutator between the density matrix and the projectors. The product of any density operator $\rho$ and a projector can be decomposed using commutator and the anti-commutator through the following Lemma.\\\\
\textbf{Lemma 10: }$P\rho = \frac{1}{2}\{\rho,P\} - \frac{1}{2}[\rho,P]$ for $\rho \in D(\X)$ for some Hilbert space $\X$ and $P\in Pos(\X)$.\\\\
\textit{Proof: }
Expanding the right-hand using the definition of anti-commutator and commutator, we get,
\begin{equation}
    \begin{aligned}
        &\frac{1}{2}\{\rho,P\} - \frac{1}{2}[\rho,P]\\
        =&\frac{1}{2}\left(\rho P +P \rho \right) - \frac{1}{2}[\rho P -P \rho]\\
        =&\frac{1}{2}(P \rho) +\frac{1}{2}(P \rho)\\
        =&P \rho 
    \end{aligned}
\end{equation}
\qed\\
Now using the \textit{Lemma 10} in \eqref{eqn5.79n}, one can have,
\begin{equation}
    \begin{aligned}
        n^*(S_{ij}) = \inf \quad & \lambda\Tr\left(\left[\rho_{AB}^{ij} - \left(\frac{1}{2}\{\rho_{AB}^{ij}, Q(0)\otimes \I \} - \frac{1}{2}[\rho_{AB}^{ij}, Q(0)\otimes \I ] \right)Q(0)\otimes \I \right.\right. \\
        & \left.\left. \qquad\qquad -\left(\frac{1}{2}\{\rho_{AB}^{ij}, (\I-Q(0)\otimes \I) \} - \frac{1}{2}[\rho_{AB}^{ij}, (\I-Q(0)\otimes \I) ] \right)\{\I-Q(0)\otimes \I\} \right]^2\right) \\
        & + (1-\lambda) \Tr\left(\left[ \rho_{AB}^{ij} - \left(\frac{1}{2}\{\rho_{AB}^{ij}, Q(\phi_A^i)\otimes \I \} - \frac{1}{2}[\rho_{AB}^{ij}, Q(\phi_A^i)\otimes \I ] \right)Q(\phi_A^i)\otimes \I \right.\right. \\
        & \left.\left. \qquad\qquad - \left(\frac{1}{2}\{\rho_{AB}^{ij}, (\I-Q(\phi_A^i)\otimes \I) \} - \frac{1}{2}[\rho_{AB}^{ij}, (\I-Q(\phi_A^i)\otimes \I) ] \right)\{\I-Q(\phi_A^i)\otimes \I\} \right]^2\right)\\
        & + \frac{\mu}{2} \Tr\left(\left[\rho_{AB}^{ij}\right]^2\right)\\
        \text{s.t.} \quad & \text{Tr}\left(\rho_{AB}^{ij} ~ CHSH(\phi_A^i,\phi_B^j)\right) = S_{ij} \\
        & \phi_A^i, \phi_B^j \in [0,\pi/2],\\
        & \rho_{AB}^{ij} \succeq 0 \\
        & \text{Tr}(\rho_{AB}^{ij}) = 1
    \end{aligned}
\end{equation}

\subsubsection{Perturbing the parameter $\phi_A^i$ in $Q(\phi_A^i)$   }
The projector $Q(\cdot)$ is of dimension $2\times2$ and orthogonal onto the one-dimensional subspace spanned by the unit vectors $v(\theta) = \begin{pmatrix} \cos\frac{\theta}{2} \\ \sin\frac{\theta}{2} \end{pmatrix}$. Hence, we can write the projectors as $
Q(\theta) = v(\theta)\,v(\theta)^T$.
If $Q(\theta)$ is a smooth function, the first-order Taylor expansion of this function about $\theta$ under a small increment $\epsilon \ll 1$ yields,
\begin{equation}
\label{eqnF.83}
Q(\theta+\epsilon) \;=\; Q(\theta) + \epsilon\,Q'(\theta) + \mathcal{O}(\epsilon^2),    
\end{equation}
where $Q'(\theta)$ is the first order derivative of the projector $Q$. As $\epsilon \to 0$, norm of the higher order terms in $\mathcal{O}(\epsilon^2)$ are bounded by $C\,\epsilon^2$ .
Thus, $Q(\phi + \epsilon)$ is linearly approximated up to first order derivative of $Q(\phi)$ with respect to $\theta$. 
\begin{equation}
    Q(\theta+\epsilon) \approx Q(\theta) + \epsilon Q'(\theta)
\end{equation}  
The first order derivative of the vector $v(\theta)$ is given as,
\begin{equation}
\label{eqnF.85}
v'(\theta) = \frac{d}{d\theta}\begin{pmatrix}\cos(\theta/2)\\ \sin(\theta/2)\end{pmatrix}
            = \begin{pmatrix} -\tfrac12\sin(\theta/2) \\[4pt] \tfrac12\cos(\theta/2) \end{pmatrix}.
\end{equation}
Now from the definition of $Q(\theta)$ and matrix product rule one can have,
\begin{equation}
Q'(\theta) \;=\; \frac{d}{d\theta}\bigl(v(\theta) v(\theta)^T\bigr)
              = v'(\theta)\,v(\theta)^T + v(\theta)\,\bigl(v'(\theta)\bigr)^T    
\end{equation}
and finally,
\begin{equation}
    Q'(\theta) \;=\; v(\theta)'v(\theta)^T + v(\theta)(v(\theta)')^T
    = \begin{pmatrix}
      -\cos(\theta/2)\sin(\theta/2) & \tfrac12(\cos^2(\theta/2)-\sin^2(\theta/2)) \\[4pt]
       \tfrac12(\cos^2(\theta/2)-\sin^2(\theta/2)) & \cos(\theta/2)\sin(\theta/2)
    \end{pmatrix}.
\end{equation}
Using the identities $\sin\theta = 2\sin(\theta/2)\cos(\theta/2)$ and $\cos\theta = \cos^2(\theta/2)-\sin^2(\theta/2)$, the above equation simplifies to the following compact form,
\begin{equation}
\label{eqnF.89}
    Q'(\theta) \;=\; \frac{1}{2}\begin{pmatrix}
    -\sin\theta & \cos\theta \\[4pt]
     \cos\theta & \sin\theta
  \end{pmatrix}
\end{equation}
Hence, the first-order Taylor approximation of $Q(\theta)$ comes out to be,
\begin{equation}
\begin{aligned}
    \,Q(\theta+\epsilon) &\approx Q(\theta) + \epsilon\,Q'(\theta) \\
&\approx \begin{pmatrix}
            \cos^2\left(\frac{\theta}{2}\right) & \cos\left(\frac{\theta}{2}\right)\sin\left(\frac{\theta}{2}\right) \\
            \cos\left(\frac{\theta}{2}\right)\sin\left(\frac{\theta}{2}\right)  & \sin^2\left(\frac{\theta}{2}\right)
        \end{pmatrix}+\frac{\epsilon}{2}\begin{pmatrix} -\sin\theta & \cos\theta \\[4pt] \cos\theta & \sin\theta \end{pmatrix}\\
        &\approx \begin{pmatrix}
            \cos^2\left(\frac{\theta}{2}\right) -\frac{\epsilon\sin(\theta)}{2} & \cos\left(\frac{\theta}{2}\right)\sin\left(\frac{\theta}{2}\right) +\frac{\epsilon\cos(\theta)}{2} \\
            \cos\left(\frac{\theta}{2}\right)\sin\left(\frac{\theta}{2}\right)+\frac{\epsilon\cos(\theta)}{2}  & \sin^2\left(\frac{\theta}{2}\right)+\frac{\epsilon\sin(\theta)}{2}
        \end{pmatrix}
\end{aligned}
\end{equation}
After establishing an approximate value of $Q(\phi +\epsilon)$ one can proceed by expressing $(Q(\phi_A^i + \epsilon)\otimes \I)\rho_{AB}^{ij}$ in first order linear approximation as,
\begin{equation}
    \begin{aligned}
        (Q(\phi_A^i + \epsilon)\otimes \I)\rho_{AB}^{ij} = (Q(\phi_A^i)\otimes \I)\rho_{AB}^{ij} + \epsilon(Q'(\phi_A^i )\otimes \I)\rho_{AB}^{ij} +\mathcal{O}(\epsilon^2)
    \end{aligned}
\end{equation}
and terms of anti-commuting and commuting operators from the result of \textit{Lemma 10} and \eqref{eqnF.83}
as,
\begin{equation}
    \begin{aligned}
(Q(\phi_A^i + \epsilon)\otimes \I)\rho_{AB}^{ij} = \frac{1}{2}\{\rho_{AB}^{ij},Q(\phi_A^i + \epsilon)\otimes \I\} -\frac{1}{2}[\rho_{AB}^{ij},Q(\phi_A^i + \epsilon)\otimes \I]   \end{aligned}
\end{equation}
We expand the commuting and the anti-commuting term individually as,
\begin{eqnarray}
        [\rho_{AB}^{ij},Q(\phi_A^i + \epsilon)\otimes \I] &=& \rho_{AB}^{ij}~Q(\phi_A^i + \epsilon)\otimes \I - Q(\phi_A^i + \epsilon)\otimes \I~\rho_{AB}^{ij}\nonumber\\
    &=& \rho_{AB}^{ij}\left(Q(\phi_A^i )\otimes \I +\epsilon(Q'(\phi_A^i )\otimes \I) + \mathcal{O}(\epsilon^2)\right)\nonumber \\
    &&-\left(Q(\phi_A^i )\otimes \I +\epsilon(Q'(\phi_A^i )\otimes \I) + \mathcal{O}(\epsilon^2)\right)\rho_{AB}^{ij}\nonumber\\
    &=&\left( \rho_{AB}^{ij}~Q(\phi_A^i )\otimes \I - Q(\phi_A^i )\otimes \I~\rho_{AB}^{ij}   \right)\nonumber \\
    &&+ ~\epsilon\left(  \rho_{AB}^{ij}~Q'(\phi_A^i )\otimes \I - Q'(\phi_A^i )\otimes \I~\rho_{AB}^{ij} \right) + \mathcal{O}\left(\epsilon^2\right)\nonumber\\
    &=&[\rho_{AB}^{ij},Q(\phi_A^i )\otimes \I] + \epsilon[\rho_{AB}^{ij}, Q'(\phi_A^i )\otimes \I] + \mathcal{O}(\epsilon^2)   
\end{eqnarray}
Similarly, for other terms we get,
\begin{equation}
\label{eqnF.94}
    \begin{aligned}
 & \{\rho_{AB}^{ij}, Q(\phi_A^i + \epsilon)\otimes \I\} =\{\rho_{AB}^{ij}, Q(\phi_A^i )\otimes \I\} + \epsilon\{\rho_{AB}^{ij}, Q'(\phi_A^i )\otimes \I\} + \mathcal{O}(\epsilon^2)\\
 &[\rho_{AB}^{ij}, (\I-Q(\phi_A^i +\epsilon )\otimes \I)] =[\rho_{AB}^{ij},(\I-Q(\phi_A^i)\otimes \I)] + \epsilon[\rho_{AB}^{ij},(\I-Q'(\phi_A^i)\otimes \I)] + \mathcal{O}(\epsilon^2)\\
    &\{\rho_{AB}^{ij},(\I-Q(\phi_A^i +\epsilon )\otimes \I)\} =\{\rho_{AB}^{ij},(\I-Q(\phi_A^i)\otimes \I)\} + \epsilon\{\rho_{AB}^{ij},(\I-Q'(\phi_A^i)\otimes \I)\} + \mathcal{O}(\epsilon^2)
    \end{aligned}
\end{equation}
Now, let us consider the following function, 
\begin{equation}
    h:(\lambda,\phi_A^i,\rho_{AB}^{ij}) \mapsto \mathbb{R}
\end{equation}
being defined as follows,
\begin{equation}
\begin{aligned}
    h(\lambda,\phi_A^i,\rho_{AB}^{ij}) &=  (1-\lambda) \Tr\left(\left[ \rho_{AB}^{ij} - \left(\frac{1}{2}\{\rho_{AB}^{ij}, Q(\phi_A^i)\otimes \I \} - \frac{1}{2}[\rho_{AB}^{ij}, Q(\phi_A^i)\otimes \I ] \right)Q(\phi_A^i)\otimes \I \right.\right. \\
 & \left.\left. \qquad\qquad - \left(\frac{1}{2}\{\rho_{AB}^{ij}, (\I-Q(\phi_A^i)\otimes \I) \} - \frac{1}{2}[\rho_{AB}^{ij}, (\I-Q(\phi_A^i)\otimes \I) ] \right)\{\I-Q(\phi_A^i)\otimes \I\} \right]^2\right)
\end{aligned}
\end{equation}

Now, one can bound the small perturbation $\epsilon$ in $\phi_A^i$ using triangular inequality as,
\begin{equation}
    \begin{aligned}
    \label{eqnF.97}
        &\Big|h(\lambda,\phi_A^i,\rho_{AB}^{ij}) -  h(\lambda,\phi_A^i+\epsilon,\rho_{AB}^{ij})\Big| \\
        \leq&\Big|h(\lambda,\phi_A^i,\rho_{AB}^{ij})\Big| + \Big|h(\lambda,\phi_A^i+\epsilon,\rho_{AB}^{ij})\Big|\\
        \leq&(1-\lambda) \Tr\left(\left[ \rho_{AB}^{ij} - \left(\frac{1}{2}\{\rho_{AB}^{ij}, Q(\phi_A^i)\otimes \I \} - \frac{1}{2}[\rho_{AB}^{ij}, Q(\phi_A^i)\otimes \I ] \right)Q(\phi_A^i)\otimes \I \right.\right. \\
        & \left.\left. \qquad\qquad - \left(\frac{1}{2}\{\rho_{AB}^{ij}, (\I-Q(\phi_A^i)\otimes \I) \} - \frac{1}{2}[\rho_{AB}^{ij}, (\I-Q(\phi_A^i)\otimes \I) ] \right)\{\I-Q(\phi_A^i)\otimes \I\} \right]^2\right)\\ 
        & +(1-\lambda) \Tr\left(\left[ \rho_{AB}^{ij} - \left(\frac{1}{2}\{\rho_{AB}^{ij}, Q(\phi_A^i+\epsilon_0)\otimes \I \} - \frac{1}{2}[\rho_{AB}^{ij}, Q(\phi_A^i+\epsilon_0)\otimes \I ] \right)Q(\phi_A^i+\epsilon_0)\otimes \I \right.\right. \\
        & \left.\left. \qquad\qquad - \left(\frac{1}{2}\{\rho_{AB}^{ij}, (\I-Q(\phi_A^i+\epsilon_0)\otimes \I) \} - \frac{1}{2}[\rho_{AB}^{ij}, (\I-Q(\phi_A^i+\epsilon_0)\otimes \I) ] \right)\{\I-Q(\phi_A^i+\epsilon_0)\otimes \I\} 
        \right]^2\right)\\
        \leq&(1-\lambda)\Big[ \Tr\left(\left[ \rho_{AB}^{ij} - \left(\frac{1}{2}\{\rho_{AB}^{ij}, Q(\phi_A^i)\otimes \I \} - \frac{1}{2}[\rho_{AB}^{ij}, Q(\phi_A^i)\otimes \I ] \right)Q(\phi_A^i)\otimes \I \right.\right. \\
  & \left.\left. \qquad\qquad - \left(\frac{1}{2}\{\rho_{AB}^{ij}, (\I-Q(\phi_A^i)\otimes \I) \} - \frac{1}{2}[\rho_{AB}^{ij}, (\I-Q(\phi_A^i)\otimes \I) ] \right)(\I-Q(\phi_A^i)\otimes \I) \right]^2 \right. \\
& \qquad\qquad + \left[ \rho_{AB}^{ij} - \left(\frac{1}{2}\{\rho_{AB}^{ij}, Q(\phi_A^i+\epsilon_0)\otimes \I \} - \frac{1}{2}[\rho_{AB}^{ij}, Q(\phi_A^i+\epsilon_0)\otimes \I ] \right)Q(\phi_A^i+\epsilon_0)\otimes \I \right. \\
  & \left.\left. \qquad\qquad - \left(\frac{1}{2}\{\rho_{AB}^{ij}, (\I-Q(\phi_A^i+\epsilon_0)\otimes \I) \} - \frac{1}{2}[\rho_{AB}^{ij}, (\I-Q(\phi_A^i+\epsilon_0)\otimes \I) ] \right)(\I-Q(\phi_A^i+\epsilon_0)\otimes \I) \right]^2\right) \Big]
   \end{aligned}
\end{equation}

To simplify the calculation, we simplify the perturbing and non-perturbing terms separately as,
\begin{equation}
    \begin{aligned}
        \text{A} &= \left[ \rho_{AB}^{ij} - \left(\frac{1}{2}\{\rho_{AB}^{ij}, P_0 \} - \frac{1}{2}[\rho_{AB}^{ij}, P_0 ] \right)P_0  - \left(\frac{1}{2}\{\rho_{AB}^{ij}, (\I-P_0) \} - \frac{1}{2}[\rho_{AB}^{ij}, (\I-P_0) ] \right)(\I-P_0) \right]^2 \\
    &=\left[  \rho_{AB}^{ij} -\frac{1}{2}\{\rho_{AB}^{ij}, P_0\}P_0 +  \frac{1}{2}[\rho_{AB}^{ij}, P_0 ] P_0 - \frac{1}{2}\{\rho_{AB}^{ij}, (\I-P_0) \}(\I-P_0) +  \frac{1}{2}[\rho_{AB}^{ij}, (\I-P_0) ] (\I-P_0)\right]^2\\
    &= \left[ \rho_{AB}^{ij} - (P_0)~\rho_{AB}^{ij} ~(P_0) - (\I-P_0)~ \rho_{AB}^{ij}~(\I-P_0) \right]^2\\
    &= \left[\rho_{AB}^{ij} ~P_0 + P_0~\rho_{AB}^{ij} -2P_0~\rho_{AB}^{ij} ~P_0 \right]^2\\
    &= \left[(\I-P_0) ~\rho_{AB}^{ij} ~ (P_0) +(P_0)~\rho_{AB}^{ij}(\I-P_0)\right]^2\\
    &= \left((\I-P_0) ~\rho_{AB}^{ij} ~ Q(\phi_A^i) \otimes \I ~\rho_{AB}^{ij}(\I-P_0) +P_0 ~\rho_{AB}^{ij}(\I-P_0)~\rho_{AB}^{ij} ~ P_0 \right)\\
    &=A_0 +A_1
    \end{aligned}
\end{equation}

and
\begin{equation}
\begin{aligned}
\text{B} &= \left[ \rho_{AB}^{ij}
- \left(
\frac{1}{2} \left\{ \rho_{AB}^{ij}, Q(\phi_A^i+\epsilon_0)\otimes \I \right\}
- \frac{1}{2} \left[ \rho_{AB}^{ij}, Q(\phi_A^i+\epsilon_0)\otimes \I \right]
\right)Q(\phi_A^i+\epsilon_0)\otimes \I \right. \\
&\quad \left. - \left(
\frac{1}{2} \left\{ \rho_{AB}^{ij}, \I - Q(\phi_A^i+\epsilon_0)\otimes \I \right\}
- \frac{1}{2} \left[ \rho_{AB}^{ij}, \I - Q(\phi_A^i+\epsilon_0)\otimes \I \right]
\right)(\I - Q(\phi_A^i+\epsilon_0)\otimes \I)
\right]^2 \\
&= \left[ \{ \rho_{AB}^{ij}, Q(\phi_A^i+\epsilon_0) \otimes \I \} - 2 Q(\phi_A^i+\epsilon_0)\otimes \I \rho_{AB}^{ij} Q(\phi_A^i+\epsilon_0)\otimes \I \right]^2 \\
&= \left[ \left(\{ \rho_{AB}^{ij}, Q(\phi_A^i+\epsilon_0) \otimes \I \}\right)^2 + \left(- 2 Q(\phi_A^i+\epsilon_0)\otimes \I \rho_{AB}^{ij} Q(\phi_A^i+\epsilon_0)\otimes \I \right)^2 \right. \\
&\quad \left. - \{ \rho_{AB}^{ij}, Q(\phi_A^i+\epsilon_0) \otimes \I \} 2 Q(\phi_A^i+\epsilon_0)\otimes \I \rho_{AB}^{ij} Q(\phi_A^i+\epsilon_0)\otimes \I \right. \\
&\quad \left. - 2 Q(\phi_A^i+\epsilon_0)\otimes \I \rho_{AB}^{ij} Q(\phi_A^i+\epsilon_0)\otimes \I \{ \rho_{AB}^{ij}, Q(\phi_A^i+\epsilon_0) \otimes \I \} \right] \\
&= \left[\left(\{\rho_{AB}^{ij},P_0\} + \epsilon_0\{\rho_{AB}^{ij},P_1\} + \mathcal{O}(\epsilon_0^2) \right)^2 + \left(4\xi \rho_{AB}^{ij} \xi \rho_{AB}^{ij} \xi \right) - \left( \{ \rho_{AB}^{ij}, \xi \} 2 \xi\rho_{AB}^{ij} \xi \right) - \left(2 \xi\rho_{AB}^{ij} \xi \{ \rho_{AB}^{ij}, \xi \} \right) \right] \\
&= \left[\left(\{\rho_{AB}^{ij},P_0\}^2+\epsilon_0 \left(\{\rho_{AB}^{ij},P_0\}\{\rho_{AB}^{ij},P_1\} + \{\rho_{AB}^{ij},P_1\}\{\rho_{AB}^{ij},P_0\}\right)  + \epsilon_0^2\{\rho_{AB}^{ij},P_1\}^2\right)  + \left(4\xi \rho_{AB}^{ij} \xi \rho_{AB}^{ij} \xi \right) \right. \\
&\quad \left. - \left( \{ \rho_{AB}^{ij}, \xi \} 2 \xi\rho_{AB}^{ij} \xi \right) - \left(2 \xi\rho_{AB}^{ij} \xi \{ \rho_{AB}^{ij}, \xi \} \right) \right] + \mathcal{O}(\epsilon_0^3)\\
&= \left[\left(\{\rho_{AB}^{ij},P_0\}^2+\epsilon_0 \left(\{\rho_{AB}^{ij},P_0\}\{\rho_{AB}^{ij},P_1\} + \{\rho_{AB}^{ij},P_1\}\{\rho_{AB}^{ij},P_0\}\right)  + \epsilon_0^2\{\rho_{AB}^{ij},P_1\}^2\right) \right. \\
&\quad \left. + 4\left(P_0\rho_{AB}^{ij}P_0\rho_{AB}^{ij}P_0 + \epsilon_0(P_1\rho_{AB}^{ij}P_0\rho_{AB}^{ij}P_0 + P_0\rho_{AB}^{ij}P_1\rho_{AB}^{ij}P_0 + P_0\rho_{AB}^{ij}P_0\rho_{AB}^{ij}P_1) \right. \right. \\
&\quad \left. \left. +\epsilon_0^2(P_0\rho_{AB}^{ij}P_1\rho_{AB}^{ij}P_1 + P_1\rho_{AB}^{ij}P_0\rho_{AB}^{ij}P_1 +P_1\rho_{AB}^{ij}P_1\rho_{AB}^{ij}P_0 +P_1\rho_{AB}^{ij}P_1\rho_{AB}^{ij}P_1 )\right) - C_1 - C_2 \right] + \mathcal{O}(\epsilon_0^3)\\
&= B_0 + \epsilon_0 B_1 +\epsilon_0^2 B_2 + \mathcal{O}(\epsilon_0^3)
\end{aligned}
\end{equation}

\begin{align*}
B_0 &= \{\rho_{AB}^{ij}, P_0\}^2 +4 P_0 \rho_{AB}^{ij} P_0 \rho_{AB}^{ij} P_0 -2 \{ \rho, P_0 \} P_0 \rho P_0 -2 P_0 \rho P_0 \{ \rho, P_0 \} \\
B_1 &= \{\rho_{AB}^{ij},P_0\}\{\rho_{AB}^{ij},P_1\} + \{\rho_{AB}^{ij},P_1\}\{\rho_{AB}^{ij},P_0\} + 4(P_1 \rho_{AB}^{ij} P_0 \rho_{AB}^{ij} P_0 + P_0 \rho_{AB}^{ij} P_1 \rho_{AB}^{ij} P_0 + P_0 \rho_{AB}^{ij} P_0 \rho_{AB}^{ij} P_1) \\
&\quad -2( \{ \rho, P_0 \} (P_0 \rho P_1 + P_1 \rho P_0) + \{ \rho, P_1 \} P_0 \rho P_0) -2( (P_0 \rho P_1 + P_1 \rho P_0) \{ \rho, P_0 \} + P_0 \rho P_0 \{ \rho, P_1 \}) \\
B_2 &= \{\rho_{AB}^{ij},P_1\}^2 + 4(P_0 \rho_{AB}^{ij} P_1 \rho_{AB}^{ij} P_1 + P_1 \rho_{AB}^{ij} P_0 \rho_{AB}^{ij} P_1 + P_1 \rho_{AB}^{ij} P_1 \rho_{AB}^{ij} P_0 +P_1 \rho_{AB}^{ij} P_1 \rho_{AB}^{ij} P_1 ) \\
&\quad -2\left( \{ \rho, P_0 \} P_1 \rho P_1 + \{ \rho, P_1 \} (P_0 \rho P_1 + P_1 \rho P_0) \right)  -2\left( P_1 \rho P_1 \{ \rho, P_0 \} + (P_0 \rho P_1 + P_1 \rho P_0) \{ \rho, P_1 \} \right)
\end{align*}
where $\xi = ( P_0 + \epsilon_0 P_1 + \mathcal{O}(\epsilon_0^2)), P_0 = Q(\phi_A^i) \otimes \I, P_1 = Q'(\phi_A^i) \otimes \I$
\begin{align*}
C_1 &= -2 \left[ \{ \rho, P_0 \} P_0 \rho P_0 + \epsilon_0 \left( \{ \rho, P_0 \} (P_0 \rho P_1 + P_1 \rho P_0) + \{ \rho, P_1 \} P_0 \rho P_0 \right)  + \epsilon_0^2 \left( \{ \rho, P_0 \} P_1 \rho P_1 + \{ \rho, P_1 \} (P_0 \rho P_1 + P_1 \rho P_0) \right) \right] + \mathcal{O}(\epsilon_0^2) \\
C_2 &= -2 \left[ P_0 \rho P_0 \{ \rho, P_0 \} + \epsilon_0 \left( (P_0 \rho P_1 + P_1 \rho P_0) \{ \rho, P_0 \} + P_0 \rho P_0 \{ \rho, P_1 \} \right) + \epsilon_0^2 \left( P_1 \rho P_1 \{ \rho, P_0 \} + (P_0 \rho P_1 + P_1 \rho P_0) \{ \rho, P_1 \} \right) \right] + \mathcal{O}(\epsilon_0^2)
\end{align*}

Inserting the explicit forms of A and B in \eqref{eqnF.97} yields,
\begin{equation}
\label{eqnF.105}
    \begin{aligned}
        &\left|h(\lambda,\phi_A^i,\rho_{AB}^{ij}) - h(\lambda,\phi_A^i+\epsilon_0,\rho_{AB}^{ij})\right| \\
\leq{}&(1-\lambda) \left[\Tr( A + B) \right] \\
\leq{}&(1-\lambda) \left[ \Tr(A_0) + \Tr(A_1) + \Tr(B_0) + \epsilon_0\Tr(B_1) + \epsilon_0^2~\Tr(B_2) \right]\\
    \end{aligned}
\end{equation}
Now, upon solving each of the term separately we have,
\begin{equation}
    \begin{aligned}
        \Tr(A_0) &= \Tr((\I-Q(\phi_A^i)\otimes \I) ~\rho_{AB}^{ij} ~ Q(\phi_A^i) \otimes \I ~\rho_{AB}^{ij}(\I-Q(\phi_A^i)\otimes \I))\\
        &= \Tr((\I-Q(\phi_A^i)\otimes \I) ~\rho_{AB}^{ij} ~ (Q(\phi_A^i)\otimes \I) ~\rho_{AB}^{ij})\\
        &= \Tr(~\rho_{AB}^{ij}~(\I-Q(\phi_A^i)\otimes \I) ~\rho_{AB}^{ij} ~ (Q(\phi_A^i)\otimes \I) )\\
        &\leq \Tr(~\rho_{AB}^{ij}~(\I-Q(\phi_A^i)\otimes \I) ~\rho_{AB}^{ij})\\
    \end{aligned}
\end{equation}
The inequality in the above equation arises from the Loewner order for two hermitian $\I$ and projector $P = (Q(\phi_A^i) \otimes \I)$ as,
\begin{equation}
    \begin{aligned}
    &(\I - P) \succeq0\\
        &P \preceq  \I\\
        & YP \preceq  Y\I \quad \text{(for Y = $\rho(\I - P)\rho \succeq 0$)}\\
        &YP \preceq  Y\\
        &Y - YP \succeq 0\\
        &\Tr(Y - YP) \geq 0\\
        &\Tr(Y) - \Tr(YP) \geq 0\\
    &\Tr(Y) \geq \Tr(YP)\\
    &\Tr(\rho(\I - P)\rho) \geq \Tr(\rho(\I - P)\rho P)
    \end{aligned}
\end{equation}
for $\rho = \rho_{AB}^{ij}$.\\
Coming back to the proof for $\Tr(A_0)$
\begin{equation}
    \begin{aligned}
        \Tr(A_0)  &\leq \Tr(~\rho_{AB}^{ij}~(\I-Q(\phi_A^i)\otimes \I) ~\rho_{AB}^{ij})\\ 
        &\leq \Tr(~\rho_{AB}^{ij}~(\I-Q(\phi_A^i)\otimes \I))\\
        &\leq \Tr(~\rho_{AB}^{ij}) - \Tr(~\rho_{AB}^{ij} (Q(\phi_A^i)\otimes \I)\\
        &\leq 1- \Tr(~\rho_{AB}^{ij} (Q(\phi_A^i)\otimes \I)\\
        &\leq 1 \quad (\text{ as $\Tr(~\rho_{AB}^{ij} (Q(\phi_A^i)\otimes \I) \geq 0$ })
    \end{aligned}
\end{equation}
Similarly,
\begin{equation}
    \begin{aligned}
        \Tr(A_1) &= \Tr((Q(\phi_A^i)\otimes \I) ~\rho_{AB}^{ij}(\I-Q(\phi_A^i)\otimes \I)~\rho_{AB}^{ij} ~ (Q(\phi_A^i)\otimes \I))\\
          &= \Tr((Q(\phi_A^i)\otimes \I) ~\rho_{AB}^{ij} ~ (\I-Q(\phi_A^i)\otimes \I) ~\rho_{AB}^{ij})\\ 
          &\leq 1
    \end{aligned}
\end{equation}

Now we can proceed towards computing other trace in a similar fashion.
\begin{equation}
    \begin{aligned}
        \Tr(B_0) &= \Tr\left(\{\rho_{AB}^{ij}, P_0\}^2 +4 P_0 \rho_{AB}^{ij} P_0 \rho_{AB}^{ij} P_0 -2 \{ \rho, P_0 \} P_0 \rho P_0 -2 P_0 \rho P_0 \{ \rho, P_0 \}\right)\\
        &= \Tr\left( (\rho_{AB}^{ij}~P_0 + P_0~\rho_{AB}^{ij})^2 +4 P_0 \rho_{AB}^{ij} P_0 \rho_{AB}^{ij} P_0 -2 \{ \rho, P_0 \} P_0 \rho P_0 -2 P_0 \rho P_0 \{ \rho, P_0 \}    \right)\\
&= \Tr\left( (\rho_{AB}^{ij}~P_0)^2 + (P_0~\rho_{AB}^{ij})^2 + \rho_{AB}^{ij}~P_0P_0~\rho_{AB}^{ij}  + P_0~\rho_{AB}^{ij}\rho_{AB}^{ij}~P_0 +4 P_0 \rho_{AB}^{ij} P_0 \rho_{AB}^{ij} P_0 -2 \{ \rho, P_0 \} P_0 \rho P_0 -2 P_0 \rho P_0 \{ \rho, P_0 \}    \right)\\  
&= \left( \Tr(\rho_{AB}^{ij}~P_0~\rho_{AB}^{ij}~P_0) + \Tr(P_0~\rho_{AB}^{ij}P_0~\rho_{AB}^{ij}) + \Tr(\rho_{AB}^{ij}~P_0~\rho_{AB}^{ij} ) + \Tr(P_0~\rho_{AB}^{ij}\rho_{AB}^{ij}~P_0) +\Tr(4 P_0 \rho_{AB}^{ij} P_0 \rho_{AB}^{ij} P_0)  \right. \\
&\quad\quad\qquad \left. -2\Tr( ( \rho~P_0~\rho P_0) -2\Tr(P_0~\rho   ~P_0 \rho P_0) -2\Tr( P_0 \rho P_0~\rho~P_0 )-2\Tr( P_0 \rho P_0~\rho  )  )  \right)\\  
&= \left( \Tr(~P_0\rho_{AB}^{ij}~P_0~\rho_{AB}^{ij}) + \Tr(P_0~\rho_{AB}^{ij}P_0~\rho_{AB}^{ij}) + \Tr(\rho_{AB}^{ij}~P_0~\rho_{AB}^{ij} ) + \Tr(P_0~\rho_{AB}^{ij}\rho_{AB}^{ij}) +4\Tr( P_0 \rho_{AB}^{ij} P_0 \rho_{AB}^{ij} )  \right. \\
&\quad\quad\qquad \left. -2\Tr( (P_0 \rho~P_0~\rho ) -2\Tr(P_0~\rho   ~P_0 \rho ) -2\Tr( P_0 \rho P_0~\rho~ )-2\Tr( P_0 \rho P_0~\rho  )  )  \right)\\ 
&= \left(  \Tr(P_0~\rho_{AB}^{ij}\rho_{AB}^{ij}) ) + \Tr(P_0~\rho_{AB}^{ij}\rho_{AB}^{ij})  -2\Tr( P_0 \rho P_0~\rho  )  )  \right)\\ 
&= 2(\Tr(P_0~(\rho_{AB}^{ij})^2) - \Tr((P_0\rho_{AB}^{ij})^2))\\
\end{aligned}
\end{equation}
Using a similar form of Loewner order for two hermetian $\I$ and projector $P$
it can be proven that $0\leq \Tr(P_0~(\rho_{AB}^{ij})^2) \leq 1 $ and $\Tr(P_0~(\rho_{AB}^{ij})^2) \geq \Tr((P_0\rho_{AB}^{ij})^2)$.
\begin{equation}
    \begin{aligned}
        &P \preceq  \I\\
        &(\rho_{AB}^{ij})^2~P \preceq  (\rho_{AB}^{ij})^2 \qquad (Y= (\rho_{AB}^{ij})^2 \succeq 0) \\
        &(\rho_{AB}^{ij})^2 - (\rho_{AB}^{ij})^2~P \succeq0\\
        &\Tr((\rho_{AB}^{ij})^2) - \Tr( (\rho_{AB}^{ij})^2~P) \geq 0\\
        &1 - \Tr( (\rho_{AB}^{ij})^2~P) \geq 0\\
        & 0\leq \Tr( (\rho_{AB}^{ij})^2~P) \leq 1 \quad (\text{since } \rho_{AB}^{ij} \succeq 0  \text{ and } P \succeq 0, \text{ lower bound by 0} )
    \\ \text{ and ,}\\
         &P \preceq  \I\\
        &(\rho_{AB}^{ij}~P~\rho_{AB}^{ij})P \preceq  (\rho_{AB}^{ij}~P~\rho_{AB}^{ij}) \qquad (Y=\rho_{AB}^{ij}~P~\rho_{AB}^{ij}\succeq 0) \\
        &\Tr(\rho_{AB}^{ij}~P~\rho_{AB}^{ij}) - \Tr((\rho_{AB}^{ij}~P)^2) \geq 0\\
        &\Tr(\rho_{AB}^{ij}~P~\rho_{AB}^{ij}) \geq \Tr((\rho_{AB}^{ij}~P)^2)  
    \end{aligned}
\end{equation}
Thus,
\begin{equation}
    \Tr(B_0) \leq 2
\end{equation}

$\Tr(B_1)$ can also be computed in a similar fashion.
\begin{equation}
    \begin{aligned}
        \Tr(B_1) &= \Tr(\{\rho_{AB}^{ij},P_0\}\{\rho_{AB}^{ij},P_1\} + \{\rho_{AB}^{ij},P_1\}\{\rho_{AB}^{ij},P_0\}
+ 4(P_1 \rho_{AB}^{ij} P_0 \rho_{AB}^{ij} P_0 + P_0 \rho_{AB}^{ij} P_1 \rho_{AB}^{ij} P_0 + P_0 \rho_{AB}^{ij} P_0 \rho_{AB}^{ij} P_1) \\
&\quad -2( \{ \rho, P_0 \} (P_0 \rho P_1 + P_1 \rho P_0) + \{ \rho, P_1 \} P_0 \rho P_0) -2( (P_0 \rho P_1 + P_1 \rho P_0) \{ \rho, P_0 \} + P_0 \rho P_0 \{ \rho, P_1 \}))\\
&= \Tr\left((\rho_{AB}^{ij}~P_0 +~P_0~\rho_{AB}^{ij})(\rho_{AB}^{ij}~P_1 + ~P_1~\rho_{AB}^{ij})\right) + \Tr\left((\rho_{AB}^{ij}~P_1 + ~P_1~\rho_{AB}^{ij})(\rho_{AB}^{ij}~P_0 +~P_0~\rho_{AB}^{ij})\right) \\
&\qquad +4~\Tr\left(P_1 \rho_{AB}^{ij} P_0 \rho_{AB}^{ij} P_0 + P_0 \rho_{AB}^{ij} P_1 \rho_{AB}^{ij} P_0 + P_0 \rho_{AB}^{ij} P_0 \rho_{AB}^{ij} P_1\right) -2\Tr\left( \{ \rho, P_0 \} (P_0 \rho P_1 + P_1 \rho P_0) + \{ \rho, P_1 \} P_0 \rho P_0\right) \\ &\qquad -2\Tr\left( (P_0 \rho P_1 + P_1 \rho P_0) \{ \rho, P_0 \} + P_0 \rho P_0 \{ \rho, P_1 \}\right)\\
&= \Tr\left(\rho_{AB}^{ij}~P_0~\rho_{AB}^{ij}~P_1\right)+ 
\Tr\left(\rho_{AB}^{ij}~P_0~P_1~\rho_{AB}^{ij}\right)+ \Tr\left(P_0~\rho_{AB}^{ij}~\rho_{AB}^{ij}~P_1\right) +\Tr\left( P_0~\rho_{AB}^{ij}~P_1~\rho_{AB}^{ij}\right) \\&\qquad +\Tr\left( \rho_{AB}^{ij}~P_1~\rho_{AB}^{ij}~P_0 \right) +\Tr\left( \rho_{AB}^{ij}~P_1~P_0~\rho_{AB}^{ij} \right) +\Tr\left( P_1~\rho_{AB}^{ij}~\rho_{AB}^{ij}~P_0 \right) +\Tr\left( P_1~\rho_{AB}^{ij}~P_0~\rho_{AB}^{ij}   \right) \\&\qquad 
+ 4\Tr\left(P_1 \rho_{AB}^{ij} P_0 \rho_{AB}^{ij} P_0 \right) +4\Tr\left(P_0 \rho_{AB}^{ij} P_1 \rho_{AB}^{ij} P_0   \right) +4\Tr\left(P_0 \rho_{AB}^{ij} P_0 \rho_{AB}^{ij} P_1   \right) \\&\qquad -2\Tr\left(\rho_{AB}^{ij}~P_0~\rho_{AB}^{ij}~P_1 \right) -2\Tr\left( \rho_{AB}^{ij}~P_0~P_1~\rho_{AB}^{ij}~P_0\right)-2\Tr\left(P_0~\rho_{AB}^{ij}~P_0~\rho_{AB}^{ij}~P_1\right)-2\Tr\left( P_0~\rho_{AB}^{ij}~P_1~\rho_{AB}^{ij} \right)\\&\qquad -2\Tr\left(\rho_{AB}^{ij}~P_1~P_0~\rho_{AB}^{ij}~P_0\right)-2\Tr\left(P_1~\rho_{AB}^{ij}~P_0~\rho_{AB}^{ij}~P_0\right)\\&\qquad -2\Tr\left( P_0~\rho_{AB}^{ij}~P_1~\rho_{AB}^{ij}\right) -2\Tr\left(P_0~\rho_{AB}^{ij}~P_1~P_0~\rho_{AB}^{ij} \right) -2\Tr\left(P_1~\rho_{AB}^{ij}~P_0~\rho_{AB}^{ij}~P_0\right) -2\Tr\left(P_1~\rho_{AB}^{ij}~P_0~\rho_{AB}^{ij}\right)\\&\qquad -2\Tr\left(P_0~\rho_{AB}^{ij}~P_0~\rho_{AB}^{ij}~P_1\right) -2\Tr\left(P_0~\rho_{AB}^{ij}~P_0~P_1~\rho_{AB}^{ij}\right)\\
&= 
\Tr\left(\rho_{AB}^{ij}~P_0~P_1~\rho_{AB}^{ij}\right)+ \Tr\left(P_0~\rho_{AB}^{ij}~\rho_{AB}^{ij}~P_1\right)  +\Tr\left( \rho_{AB}^{ij}~P_1~P_0~\rho_{AB}^{ij} \right) +\Tr\left( P_1~\rho_{AB}^{ij}~\rho_{AB}^{ij}~P_0 \right)
 +4\Tr\left(P_0 \rho_{AB}^{ij} P_1 \rho_{AB}^{ij} P_0   \right)  \\&\qquad  -2\Tr\left( \rho_{AB}^{ij}~P_0~P_1~\rho_{AB}^{ij}~P_0\right)-2\Tr\left( \rho_{AB}^{ij}~P_0~\rho_{AB}^{ij}~P_1 \right) -2\Tr\left(\rho_{AB}^{ij}~P_1~P_0~\rho_{AB}^{ij}~P_0\right) \\&\qquad -2\Tr\left( \rho_{AB}^{ij}~P_0~\rho_{AB}^{ij}~P_1\right) -2\Tr\left(P_0~\rho_{AB}^{ij}~P_1~P_0~\rho_{AB}^{ij} \right)  -2\Tr\left(P_0~\rho_{AB}^{ij}~P_0~P_1~\rho_{AB}^{ij}\right)\\
 &= 
\Tr\left(~\rho_{AB}^{ij}~\rho_{AB}^{ij}~P_0~P_1\right)+ \Tr\left(P_0~\rho_{AB}^{ij}~\rho_{AB}^{ij}~P_1\right)  +\Tr\left( P_0~\rho_{AB}^{ij}~\rho_{AB}^{ij}~P_1 \right) +\Tr\left( ~\rho_{AB}^{ij}~\rho_{AB}^{ij}~P_0~P_1 \right)
 +4\Tr\left(\rho_{AB}^{ij} P_0 \rho_{AB}^{ij} P_1   \right)  \\&\qquad  -2\Tr\left(~\rho_{AB}^{ij}~P_0~ \rho_{AB}^{ij}~P_0~P_1\right)-2\Tr\left( \rho_{AB}^{ij}~P_0~\rho_{AB}^{ij}~P_1 \right) -2\Tr\left(~P_0~\rho_{AB}^{ij}~P_0~\rho_{AB}^{ij}~P_1\right) \\&\qquad -2\Tr\left( \rho_{AB}^{ij}~P_0~\rho_{AB}^{ij}~P_1\right) -2\Tr\left(~P_0~\rho_{AB}^{ij}~P_0~\rho_{AB}^{ij}~P_1 \right)  -2\Tr\left(~\rho_{AB}^{ij}~P_0~\rho_{AB}^{ij}~P_0~P_1\right)\\
 &= 2~\Tr\left(~\rho_{AB}^{ij}~\rho_{AB}^{ij}~P_0~P_1\right)+ 2~\Tr\left(P_0~\rho_{AB}^{ij}~\rho_{AB}^{ij}~P_1\right)  
-4\Tr\left(~\rho_{AB}^{ij}~P_0~\rho_{AB}^{ij}~P_0~P_1\right)-4\Tr\left(~P_0~\rho_{AB}^{ij}~P_0~\rho_{AB}^{ij}~P_1\right) 
\end{aligned}
\end{equation}
The four trace term can't be solve as the earlier three cases. $P_1=Q'(\phi_A^i) \times \I = Q'(\theta) \;=\; \frac{1}{2}\begin{pmatrix}
 -\sin\theta & \cos\theta \\\cos\theta & \sin\theta
\end{pmatrix}$ 
is a Hermitian operator but not a Positive semi-definite operator(eigenvalues are $\pm \frac{1}{2})$.
\begin{equation}
    \begin{aligned}
        &(\I -P_1) \succeq 0 \quad \left(\text{ eigenvalues are } \frac{1}{2} \text{ and } \frac{3}{2}\right)\\
        & \I \succeq P_1 \\
        & \rho_{AB}^{ij}~\rho_{AB}^{ij}~P_0~ \nsucceq ~\rho_{AB}^{ij}~\rho_{AB}^{ij}~P_0~P_1 \quad \left( Y = ~\rho_{AB}^{ij}~\rho_{AB}^{ij}~P_0~ \text{ does not commute with } P_1\right)
    \end{aligned}
\end{equation}
Since $~\rho_{AB}^{ij}~\rho_{AB}^{ij}~P_0~$ and $P_1$ does not commute we can't use the above partial ordering but since $P_1$ is hermitian we can have spectral double sided spectral bound.
\begin{equation}
    \lambda_{min}(H)~\I \preceq H \preceq \lambda_{max}(H)~\I
\end{equation}
Equivalently we can write,
\begin{equation}
\lambda_{min}(H)~\Tr(Y) \leq \Tr (YH) \leq \lambda_{max}(H)~\Tr(Y)     
\end{equation} 
for $Y \succeq 0$. Alternatively, $(\I-P_1)$ is a Hermitian so we can the following spectral bound.
\begin{equation}
\begin{aligned}
    & \lambda_{min}((\I-P_1))~\Tr(Y) \leq \Tr (Y(\I-P_1)) \leq \lambda_{max}((\I-P_1))~\Tr(Y)\\
     &\frac{1}{2} \Tr(Y) \leq \Tr(Y) - \Tr(YP_1) \leq \frac{3}{2} \Tr(Y)\\
\end{aligned}
\end{equation}
The above relation requires $Y$ to be a positive semi definite matrix. This possess a serious flaw. As per the given setting this requires the complete knowledge of the density matrix and the projector operator which compromises the generality of the proof.
Alternatively we can can be upper bounded using the Holder inequality \cite{watrous2018theory} as,
\begin{equation}
\begin{aligned}
&\left|\Tr\left(~\rho_{AB}^{ij}~\rho_{AB}^{ij}~P_0~P_1\right)\right|\\
&\leq \Big\| ~\rho_{AB}^{ij}~\rho_{AB}^{ij}~P_0 \Big\|_1 \Big\| P_1\Big\|_{\infty}\\
&\leq \frac{1}{2} \Big\| (\rho_{AB}^{ij})^2 \Big\|_1 ~\Big\| P_0 \Big\|_{\infty}\\
&\leq \frac{1}{2} \Tr(\rho_{AB}^{ij})\\
&\leq \frac{1}{2}
\end{aligned}
\end{equation}
Similarly,
\begin{equation}
    \begin{aligned}
        &\left| \Tr\left(P_0~\rho_{AB}^{ij}~\rho_{AB}^{ij}~P_1\right)\right| \\
        &\leq \Big\|P_0~\rho_{AB}^{ij}~\rho_{AB}^{ij} \Big\|_1 \Big\| P_1 \Big\|_{\infty}\\
        &\leq \frac{1}{2}
    \end{aligned}
\end{equation}
Also,
\begin{equation}
\begin{aligned}
&\left|\Tr\left(~\rho_{AB}^{ij}~P_0~\rho_{AB}^{ij}~P_0~P_1\right)\right|\\
&\le \big\|\rho_{AB}^{ij}P_0\rho_{AB}^{ij}\big\|_1 \,\big\|P_0P_1\big\|_\infty
\\
&\le \big\|\rho_{AB}^{ij}\,P_0\big\|_2 \,\big\|\rho_{AB}^{ij}\big\|_2 \,\big\|P_0P_1\big\|_\infty
\\
&\le \big\|\rho_{AB}^{ij}\big\|_2 \,\big\|\rho_{AB}^{ij}\big\|_2 \,\|P_0\|_\infty\,\|P_1\|_\infty
\\
&= \|\rho_{AB}^{ij}\|_2^2 \,\|P_0\|_\infty\,\|P_1\|_\infty
= \Tr\!\big((\rho_{AB}^{ij})^2\big)\cdot 1\cdot\frac12
\\
&\le \frac12.
\end{aligned}    
\end{equation}
Finally for \(\Tr\big(P_0\rho_{AB}^{ij}P_0\rho_{AB}^{ij}\,P_1\big)\)
\begin{equation}
    \begin{aligned}
& \left|\Tr\big(P_0\rho_{AB}^{ij}P_0\rho_{AB}^{ij}\,P_1\big)\right|\\
&\le \big\|P_0\rho_{AB}^{ij}P_0\rho_{AB}^{ij}\big\|_1\,\big\|P_1\big\|_\infty\\
&\le \big\|P_0\rho_{AB}^{ij}\big\|_2\,\big\|P_0\rho_{AB}^{ij}\big\|_2\,\big\|P_1\big\|_\infty\\
&\le \big\|\rho_{AB}^{ij}\big\|_2^2\,\|P_0\|_\infty^2\,\big\|P_1\big\|_\infty\\
&\le \Tr\!\big((\rho_{AB}^{ij})^2\big)\cdot 1\cdot\frac12\\
&\le \frac12.
\end{aligned}
\end{equation}
Coming back to evaluating $\Tr(B_1)$
\begin{equation}
\Tr(B_1) = 2x + 2y - 4u - 4v, \quad |x|, |y|, |u|, |v| \le \tfrac12,
\end{equation}
the worst-case bounds are
\begin{equation}
\max \Tr(B_1) = 2\!\left(\tfrac12\right) + 2\!\left(\tfrac12\right) - 4\!\left(-\tfrac12\right) - 4\!\left(-\tfrac12\right) = 6,
\end{equation}
\begin{equation}
\min \Tr(B_1) = 2\!\left(-\tfrac12\right) + 2\!\left(-\tfrac12\right) - 4\!\left(\tfrac12\right) - 4\!\left(\tfrac12\right) = -6.
\end{equation}
So,
\begin{equation}
{|\Tr(B_1)| \le 6}.
\end{equation}
A tighter bound exists but for that an additional assumption is needed. 
\begin{equation}
\text{supp}(\rho) \subseteq \text{ran}(P_0)
\end{equation}
Support of $\rho$ is the span of all eigenvectors with non-zero eigenvalues. Range of the projector $P_0$ is span of all nonzero columns of $P_0$.
Following two relation can directly be followed from the given assumption.
\begin{equation}
    \begin{aligned}
        &\rho |\psi\rangle = \lambda |\psi\rangle \quad \forall \quad (\lambda >0) \quad \implies \psi \in \text{ran}(P_0)\\
        &P_0 |\psi \rangle = |\psi\rangle \quad \forall|\psi\rangle \in \text{supp}(\rho) \quad (P_0 \text{ act as identity operator to } \psi)
    \end{aligned}
\end{equation}
 For any vector $|\phi\rangle$, decompose $\rho |\phi\rangle$:
    \begin{equation}
        \begin{aligned}
        \rho |\phi\rangle &= \sum_{\lambda_k > 0} \lambda_k \langle\psi_k|\phi\rangle |\psi_k\rangle + \sum_{\lambda_k = 0} 0 \cdot \langle\psi_k|\phi\rangle |\psi_k\rangle\\
    &= \sum_k \lambda_k \langle\psi_k|\phi\rangle|\psi_k\rangle, \quad \lambda_k > 0, \quad |\psi_k\rangle \in \text{supp}(\rho)\\
    P (\rho |\phi\rangle) &= P \left( \sum_k \lambda_k \langle\psi_k|\phi\rangle|\psi_k\rangle \right)\\ &= \sum_k \lambda_k \langle\psi_k|\phi\rangle P |\psi_k\rangle \\ &= \sum_k \lambda_k \langle\psi_k|\phi\rangle|\psi_k\rangle \\ &= \rho |\phi\rangle \quad \forall |\phi\rangle\\
    P \rho &= \rho
        \end{aligned}
    \end{equation}
\begin{equation}
    \begin{aligned}
\rho\dagger &=(P\rho)\dagger\\ &= \rho^\dagger P^\dagger \\ 
    & = \rho  P \\
     &= \rho    
    \end{aligned}
\end{equation}
From the last two equation,
\begin{equation}
    \rho = P\rho =\rho P = P \rho P 
\end{equation}
Now using the above relation $\Tr(B_1)$ bound can be tighten using the third term $\Tr(\rho_{AB}^{ij} P_0 \rho_{AB}^{ij} P_0 P_1).$
\begin{equation}
    \begin{aligned}
& \Tr(\rho_{AB}^{ij} P_0 \rho_{AB}^{ij} P_0 P_1) \\
&= \Tr((P_0 \rho_{AB}^{ij} P_0) P_0 (P_0 \rho_{AB}^{ij} P_0) P_1) \quad (\text{substitute } \rho_{AB}^{ij} = P_0 \rho_{AB}^{ij} P_0) \\
&= \Tr(P_0 (\rho_{AB}^{ij} P_0 \rho_{AB}^{ij}) P_0 P_1) \\
&= \Tr((\rho_{AB}^{ij} P_0 \rho_{AB}^{ij}) P_0 P_1 P_0)\\
&= \Tr((\rho_{AB}^{ij} P_0 \rho_{AB}^{ij})(P_0 P_1 P_0)).
    \end{aligned}
\end{equation}
Now, \(P_0 = Q(\theta)\), \(P_1 = Q'(\theta)\) we have,
\begin{equation}
    \begin{aligned}
        &P_0^2 = P_0\\
        &P_0P_0 = P_0\\
        &P_1P_0 + P_0P_1 = P_1\\
        &P_0P_1P_0 + P_0P_0P_1 = P_0P_1\\
        &P_0P_1P_0 +P_0P_1 = P_0P_1\\
        &P_0P_1P_0 = 0
    \end{aligned}
\end{equation}
thus, 
\begin{equation}
\Tr(B_1) = 2x + 2y - 4v, \quad |x|, |y|, |v| \le \tfrac12
\quad\Rightarrow\quad
{|\Tr(B_1)| \le 4}.
\end{equation}

Since $\epsilon_0 >0$ is a very small quantity, $\epsilon_0^2 << \epsilon_0$ we have
$B = B_0 + \epsilon_0B_1 + \mathcal{O}(\epsilon_0^2)$.
Now, from \eqref{eqnF.105},
\begin{equation}
    \begin{aligned}
        &\left|h(\lambda,\phi_A^i,\rho_{AB}^{ij}) - h(\lambda,\phi_A^i+\epsilon_0,\rho_{AB}^{ij})\right| \\
\leq{}&(1-\lambda) \left[\Tr( A + B) \right] \\
\leq{}&(1-\lambda) \left[ \Tr(A_0) + \Tr(A_1) + \Tr(B_0) + \epsilon_0\Tr(B_1)  \right]\\
\leq{}&(1-\lambda)[1+1+2+\epsilon_04]\\
\leq{}&(1-\lambda)[4+4\epsilon_0+\mathcal{O}(\epsilon_0^2)]
    \end{aligned}
\end{equation}
The above quantity is $\mathcal{O}(1)$ to make it $\mathcal{O}(\epsilon)$ we can use the result $\rho = P\rho = \rho P = P \rho P$.
\begin{equation}
    \begin{aligned}
        \Tr(A_0) &= \Tr(~\rho_{AB}^{ij}~(\I-P_0) ~\rho_{AB}^{ij})\\
        &= \Tr(~(\I-P_0) ~\rho_{AB}^{ij})\\
        &= \Tr(~(\I-P_0) P_0~\rho_{AB}^{ij}~P_0)\\
        &= \Tr(P_0~\rho_{AB}^{ij}~P_0 - P_0^2~\rho_{AB}^{ij}~P_0)\\
        &= 0
    \end{aligned}
\end{equation}
Similarly,
\begin{equation}
    \begin{aligned}
        \Tr(B_0) &= 2(\Tr(P_0~\rho_{AB}^{ij}~\rho_{AB}^{ij}) - \Tr((P_0\rho_{AB}^{ij}P_0\rho_{AB}^{ij})))\\
&= 2(\Tr(P_0\rho_{AB}^{ij}~P_0\rho_{AB}^{ij}P_0) - \Tr((P_0~\rho_{AB}^{ij}~P_0~\rho_{AB}^{ij})))\\
&=2(\Tr(P_0\rho_{AB}^{ij}P_0\rho_{AB}^{ij}) - \Tr((P_0~\rho_{AB}^{ij}~P_0~\rho_{AB}^{ij})))\\
&=0
    \end{aligned}
\end{equation}
where $P_0 = Q(\phi_A^i)\otimes\I$
Thus finally,
\begin{equation}
\label{eqnF.145}
\boxed{    \begin{aligned}
        &\left|h(\lambda,\phi_A^i,\rho_{AB}^{ij}) - h(\lambda,\phi_A^i+\epsilon_0,\rho_{AB}^{ij})\right| \\
\leq{}&(1-\lambda)[4\epsilon_0+\mathcal{O}(\epsilon_0^2)]
    \end{aligned}}
\end{equation}

\subsubsection{Modified optimization problem incorporating the change in Alice's angle}
Now we are at a point where we can lower bound the maximal error in $k^{th}$ segment centered around $\phi_{A_k}^i$. Using Eqs. \eqref{eqnF.145} and \eqref{eqn5.70n} in the modified objective function given in \eqref{eqn5.26}, we get the modified optimization problem as,

\begin{equation}
\boxed{
    \begin{aligned}
n^*(S_{ij}) \geq \inf \quad & \lambda \left\| \rho_{AB}^{ij} - \Lambda_0[\rho_{AB}^{ij}] \right\|_F^2 + (1-\lambda)\left\| \rho_{AB}^{ij} - \Lambda_1[\rho_{AB}^{ij}] \right\|_F^2 \\&+ \frac{\mu}{2} \|\rho_{AB}^{ij}\|_F^2 \text{ }-2(1-\lambda)\epsilon_0 \\
\text{s.t.} \quad &\text{Tr}\left(\rho_{AB}^{ij} ~CHSH(\phi_A^i,\phi_B^j)\right) = S_{ij} \text{ }-2\epsilon_0 \\
& \phi_A^i, \phi_B^j \in [0,\pi/2],\\
    &\rho_{AB}^{ij} \succeq 0 \\ &\text{Tr}(\rho_{AB}^{ij}) = 1
\end{aligned}
}
\end{equation}
By averaging multiple instances or scenarios in the optimisation problem, the maximum deviation of the objective function's value from its expected or central tendency is reduced. Specifically, in this case, the maximum deviation has been halved from $\epsilon_0 = 4$ to $\epsilon_0' = 2$. This reduction occurs because averaging tends to smooth out extreme values. Given that the optimal value lies within a symmetric interval of $\phi_{A_k}^i \pm \epsilon_0$, averaging over multiple such intervals can lead to a smaller overall range of uncertainty for the optimal value.

\section{Creating a convex hull from all two qubit functions $C^*(S)$ for all $S \in (2,2\sqrt{2}]$ }
In Eq. \eqref{eqn89}, it had been shown that the function $C^{*}(S)$ can be lower bounded as an integral of the function of $C^{*}(S)$ over two qubit blocks, $C^*_{\mathbb{C}^{4 \times 4}}(S')$, over the range $S \in (2,2\sqrt{2}]$, i.e.,
$C^*(S)  \geq \int_{S'=2}^{2\sqrt{2}}\eta(dS')\cdot C^*_{\mathbb{C}^{4 \times 4}}(S') $ such that $\quad \eta([2,2\sqrt{2}]) \leq 1, \quad
        \eta \geq 0, \quad \text{and}
        \int_{S'=2}^{2\sqrt{2}}  \eta(dS')S' = S$.
Moreover, it has been showed that each such two qubit function $C^*_{\mathbb{C}^{4 \times 4}}(S')$ can further be lower bounded through a strongly convex objective function $n^*(S_{ij})$, using a modified Pinsker's inequality\cite{Schwonnek_2021} and the results from Eqs. \eqref{eqnD.6} \eqref{eqn5.26}.
Now given values of $C^*_{\mathbb{C}^{4 \times 4}}(S)$ for all $S\in(2,2\sqrt{2}]$, one needs a convex function, say, ${\overline{C}(S)}$\cite{Schwonnek_2021} that would essentially give,
\begin{equation}
    C^*_{\mathbb{C}^{4 \times 4}}(S_{ij}) \geq {\overline{C}(S_{ij})}
\end{equation} for $ij^{th}$ block.
Thus incorporating this into \eqref{eqn89} as,
\begin{equation}
    \begin{aligned}
         C^*(S) 
        &\geq \int_{S'=2}^{2\sqrt{2}}\eta(dS')\cdot {\overline{C}(S')}\\
        s.t \quad &\eta([2,2\sqrt{2}]) \leq 1 \\
        &\eta \geq 0\\
        &\int_{S'=2}^{2\sqrt{2}} \eta(dS')S' = S
    \end{aligned}
\end{equation}
gives \begin{equation}
    C^*(S) \geq {\overline{C}(S)}
\end{equation}
The final task is to prove the existence of one such function ${\overline{C}(S)}$ that lower bounds $C^*(S)$.\\\\
\textit{Theorem 6: }$n^*(S)$ is a valid lower bound for $C^*(S)$.\\\\
\textit{Proof: }
\subsubsection{ Strong Convexity of the Objective Function}
The optimization problem defining \(n^*(S)\) is given as,
\begin{equation}
\begin{aligned}
&n^*(S_{ij}) = \inf \left\{ 
\lambda \left\| \rho_{AB}^{ij} - \Lambda_0\left[\rho_{AB}^{ij}\right] \right\|_F^2 
+ (1-\lambda)\left\| \rho_{AB}^{ij} - \Lambda_1\left[\rho_{AB}^{ij}\right] \right\|_F^2 
+ \frac{\mu}{2} \left\| \rho_{AB}^{ij}\right\|_F^2 
\right\}\\    
&s.t. ~ \text{Tr}\left(\rho_{AB}^{ij}~ CHSH(\phi_A^i, \phi_B^j)\right) = S_{ij}.
\end{aligned}
\end{equation}
Now, the given objective function involving the Frobenius norm \(\left\| \cdot \right\|_F^2\) is strongly convex in \(\rho\). The term \(\frac{\mu}{2} \left\| \rho_{AB}^{ij} \right\|_F^2\) introduces \(\mu\)-strong convexity, and the linear constraints of the optimization problem preserve the strong convexity on the feasible set.
For fixed \(S_{ij}\), the objective function is strongly convex in \(\rho\). The parameterized problem's solution function \(n^*(S_{ij})\) inherits convexity. Moreover, strong convexity implies quadratic dependence on the perturbations in \(S_{ij}\).
Thus, \(n^*(S)\) is {strongly convex} in \(S\) for $S\in(2,2\sqrt{2}]$
\subsubsection{ Application of Jensen's inequality}
For the measure \(\eta\) with,
\begin{equation}
\int_{2}^{2\sqrt{2}} \!\eta(dS') = 1, \quad \eta \geq 0, \quad \int_{2}^{2\sqrt{2}} \!\eta(dS') S' = S    
\end{equation}
Now application of Jensen's inequality for strongly convex functions yields,
\begin{equation}
\int_{2}^{2\sqrt{2}} \!\eta(dS') n^*(S') \geq n^*\left(\int_{2}^{2\sqrt{2}} \!\eta(dS') S'\right) + \frac{\mu}{2} ~\text{Var}_\eta(S')    
\end{equation}
Since \(\text{Var}_\eta(S') \geq 0\),
\begin{equation}
\int_{2}^{2\sqrt{2}} \!\eta(dS') n^*(S') \geq n^*(S).    
\end{equation}
Substituting \(\overline{C}(S') = n^*(S')\) into the original inequality, we get,
\begin{equation}
C^{*}(S) \geq \int_{2}^{2\sqrt{2}} \!\eta(dS') n^*(S') \geq n^*(S)
\end{equation}
Therefore, \(\overline{C}(S) = n^*(S)\) is a valid convex lower bound.
\qed
\bibliographystyle{alpha}
\bibliography{references}

\end{document}